\documentclass[sigconf]{acmart}
\usepackage[utf8]{inputenc}
\usepackage{booktabs,longtable}  
\usepackage{tabularx}    
\usepackage{ragged2e}    
\newcolumntype{Y}{>{\RaggedRight\arraybackslash}X} 
\definecolor{canvasJR}{RGB}{237, 225, 197}   
\definecolor{canvasSL}{RGB}{229, 218, 242}   
\definecolor{canvasPD}{RGB}{245, 220, 228}  
\definecolor{canvasFG}{RGB}{225, 238, 225} 
\definecolor{codingBeige}{RGB}{240,232,216}
\usepackage{subcaption}
\renewcommand{\arraystretch}{1.2} 
\usepackage{enumitem}
\usepackage{tikz}
\usetikzlibrary{arrows.meta,positioning,fit,shadows.blur}
\usetikzlibrary{positioning,arrows.meta}
\usepackage{colortbl}
\usepackage{array}
\definecolor{gold}{RGB}{245,200,90}
\newcolumntype{P}{>{\raggedright\arraybackslash}p{3cm}}
\usetikzlibrary{matrix}
\newif\ifhighlight
\usepackage{soul}            
\usepackage{marginnote}      
\usepackage{etoolbox}  
\definecolor{revhl}{RGB}{255,245,150}  
\sethlcolor{revhl}

\AtBeginDocument{%
  }

\highlighttrue

\copyrightyear{2026}
\acmYear{2026}
\setcopyright{cc}
\setcctype{by}
\acmConference[CHI '26]{Proceedings of the 2026 CHI Conference on Human Factors in Computing Systems}{April 13--17, 2026}{Barcelona, Spain}
\acmBooktitle{Proceedings of the 2026 CHI Conference on Human Factors in Computing Systems (CHI '26), April 13--17, 2026, Barcelona, Spain}

\acmDOI{10.1145/3772318.3790719}
\acmISBN{979-8-4007-2278-3/2026/04}

\begin{document}

\title{Reimagining Wearable AR Gesture Design: Physical Therapy Reasoning in Everyday Contexts}

\author{Wei Wu}
\authornote{These authors contributed equally to this work.}
\affiliation{%
  \institution{Ghost Lab, Northeastern University}
  \city{Boston}
  \state{Massachusetts}
  \country{United States}
}
\email{wu.w4@northeastern.edu}

\author{Binyan Xu}
\authornotemark[1]
\affiliation{%
  \institution{Bouvé College of Health Sciences, Northeastern University}
  \city{Boston}
  \state{Massachusetts}
  \country{United States}
}
\email{xu.biny@northeastern.edu}

\author{Soonhyeon Kweon}
\affiliation{%
  \institution{Bouvé College of Health Sciences, Northeastern University}
  \city{Boston}
  \state{Massachusetts}
  \country{United States}
}
\email{kweon.s@northeastern.edu}

\author{Yujie Wang}
\affiliation{%
  \institution{Fine Art Department, Zhengzhou University}
  \city{Zhengzhou}
  \state{Henan}
  \country{China}
}
\email{wyj20080221@stu.zzu.edu.cn}

\author{Leanne Chukoskie}
\affiliation{%
  \department{Bouvé College of Health Sciences; College of Arts, Media and Design}
  \institution{Northeastern University}
  \city{Boston}
  \state{Massachusetts}
  \country{United States}
}
\email{l.chukoskie@northeastern.edu}

\author{Casper Harteveld}
\affiliation{%
  \institution{College of Arts, Media and Design, Northeastern University}
  \city{Boston}
  \state{Massachusetts}
  \country{United States}
}
\email{c.harteveld@northeastern.edu}

\begin{abstract}
Lightweight Augmented Reality (AR) glasses are entering everyday use, expanding the scope of interaction design beyond short sessions. Yet, most gesture vocabularies still stem from VR headsets or early AR goggles, prioritizing recognizer accuracy while neglecting fatigue and social legibility. We worked with physical therapists (PTs) to reimagine gestures for daily AR, leveraging their expertise in safe, sustainable movement. From a review of 104 applications, we derived 15 common gesture intents and built an on-device generator. Ten licensed PTs (\textit{M}=14.8 years of practice) then shaped these intents through three rounds: unaided gestures, PT-guided substitutions, and stage-aware card sorting. We contribute: (1) a PT-informed gesture translation method; (2) the Everyday-AR Golden Ergonomic Canvas; and (3) a stage-aware social legibility framework showing how gesture suitability shifts with social readability. Together, these contributions offer a recognizer-agnostic reference frame for constructing sustainable and socially coherent gesture vocabularies for lightweight AR glasses.

\end{abstract}



\begin{teaserfigure}
  \centering
  \includegraphics[width=\textwidth]{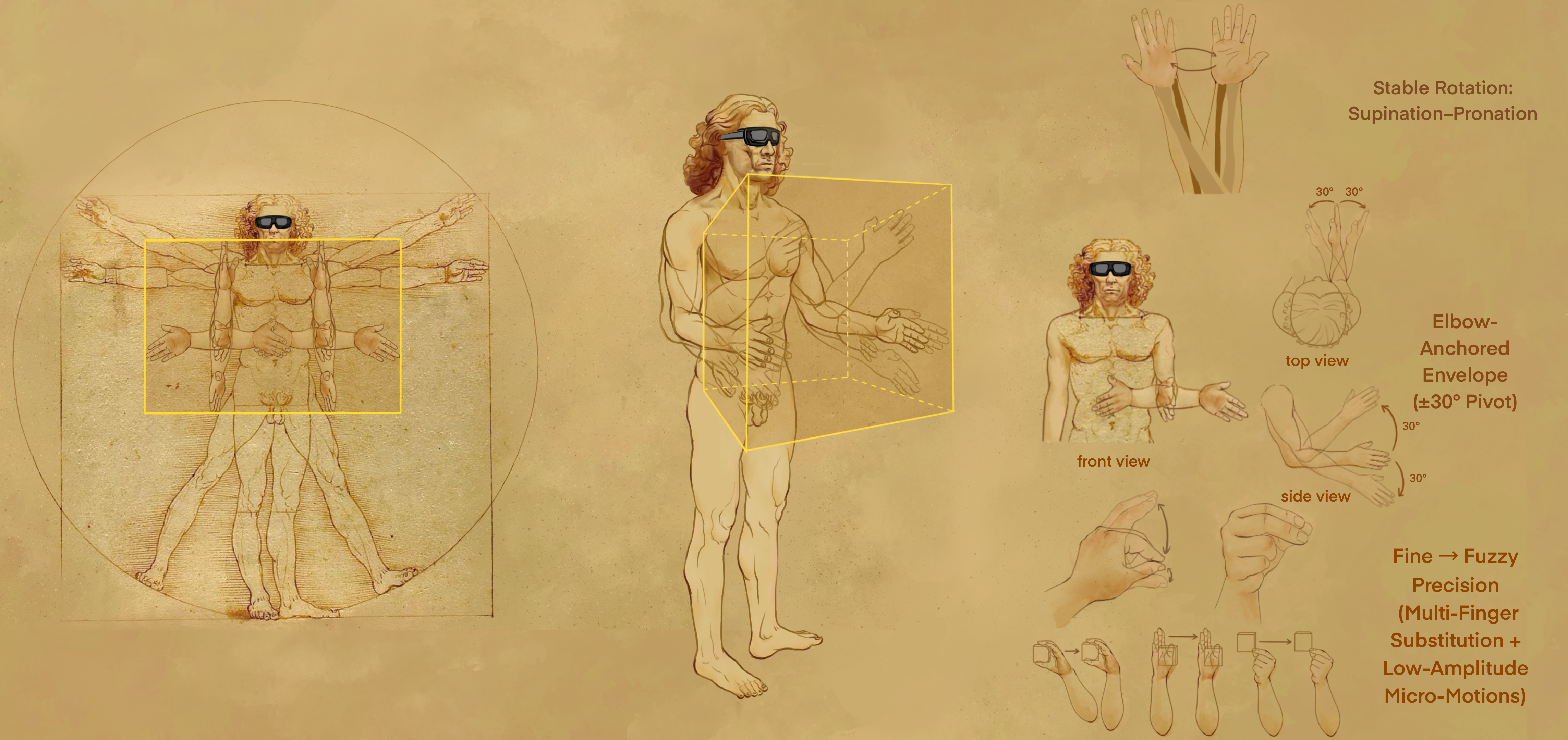}
  \caption{This illustration pays homage to Leonardo da Vinci’s Vitruvian Man by reinterpreting it as an Everyday-AR Golden Ergonomic Canvas: a unified depiction of the chest-level workspace, the elbow-anchored ±30° gesture envelope, and the fine–gross motor integration principles identified by physical therapists as foundational for sustainable, socially coherent gesture design in wearable AR. By embedding these rules within a classical anatomical form, the canvas also gestures toward a broader shift—how the arrival of everyday AR subtly reshapes long-standing ergonomic assumptions inherited from centuries of human–body representation.}
  \Description{Vitruvian-inspired ergonomic canvas for everyday AR gesture design.}
\end{teaserfigure}

\maketitle

\section{Introduction}

Extended reality (XR)—encompassing virtual, augmented, and mixed reality—has long been heralded as a transformative medium for learning, work, and entertainment, promising immersive and embodied experiences beyond traditional screens~\cite{marr2021extended,dutta2024immersive,aurelia2024immersive}. Studies highlight XR’s potential across behavioral science, rehabilitation, and education~\cite{schuermans2022msk,morimoto2022spinexr,burke2025higheredxr}. Yet, despite two decades of progress, the dominant goggles-and-headset paradigm has remained tethered to controlled, niche settings—typically youthful, affluent, able-bodied users performing short-duration tasks. Its bulk, isolation, and reliance on fine-motor-centric interaction vocabularies have systematically limited everyday, mobile, and socially embedded use~\cite{alhakamy2024xrindustry,ali2023xrhealthcare,luo2024xrealityenglish}.

Lightweight \emph{AR glasses} are emerging as the most promising gateway for bringing XR into everyday contexts.
Unlike passthrough HMDs, which simulate AR via video feeds, or HUD-based ``AI glasses’’ that overlay notifications without spatial registration~\cite{Fuchs1998OVST,RayBanMetaAI,MetaAIGlassesSite}, optical AR glasses anchor graphics directly into the world while preserving peripheral awareness. Their small form factor, silent operation, and on-face spatial stability open possibilities for mobile work, social communication, rehabilitation, and casual play~\cite{snapAWE2025,reutersSnap2025,xu2025ptmovementlogics,wu2025ghostgait}. Crucially, lightweight optical AR glasses do not merely shrink VR; they introduce a fundamentally different interaction condition: long-duration, low-friction, socially exposed, often one-handed embodied input. However, despite hardware advances, these devices still lack a viable interaction grammar for everyday use.

The field is approaching a critical inflection point. At AWE~2025, Snap announced that next-generation Spectacles will ship to consumers in 2026~\cite{snapAWE2025,reutersSnap2025}, while Meta positioned Orion as ``the future of wearables,'' with a new developer kit expected at Meta Connect~2025~\cite{metaOrion,uploadvrOrion2025}. These signals reveal a widening gap: hardware is accelerating toward everyday adoption, while interaction paradigms remain anchored in VR and enterprise logic—optimized for recognition robustness (i.e., maximizing recognizer stability and visibility) rather than all-day comfort, social readability, or sustainable repetition~\cite{Wobbrock2009UserDefined,Annenberg1989Gesture,Ashbrook2010MAGIC}. Although recent consumer AR products (e.g., Vision Pro, Ray-Ban Meta) have introduced more diverse gesture styles—including resting-posture microgestures and wrist-based flicks—there remains no coherent, ergonomically grounded interaction grammar for lightweight optical AR glasses~\cite{metaNature2025}. In this paper, we specifically target spatial, world-anchored AR interfaces on optical see-through glasses, rather than 2D GUI overlays or front-facing air-tap systems. Within this scope, many available gestures still reflect VR-era and industrial AR conventions: large mid-air sweeps optimized for recogniser visibility, pinches shaped by occlusion constraints, and microgestures introduced in response to limitations in hand-tracking fidelity~\cite{pfeuffer2024gazepinch}. This fragmented design landscape underscores the need for a sustainable, socially legible gesture framework as AR glasses transition toward everyday use. 
Prior gesture design work spans expert-driven mappings and computation-driven authoring~\cite{Burnett2013Swipe,Hinckley2011SensorSynaesthesia}, 
as well as user-elicited “natural” gestures~\cite{Wobbrock2009UserDefined,Garzotto2013Touchless}. These studies generated valuable corpora, but rarely addressed two challenges unique to everyday AR:  
(1) endurance and joint safety, as gestures must be repeated across long sessions~\cite{HincapieRamos2014CE}; and  
(2) social visibility, as wearers navigate frontstage, backstage, and off-stage situations~\cite{Rico2010Social,Serrano2014HandToFace,Goffman1959}.  
What remains missing is a principled way to \emph{translate} intuitive but fragile gestures—often surfaced in elicitation studies—into biomechanically stable, range-safe, and socially coherent forms suitable for diverse everyday settings~\cite{pfeuffer2024gazepinch}.

To address this gap, we reframe gesture design for AR glasses around two intersecting dimensions: ergonomic sustainability and stage-aware social legibility. We engage licensed physical therapists (PTs) as expert design partners. Beyond rehabilitation, PTs specialize in diagnosing unsafe or fatiguing movement patterns and prescribing function-preserving substitutions that protect joints, minimize load, and maintain intended action. Their reasoning is uniquely suited to authoring gestures that remain viable as AR glasses leave controlled labs and enter workplaces, transit, clinics, and homes.

We developed a three-stage, PT-informed elicitation pipeline. First, we reviewed 104 publicly listed Spectacles applications, coding their gesture vocabularies to distill 15 recurring \emph{gesture intents}. Second, we implemented these intents in an on-device generator to elicit intuitive actions without recognizer bias. Third, ten licensed PTs (M~=~14.8~years’ practice) iteratively refined these gestures through intuitive execution, ergonomic substitution, and a card-sorting activity through which PTs’ reasoning revealed six emergent everyday contexts (front/back/off-stage × individual/public)~\cite{xia2022iterative}. This pipeline reframes gesture elicitation not as a search for the “most natural’’ gesture, but as a negotiation between intent, joint safety, and social performance.

From this investigation, we advance gesture design for lightweight AR glasses along three dimensions.

First, we offer a PT-informed gesture translation
method that integrates licensed physical therapists as expert diagnosticians, enabling systematic identification of fatigue risks and graded substitution of intuitive but fragile mid-air gestures. This method provides a replicable, recognizer-agnostic pipeline for reorganizing gestures around joint safety,
proximal stability, and sustainable repetition.

Second, our qualitative analysis surfaces four recurrent clinical principles—Joint–Rotation Substitution, Shoulder-Line Workspace Constraint, Proximal–Distal Stability, and Fine–Gross Motor Integration—that characterize
how therapists stabilize the proximal chain, redistribute load, and refine gesture sequencing. We operationalize these principles through the
\textit{Everyday-AR Golden Ergonomic Canvas}, a chest-level, elbow-anchored workspace synthesizing repeated clinical corrections into an interpretable scaffold for constructing sustainable gesture vocabularies.

Third, by integrating dramaturgical theory with PT reasoning, we develop a
stage-aware gesture sorting protocol that links ergonomic stability with social
legibility across six everyday AR contexts. This coupling highlights how
gesture amplitude, posture, and expressivity shift with impression management concerns as AR use transitions from private to public settings.

Together, these contributions position gesture design for AR glasses as an inquiry into how bodies, technologies, and audiences co-produce sustainable and socially attuned performances, offering actionable guidance for constructing mid-air interaction vocabularies that extend beyond controlled laboratory environments.
\section{Background and Related Work}

\subsection{Background: From Enterprise AR to Everyday AR Interaction}

Since the debut of Google Glass (2013), head-worn augmented reality has undergone several adoption waves across consumer, enterprise, and clinical domains. Google Glass initially represented the \emph{ideal of AR as lightweight, socially acceptable eyewear}, but early limitations in sensing, battery life, and privacy acceptance constrained its public uptake~\cite{time2015glass,glasswiki}. After the Explorer program ended amid technical and privacy concerns~\cite{time2015glass,glasswiki}, Google Glass shifted into limited enterprise use, largely in logistics settings~\cite{schan2019smartglasses,gajsek2019smart,murauer2019fullshift}. As spatial mapping, tracking, and interaction capabilities expanded, AR devices were gradually forced away from this glasses-like ideal and toward bulkier, goggle-style headsets capable of supporting wider field-of-view optics, depth sensing, thermal dissipation, and onboard computation. This transition reflected engineering necessity but also widened the gap between AR’s \emph{conceptual promise as everyday eyewear} and its \emph{practical trajectory as workplace hardware}.

Enterprise-focused devices such as Microsoft HoloLens~2 and Magic Leap~2 further consolidated articulated hand tracking, eye tracking, and robust voice-based commands into interaction stacks optimized for structured, task-focused environments~\cite{mslearn_hands,mslearn_eyes,matveiuk2019xr,booth2019af,verge2022ml2,ml22022pr,zari2023magic,caruso2021eye,andersson2020developing}. While these devices demonstrated clear value, especially during the COVID-19 expansion of telehealth and telementoring~\cite{ong2021xrtelehealth,dinh2023artelemed,jms2024mror,nickel2022telestration}, their interaction vocabularies \textbf{remained anchored} in \emph{goggle-era design}: coarse, highly visible mid-air gestures and gaze+voice input suited for professional, stationary contexts.

In parallel, lighter-weight AR glasses have begun reshaping expectations for on-face computing in everyday environments. Snap’s creator-focused Spectacles (2021) introduced spatial AR authoring in outdoor and social settings~\cite{snap2021spectacles}, with subsequent iterations reinforcing this trajectory toward everyday AR~\cite{snap2026specs}. More visible momentum in daily life has come from \emph{AI glasses} (e.g., Meta Ray-Ban), which popularize on-face voice and AI assistance~\cite{meta2024rbai}. While these devices lack spatial anchoring and therefore fall outside our scope of AR interaction, their uptake reframes public expectations for eyewear computing. At the same time, Meta’s Orion prototypes signal a path toward truly see-through AR in familiar eyewear form factors~\cite{metaOrionTech,metaOrionNews,kaifosh2025neurointerface}. Recent HCI work echoes this shift, framing ``everyday AR'' as context-aware, socially embedded interaction paradigms~\cite{bowman2021everyday,bonner2023filters,tran2025wearable,simeone2023everyday}.

As AR shifts from controlled workplaces into offices, transit, schools, and parks, interaction demands fundamentally change. Enterprise deployments historically emphasized robustness under noise, clutter, and time pressure~\cite{haider2025ar,alrawi2025remote}, favoring gaze+voice commands and coarse mid-air gestures designed for recognizer reliability~\cite{mslearn_hands,mslearn_eyes}. Everyday AR instead requires interaction that is quieter, subtler, fatigue-resistant, socially legible, and appropriate while walking, conversing, or carrying objects.

Prior AR/VR work identifies several embodied tensions that intensify in everyday settings: mid-air input induces rapid shoulder fatigue and deteriorating fine-motor endurance~\cite{HincapieRamos2014CE,mizrahi2020neuromechanical}; occlusion and sensing limits motivate wrist- or finger-worn alternatives and finer-grained articulation~\cite{frl2021emg,li2025vibring,nguyen2023hands}; and public use raises dramaturgical concerns, as gesture visibility and acceptability depend on audience and setting~\cite{Rico2010Social,Serrano2014HandToFace}. These pressures also drive interest in multimodal, low-amplitude, or silent input, such as sEMG-based text entry.

Alongside these developments, several manufacturers have introduced \emph{fine-grained sensing alternatives}—smart rings, wrist-worn EMG bands, and low-amplitude microgestures—to reduce reliance on large VR-style arm motions. These modalities represent meaningful progress, but fine-motor input alone cannot satisfy the full range of everyday AR interaction requirements. Fine gestures fatigue quickly without proximal support, are difficult to perform while walking or multitasking, and often lack the \emph{social readability} required in public environments. Moreover, many world-anchored AR tasks---such as adjusting spatial objects, establishing reference frames, or stabilizing posture during mobile use---cannot be completed solely through microgestures. As physical therapists in our study repeatedly emphasized, sustainable interaction depends on a coordinated blend of \emph{fine and gross motor actions}, not the elimination of gross movement.

Taken together, these developments underscore a central challenge: as AR transitions into everyday life, gesture vocabularies inherited from the VR and industrial AR eras must be revisited and redesigned to support ergonomic sustainability, social legibility, and context-sensitive use. This need motivates our PT-informed investigation into gesture translation for lightweight, everyday AR glasses.

\subsection{Methods for Gesture Design}
Prior gesture vocabulary design work largely falls into three families: expert-driven, computation-driven, and user-driven (elicitation). 
First, \emph{expert-driven} methods rely on interaction or domain experts who, drawing on practice-based intuition, map physical actions to system commands; these are common in automotive and mobile contexts that combine gestures with touch~\cite{Burnett2013Swipe,Garzotto2013Touchless,Hinckley2011SensorSynaesthesia}. 
While efficient and cohesive, such methods have been criticized for sacrificing \emph{discoverability and intuitiveness}, making gesture vocabularies less friendly to novices~\cite{Wobbrock2009UserDefined}. 
Second, \emph{computation-driven} approaches use recognition and authoring tools to prioritize gestures that are easy for systems to detect and for designers to specify (e.g., demonstration-based prototyping and motion-feature toolchains)~\cite{Annenberg1989Gesture,Ashbrook2010MAGIC,Dey2004aCAPpella,Lu2013GestureStudio}; however, algorithmic pipelines often fail to account for socio-technical factors such as \emph{learnability, transferability, and social acceptability}. 
Third, \emph{user-driven} (elicitation) methods marshal “wisdom of the crowd’’ through participatory prompts and contextual tasks to derive vocabularies from observed behavior and perceived natural mappings; these typically excel in intuitiveness and discoverability but may overlook long-term \emph{ergonomic} suitability or cross-session consistency~\cite{Nielsen2004Procedure,Pyryeskin2012Comparing,Wobbrock2009UserDefined}.

Building on this landscape, we adopt a user-elicitation paradigm~\cite{Wobbrock2009UserDefined} while positioning \emph{physical therapists (PTs) as informed experts} on human movement and capability. 
Without presuming prior AR experience, we first invited PTs to propose \emph{natural, intuitive} gestures while thinking aloud about body mechanics. 
Next, PTs re-articulated and refined these gestures using \emph{clinical language and logic} to emphasize safety, endurance, and suitability for repeated use. 
Finally, we organized a classification exercise (inspired by card-sorting techniques) in which PTs grouped candidate gestures across different \emph{task--posture--setting} combinations, discussing not only \emph{usability} and \emph{social acceptability}, but also their \emph{progression potential}---that is, how a gesture could be scaled in range or complexity as users gain strength, skill, or situational confidence. 
This three-stage, PT-guided elicitation pathway balances \emph{intuitive discoverability} with \emph{clinically grounded, parameterizable progression}, yielding a transferable, explainable, and safety-bounded baseline for gesture vocabularies in everyday AR.

To further structure this analysis, we draw on an iterative, \emph{factor-centric} framework proposed in prior work~\cite{xia2022iterative}, which synthesizes 13 situational, cognitive, physical, and system factors that shape gesture selection. Although originally not developed for \emph{everyday} AR settings, this framework provides a useful scaffold for articulating why PT-informed revisions matter. In particular, we foreground Context, Social Acceptability, and Ergonomics, complemented by PT-informed substitutions and place-sensitive adaptations, to anchor gesture evaluation within the realities of public and private settings.

\subsection{Embodied Interaction and Soma-Based Perspectives on Movement}

Beyond technical or ergonomic framings, work in \emph{embodied interaction} and 
\emph{soma design} highlights that gestures are not merely input events but lived, 
expressive bodily actions. Early HCI explorations, such as Soft(n)’s investigation 
into somaesthetic touch~\cite{hook2009softn} and Schiphorst’s movement-centered 
design methods~\cite{schiphorst2013moving}, foreground bodily experience, rhythm, 
and sensory awareness as primary design materials. Höök and colleagues~\cite{hook2017embodied} further 
articulate principles for “embodied being-in-the-world,” arguing that technologies 
should engage with users’ perceptual, affective, and kinesthetic experience rather 
than treating the body as a neutral control surface.

These perspectives resonate with our findings in two ways. First, they emphasize 
that gestures---especially in public or semi-public settings---carry expressive and 
dramaturgical qualities, influencing how users negotiate visibility, privacy, and 
identity during everyday AR use. Second, somaesthetic work stresses attunement and 
intentionality in movement, which parallels the clinical reasoning applied by 
physical therapists (PTs) in our study: PTs intentionally modulate joint coupling, 
base-of-support, and amplitude to scaffold sustainable, meaningful, and repeatable 
action.

However, while soma-based approaches illuminate the experiential qualities of 
movement, they seldom prescribe systematic pathways for translating fragile or 
socially awkward motions into safer, joint-stable alternatives suited for everyday 
wearable AR contexts. Our PT-guided elicitation and substitution pathway therefore 
complements soma design by operationalizing embodied principles into 
\emph{reproducible, ergonomically grounded} gesture authoring for daily use.

\subsection{Ergonomics and Clinical Perspectives on AR Gesture Design}

Mid-air input imposes cumulative load on the shoulder and upper limb. 
The \emph{Consumed Endurance} metric~\cite{HincapieRamos2014CE} quantified fatigue and inspired posture-aware planning. 
Later studies caution against overhead or cross-body motions for prolonged use and highlight strain from unsupported fine-motor repetition~\cite{mizrahi2020neuromechanical}. 
Industrial ergonomics similarly advises stabilizing joints and distributing load~\cite{feg2020ergonomics}, while social acceptability further constrains the amplitude and locus of movement in public~\cite{Rico2010Social,Serrano2014HandToFace}. 
Together, this work diagnoses where fatigue and awkwardness arise, but rarely prescribes how to systematically redesign gesture vocabularies through structured substitution.

Rehabilitation research provides a complementary lens. 
Beyond the basic fine–gross distinction (e.g., PDMS-2, BOT-2~\cite{bruininks2005bot2,casesmith2014occupational}), therapists attend to how movements are sequenced, scaled, and progressed over time, balancing safety with engagement~\cite{stergiou2006optimal,schwartz2022attunement}. 
HCI collaborations such as \emph{Spellcasters}~\cite{duval2022spellcasters}
show how clinical reasoning can inform playful or productive gestures, and \emph{Ability-Based Design} argues for starting from users’ actual abilities rather than idealized bodies~\cite{wobbrock2011ability}. 
Yet, PT-informed motor frameworks are seldom embedded across the full gesture-design pipeline—from elicitation, through classification, to evaluation—or explicitly tied to social staging and session length.

Across these strands, two gaps remain clear.  
First, there is a need for a \emph{factor-centric, place-sensitive pipeline} that links ergonomic safety, social acceptability, and contextual visibility when authoring gestures for everyday AR.  
Second, there is a lack of a principled \emph{substitution framework} to re-author fragile or fatiguing motions into joint-stable, socially legible alternatives.  
Addressing these gaps motivates our study, which integrates physical therapists’ movement expertise into gesture elicitation and analysis, examines how substitutions perform across public and private contexts, and evaluates them for ergonomic sustainability and social clarity.
\section{Methods}

\subsection{Study Overview}

We conducted a three-stage, in-person study to develop clinically grounded, 
context-sensitive gesture vocabularies for everyday AR glasses. 
Stage~1 built a corpus of \emph{gesture intents}—high-level interaction goals such as 
\emph{Rotate}, \emph{Throw}, or \emph{Delete}—from existing AR applications. 
Stage~2 deployed an on-device prototype presenting these intents as animated cues, enabling 
gesture elicitation without biasing participants toward any existing recognizer or platform. 
Stage~3 invited licensed physical therapists (PTs) to reinterpret the intents across everyday 
contexts, using clinical reasoning to generate sustainable and socially legible alternatives.

PTs were selected not as stand-ins for general users, but as 
\emph{movement-safety experts}.  
Unlike ergonomics or sports-science specialists, who emphasize universal rules, PTs treat 
movement as an individualized, contextually variable activity grounded in anatomy, 
kinesiology, and fatigue management~\cite{APTAeducation2020,WHOrehab2017}. 
Their expertise allowed us to examine why certain ``natural'' gestures are fragile and how 
they can be re-authored into durable, safe, and context-adaptive forms suitable for 
everyday AR.
The Methods section proceeds as follows: we first describe how the gesture-intent 
taxonomy was derived (Section~3.2), then detail the prototype used for elicitation, 
followed by participant recruitment, study procedure, and our two-pass analysis.

\subsection{Gesture Corpus}

To ground elicitation in real usage, we first reviewed 104 publicly listed 
Snap Spectacles AR applications (August~2025). 
Two authors independently coded each app’s primary category 
(system, utility, experience, game) and the presence of high-level 
\emph{gesture intents}—interaction goals such as \emph{Rotate}, 
\emph{Throw}, or \emph{Delete}. 

For each application, we identified the \emph{gesture intent} as the 
functional goal of the interaction independent of its concrete gesture form, 
and compared these intents against established 2D (touch/pen) and 3D (VR/AR) 
interaction taxonomies.  
Because everyday AR glasses blend both 2D system commands 
(e.g., \emph{Select}, \emph{Delete}, \emph{Switch}) and 3D spatial 
manipulations (e.g., \emph{Move a lot}, \emph{Rotate}, \emph{Connect}), 
no existing gesture set was sufficient on its own.

To avoid bias toward any single platform vocabulary, we treated the 
Spectacles corpus as an ecological sample of what contemporary head-worn AR 
asks users to do, and performed a light conceptual cross-check against 
publicly documented interactions from other AR ecosystems (e.g., HoloLens, 
Xreal, visionOS) to ensure that no major functional categories were omitted.  
This cross-check was not a full cross-platform corpus analysis; rather, it 
served as a sanity check confirming that our distilled intents capture the 
major classes of actions expected in current head-worn AR systems.

From these data, we distilled fifteen intents spanning object manipulation, 
system commands, and body/spatial actions (Table~\ref{tab:intents}); 
corpus-level prevalence statistics are provided in 
Appendix~\ref{appendix:corpus-prevalence}.

\begin{table}[h]
\centering
\caption{Gesture intent taxonomy distilled from the corpus.}
\label{tab:intents}
\begin{tabular}{|p{0.38\linewidth}|p{0.5\linewidth}|}
\hline
\textbf{Category} & \textbf{Gesture Intents} \\
\hline
\textbf{Object Manipulations} & Move a little; Move a lot; Rotate; Shrink; Enlarge; Connect; Separate \\
\hline
\textbf{System Commands} & Select; Delete; Switch; Open; Close; Draw \\
\hline
\textbf{Body/Spatial Actions} & Throw; Dodge \\
\hline
\end{tabular}
\end{table}

\subsection{System for Embodied Gesture Design}
\textit{Lightweight AR Glasses – Snap Spectacles.}
Building on this taxonomy, we implemented a custom elicitation system in 
\emph{Lens Studio~5} and deployed it on \emph{Snap Spectacles}, a lightweight, 
standalone, see-through AR glasses platform developed by Snap Inc. 
We chose Spectacles as our testbed because they combine an untethered, 
glasses-class form factor with an on-device stereoscopic display and a 
comparatively mature developer ecosystem. At the time of the study, we treated 
Spectacles as a proxy for lightweight AR glasses more broadly: Meta Orion was 
not yet released, and Xreal devices depended on external computing and 
functioned primarily as tethered displays.

\textit{System Configuration.}
Spectacles run Snap OS~2.0 and deliver AR experiences as \emph{Lenses}—
lightweight application units akin to mobile apps—created in Lens Studio~5. The developer version 
used in our study (the “Spectacles~’24” platform) provides an approximately 
$46^\circ$ diagonal field of view and weighs about 226\,g.

\textit{System Design for Embodied Gesture Elicitation.}
For each gesture intent, the glasses displayed a 10\,s cube animation 
(e.g., a rotating cube for \emph{Rotate}) with an overlaid text prompt 
(“Gesture Intent: Rotate”). The cube acted as a persistent \emph{spatial anchor}, 
enabling PTs to reason about gestures in relation to a 3D object while avoiding 
bias from built-in recognizers. All built-in gesture shortcuts were disabled. 
Participants saw a fixed instruction (“say \emph{next} to change”), while the 
researcher advanced the sequence via a paired phone. Each 
animation looped three times, and the 15 intents were presented in randomized 
order. Spectacles recorded first-person video and audio so participants could 
focus on performing and reflecting rather than on operating the device.

\begin{figure*}[h]
  \centering
  \begin{subfigure}{0.55\linewidth}
      \includegraphics[width=\linewidth]
      {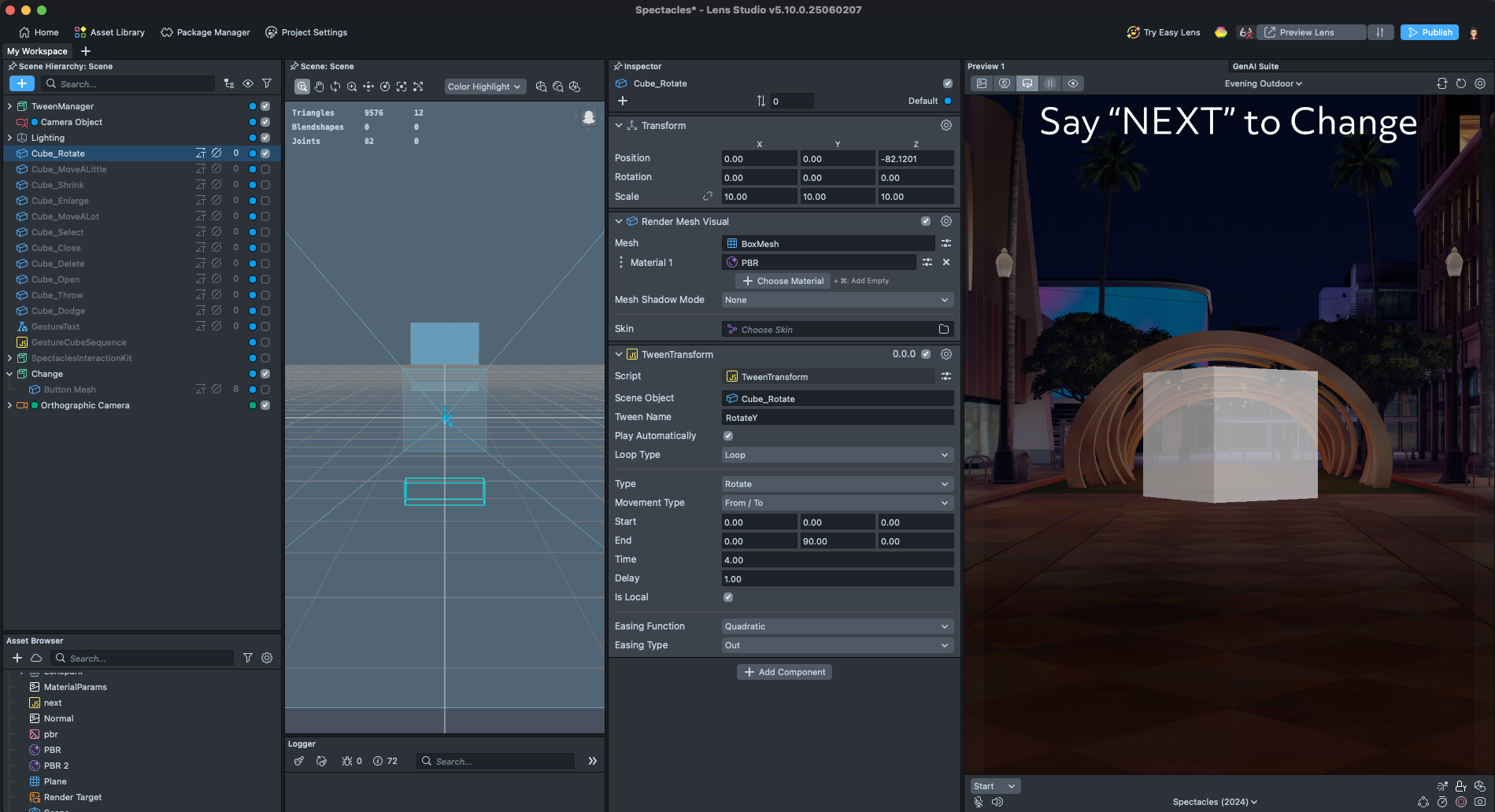}
      \caption{Authoring animations in Lens Studio~5.}
  \end{subfigure}
  \hfill
  \begin{subfigure}{0.4\linewidth}
      \includegraphics[width=\linewidth]{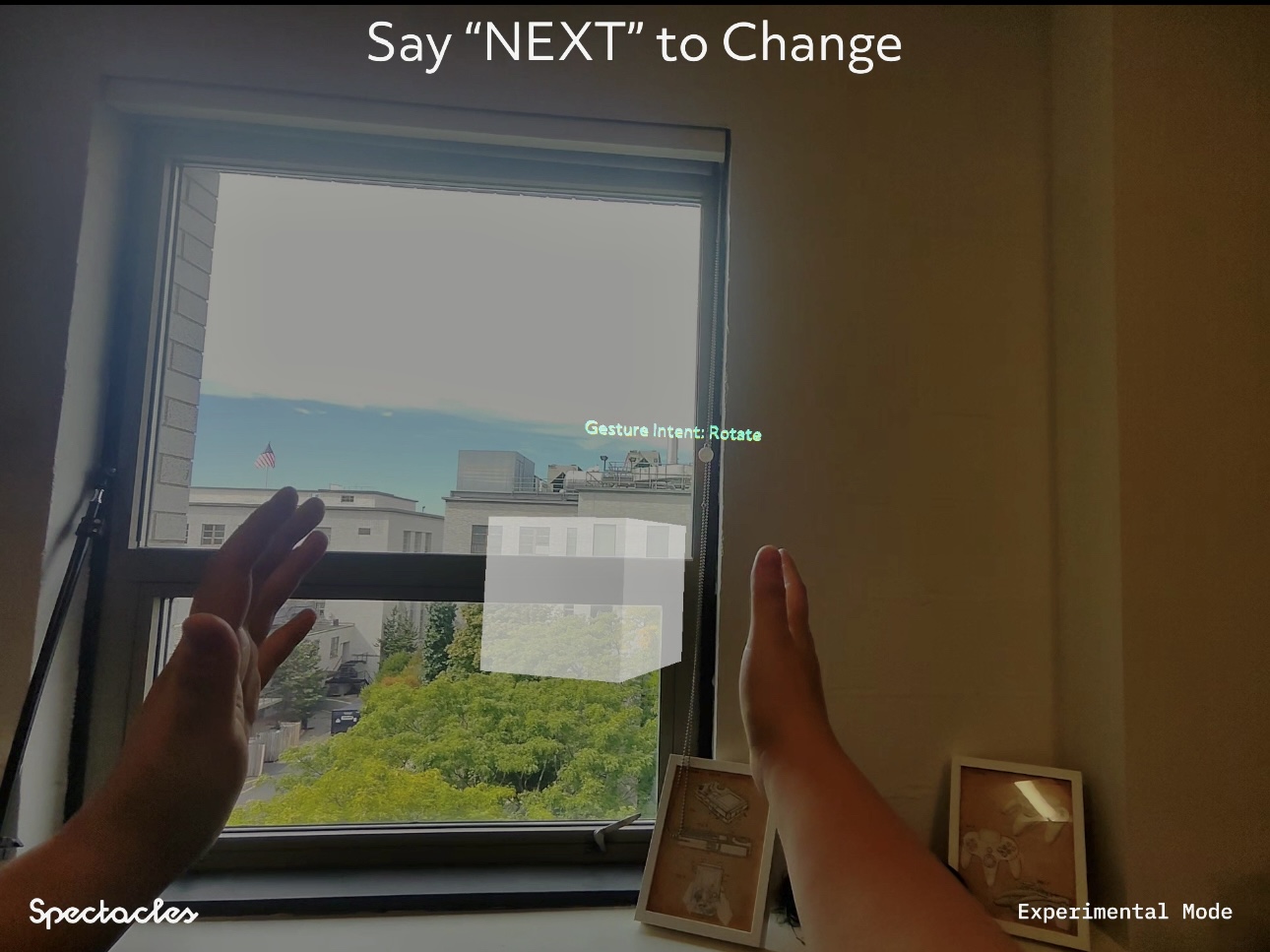}
      \caption{Rendering the “Rotate’’ cue on Spectacles.}
  \end{subfigure}
  \caption{Prototype for gesture elicitation: (a) authoring in Lens Studio; (b) rendering the \emph{Rotate} cue.}
  \Description{The figure shows a prototype system for gesture elicitation on lightweight augmented reality glasses. (a) The left image depicts the Lens Studio~5 authoring environment, where a semi-transparent cube is placed in front of the user and animated to rotate in place as a visual cue for the “Rotate” gesture intent. The interface includes a 3D scene view, an object hierarchy, and parameter panels for animation control. (b) The right image shows the same rotating cube rendered on Snap Spectacles from a first-person perspective, overlaid in the user’s physical environment with instructional text. The cube remains spatially anchored in front of the user, serving as a persistent reference during gesture execution.}
  \label{fig:rotate-cube}
\end{figure*}

Figure~\ref{fig:rotate-cube} shows a representative cue; additional example screenshots are included in Appendix~\ref{appendix:cue-gallery}.

\begin{table*}[t]
\centering
\caption{Participant demographics and XR familiarity.}
\label{tab:participants}
\renewcommand{\arraystretch}{1.4}  

\resizebox{\linewidth}{!}{
\begin{tabular}{|c|c|c|c|c|c|c|c|c|}
\hline
\textbf{PT ID} & \textbf{Gender} & \textbf{Age} &
\shortstack{\textbf{Years of}\\ \textbf{Experience}} &
\textbf{Specialty} & \textbf{Setting} &
\textbf{AR Familiarity} & \textbf{VR Familiarity} & \textbf{Digital Rehab Tools} \\
\hline
01 & Female & 60+ & 37 & Ortho; Education & Outpatient; PT Professor & None & None & None \\
02 & Male & 20--30 & 1 & Ortho & Outpatient & None & None & None \\
03 & Male & 20--30 & 1 & Ortho; Pediatrics & Outpatient & Seen Demos & Used once/twice & Mobile Apps \\
04 & Male & 20--30 & 1 & Sports Medicine & Outpatient & Seen Demos & Used once/twice & Mobile Apps \\
05 & Female & 46--60 & 34 & Neuro; Pediatrics & Hospital PT Services & Seen Demos & Seen Demos & Kinect, Wii Fit, Wearables \\
06 & Male & 20--30 & 1 & Sports Medicine & Elite Sports & Seen Demos & Seen Demos & Mobile Apps, Wearables \\
07 & Female & 31--45 & 7 & ICU PT & CVP \& AHF Care & Seen Demos & Seen Demos & Mobile Apps \\
08 & Male & 20--30 & 1 & Ortho & Outpatient & Seen Demos & Seen Demos & Mobile Apps \\
09 & Male & 60+ & 36 & Ortho & PT Clinic Owner & None & None & None \\
10 & Female & 46--60 & 28 & Ortho & PT Clinic Owner & None & Seen Demos & Mobile Apps \\
\hline
\end{tabular}}
\end{table*}
\subsection{Participants}

We recruited ten licensed physical therapists (PTs), each with a doctoral-level Doctor of Physical Therapy (DPT) degree and active U.S.\ state licensure. Participants ranged in age from their 
20s to 60s (see Table~\ref{tab:participants}) and represented a broad distribution of clinical experience (1–37 years; $M = 14.8$), specialties (orthopedics, sports medicine, 
neurorehabilitation, pediatrics, ICU care), and practice settings (outpatient clinics, hospital PT services, elite sports, private clinics, and academia).  Because most had minimal or no experience with AR/VR, they provided unbiased, 
clinically grounded interpretations rather than technology-specific expectations. 

All participants completed a single 90–120\,min in-person session and were compensated \$20 per hour. 

\subsection{Procedure}

The study consisted of three progressively structured rounds designed to build familiarity with AR cues, surface intuitive gesture tendencies, and elicit clinically grounded substitutions and context-sensitive variants (see Appendix~\ref{app:pt-protocol}). Short breaks were offered as needed.

\begin{itemize}
\item \textbf{Round 1 (Intuitive Execution).}  
Wearing Spectacles, participants viewed all 15 gesture intents, each presented as a 10\,s cube animation with a text label (“Gesture Intent: Rotate”). Each animation \emph{continuously looped} until the participant verbally requested “next,” allowing time to settle into the cue’s temporal rhythm and experiment with multiple gesture forms. Participants performed whatever gesture first came to mind—refining it across loops—and explained their reasoning in a think-aloud manner while the researcher advanced the sequence.

\item \textbf{Round 2 (PT-Grounded Substitution).}  
Still wearing Spectacles, participants revisited the same 15 intents, again shown in randomized order using the identical looping-cue mechanism. For each intent, PTs reconsidered the gesture from a clinical perspective and re-authored fragile, effortful, or fatigue-prone forms into safer, sustainable alternatives. 

\item \textbf{Round 3 (Context Elicitation and Card Sorting).}  
In the final round, participants no longer wore AR glasses. Each gesture intent was printed on an individual card. Drawing on their clinical expertise—especially their experience evaluating activities of daily living (ADLs)—PTs identified realistic contexts in which each gesture might occur (e.g., commuting, office work, cooking, walking outdoors). They then completed a structured card-sorting task, grouping intents by context and explaining when and why gesture forms should shift across settings.
\end{itemize}

\subsection{Data Sources and Analysis}

Our dataset consisted of four sources:  
(1) fixed-position video capturing full-body movement,  
(2) first-person audio/video from Spectacles,  
(3) structured observer notes (AR developer, licensed PT, HCI researcher), and  
(4) post-session reflections documented immediately after each session.  
Observers recorded gesture forms, participant rationales, markers of physical risk
(e.g., distal overuse, awkward wrist torque), and any context-dependent substitutions
mentioned spontaneously.

All materials were transcribed and coded in NVivo using a two-stage analysis.

\textbf{Pass~1 (Gesture and Clinical Coding).}  
We coded gesture-level properties such as fine- and gross-motor emphasis, range of motion,
joint anchoring, posture, and repetition frequency.  
Clinical reasoning codes captured fatigue sources, population-specific constraints, joint loading,
risk flags, and the substitution logic used by PTs to re-author fragile gestures while
preserving functional intent.

\textbf{Pass~2 (Ergonomic Dimension Building and Principle Abstraction).}  
Building on Pass~1, we revisited Round~1--2 field notes and think-aloud transcripts and iteratively clustered related codes to consolidate them into eight higher-order ergonomic dimensions: \emph{Posture Envelope}, \emph{Range-of-Motion Band}, \emph{Joint Load \& Fatigue}, \emph{Stability \& Support}, \emph{Visual Occlusion}, \emph{Social Visibility}, \emph{Cognitive Layer}, and \emph{Safety Limits}.
Through cycles of comparison, memoing, and abductive, theory-informed interpretation with our PT co-author, we further grouped these dimensions into four PT-informed biomechanical principles and summarized them in the Everyday-AR Golden Ergonomic Canvas representation used in our results (Figure~\ref{fig:canvas} and see Appendix~\ref{app:ergonomic-table} for extended materials).

Three coders independently annotated all transcripts and videos.  
Disagreements were resolved through discussion, yielding substantial agreement (Cohen's $\kappa = 0.82$).  
This combined analysis allowed us to determine when gesture families should shift (e.g., expressive $\rightarrow$ subtle, distal $\rightarrow$ proximal) and which PT-informed substitutions best support sustainable, socially coherent everyday AR interaction.

\subsection{Ethics, Validity, and Scope}
This study received IRB approval from Northeastern University, and written informed consent was obtained from all participants. All recordings were stored on encrypted drives, and participant identities were anonymized; bystander capture was minimized 
and faces were cropped during transcription. De-identified codebooks and example stimuli will be shared upon request.

Credibility was supported through triangulation across data sources, multiple coders, and prompted clarification during think-aloud sessions. Our observer team (licensed PT, HCI 
researcher, and AR developer) ensured balanced perspectives throughout data collection.

The goal of the study was to develop \emph{design guidance} for sustainable everyday-AR gestures, rather than to benchmark gesture-recognition performance. PTs contributed as movement-safety experts, helping translate intuitive but fragile gestures into durable, clinically coherent alternatives suitable for everyday AR use.

\section{Results}

Our analysis revealed that physical therapists (PTs) consistently reorganized participants’ intuitive gestures around four recurring biomechanical principles: \textbf{Joint-Rotation Substitution}, \textbf{Shoulder-Line Workspace Constraint}, \textbf{Proximal–Distal Stability}, and \textbf{Fine–Gross Motor Integration}. These principles describe how gestures must be reshaped to remain sustainable, socially coherent, and feasible for everyday AR use.

We structure the results around three layers of this reasoning. First, we detail the four principles and show how they explain why many ``natural’’ gestures from Round~1 were biomechanically fragile or fatiguing (Section~\ref{sec:principles}). Second, we present the PT-guided \emph{gesture substitution patterns} formulated in Round~2
(Section~\ref{sec:substitution-patterns}), illustrating how fragile gestures were systematically re-authored into safer, more repeatable forms. Third, we synthesize these patterns into the \emph{Everyday-AR Golden Ergonomic Canvas}, a chest-level, elbow-anchored workspace that unifies therapist reasoning about sustainable joint sequencing (Section~\ref{sec:golden-canvas}). Finally, we report findings from Round~3 (Section~\ref{sec:everyday-contexts}), where PTs organized gesture intents across six dramaturgical everyday contexts, clarifying how social visibility, attentional demand, and bodily constraints shape gesture appropriateness in public, semi-public, and private settings.

\subsection{Four PT-Informed Biomechanical Principles}
\label{sec:principles}
\begin{figure*}[h]
  \centering
  \includegraphics[width=\textwidth]{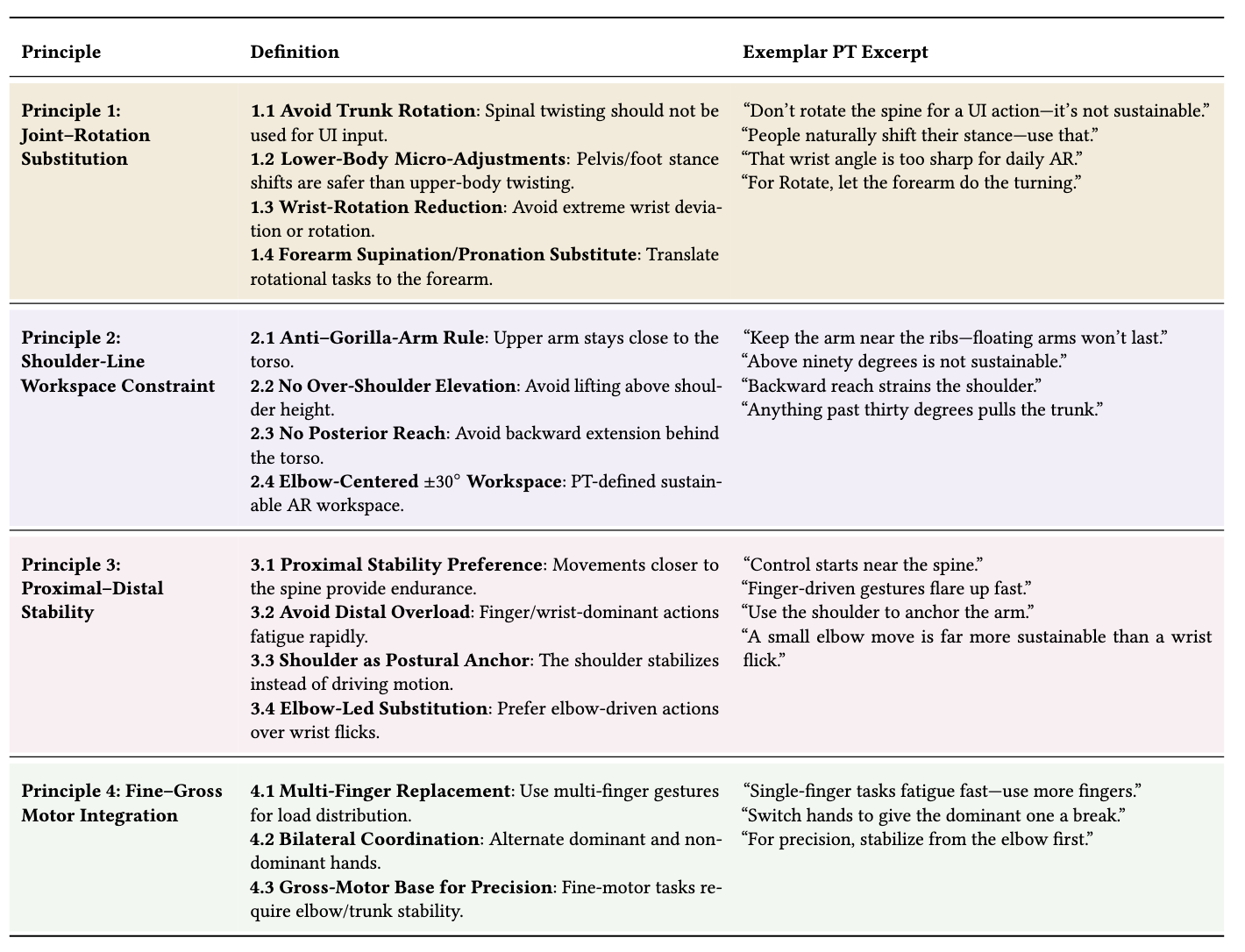}
  \caption{PT-informed gesture design code book used in open coding (Round~1--2).
  Codes consolidate four biomechanical principles and articulate how PTs evaluated gesture sustainability and substitution.}
  \Description{A color-coded three-column code book summarizing PT-informed gesture design principles. The left column lists four biomechanical principles. The middle column describes sub-principles defining sustainable gesture constraints. The right column presents exemplar quotes from physical therapists explaining fatigue, stability, and substitution strategies in wearable AR gestures.}
  \label{fig:table3}
\end{figure*}

Iterative open coding of all Round~1--2 think-aloud sessions revealed four recurring biomechanical principles that therapists used to judge whether a gesture was sustainable, risky, or socially workable. These principles
synthesize the clinical reasoning patterns that PTs repeatedly invoked across sessions and also structure the case-driven subsections that follow. To ensure analytic transparency, we place the full coding book used to derive these principles (Cohen’s~$\kappa$ = 0.82) directly in the main text (Figure~\ref{fig:table3} ). We briefly enumerate the four principles here before
discussing each in detail:
\begin{itemize}[leftmargin=1.3em]
    \item \textbf{Joint–Rotation Substitution} — redirect rotation away from the
    spine and wrist toward safer forms such as forearm supination/pronation and
    stance-based micro-adjustments.

    \item \textbf{Shoulder-Line Workspace Constraint} — constrain everyday AR
    interaction to a chest-level, elbow-led $\pm30^\circ$ envelope and avoid
    arm abduction, overhead reach, and “gorilla-arm’’ postures.

    \item \textbf{Proximal–Distal Stability} — shift effort toward proximal
    joints for endurance and minimize distal-dominant gestures that require
    static stabilization (e.g., finger or wrist isolation).

    \item \textbf{Fine–Gross Motor Integration} — support precision actions with
    a stable gross-motor base using multi-finger grips, bilateral alternation,
    or elbow-supported micro-movements.
\end{itemize}

\subsubsection{\colorbox{canvasJR!65}{\strut Principle 1: Joint–Rotation Substitution}}

Rotation-intensive gestures consistently triggered problematic movement
patterns in Round~1: participants twisted their torsos, externally rotated
their shoulders, or snapped their wrists depending on which joint felt most
intuitive. PTs uniformly discouraged these strategies in Round~2, citing
spinal torque, rotator-cuff strain, and rapid fatigue in the wrist’s small
stabilizers. Across sessions, therapists converged on a simple rule:
rotation should be generated by the forearm, while the shoulder,
wrist, and spine remain stable.

\begin{figure}[h]
  \centering
  \subfloat[Baseball-style overhand throw.]{%
    \includegraphics[width=0.28\linewidth]{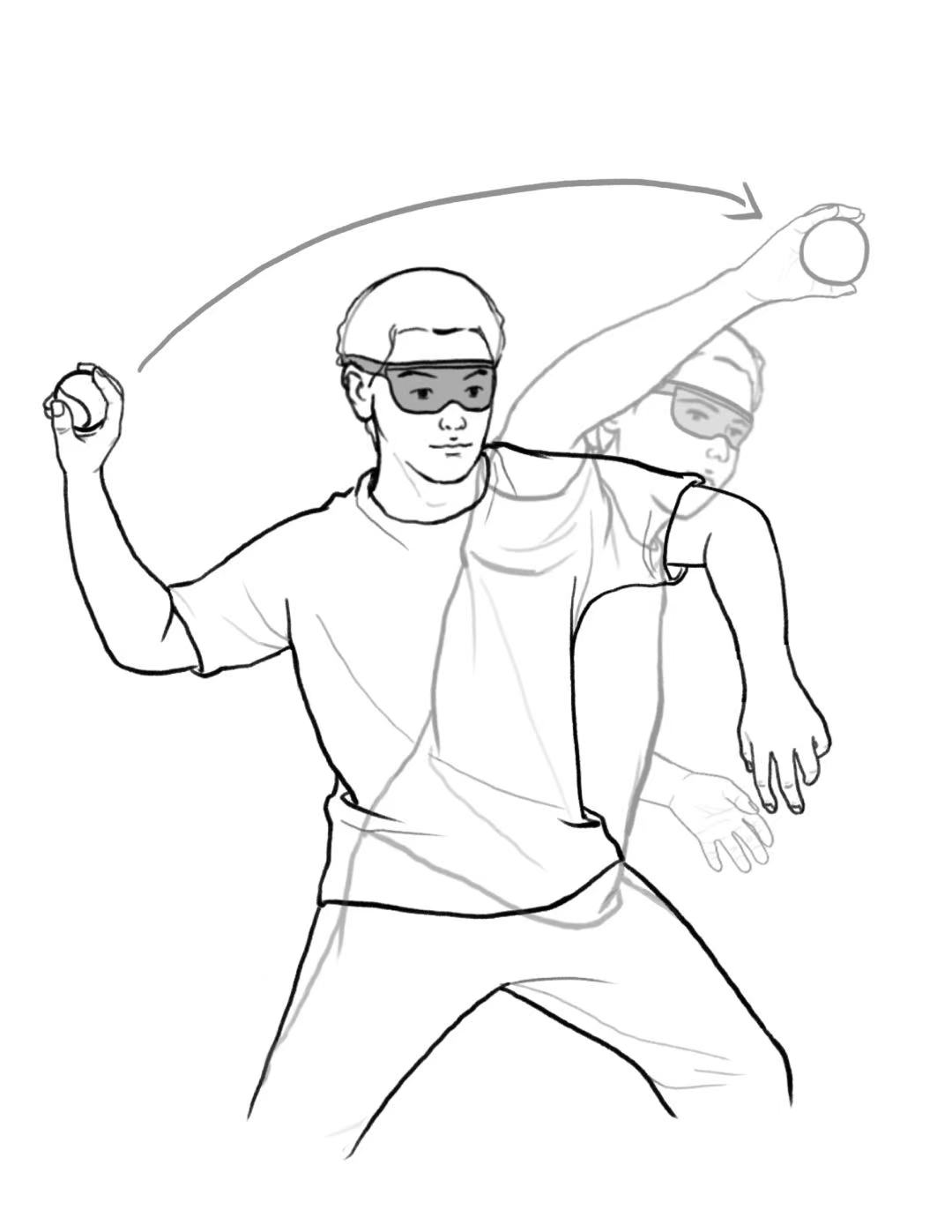}}
  \hfill
  \subfloat[Basketball-style push shot.]{%
    \includegraphics[width=0.28\linewidth]{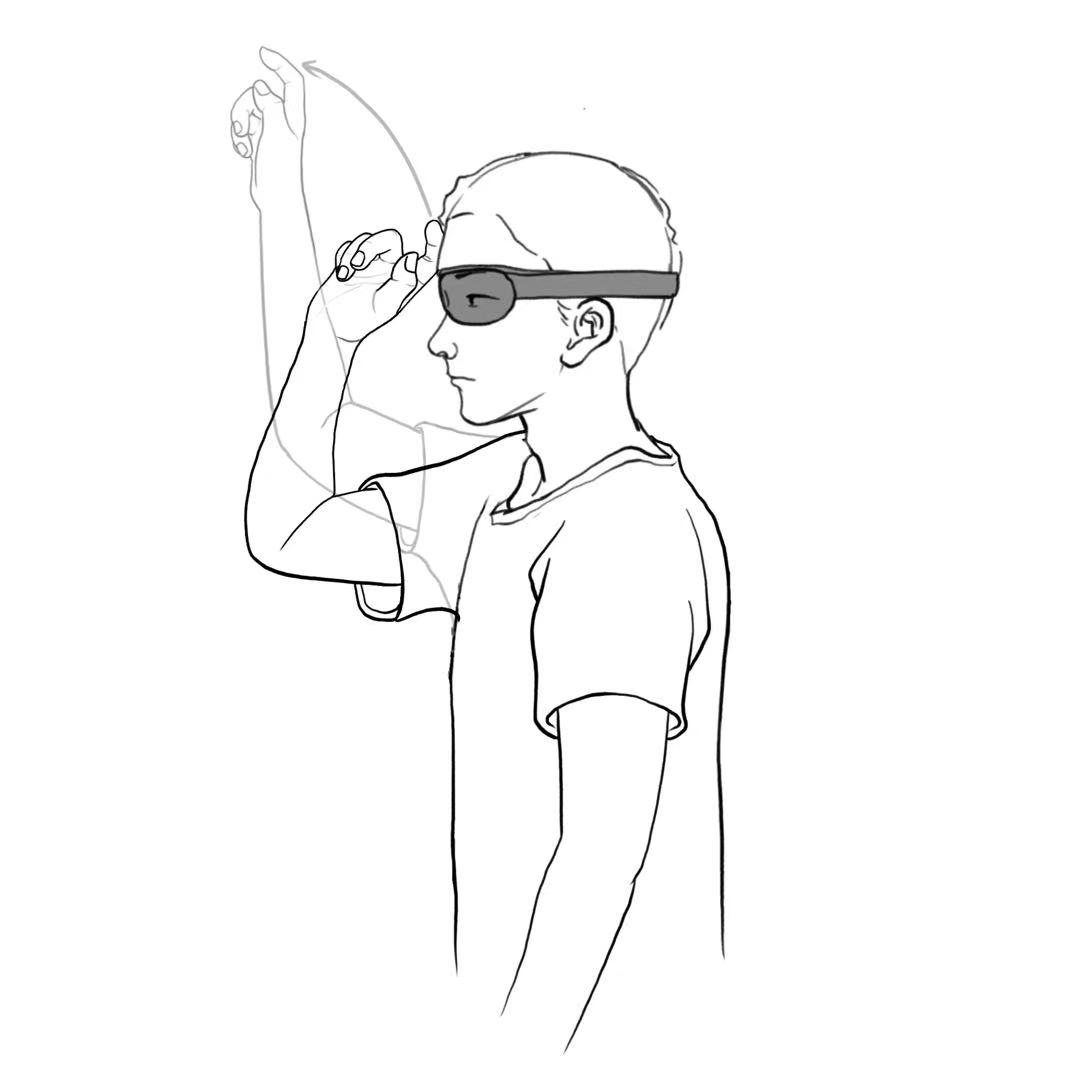}}
  \hfill
  \subfloat[Underhand toss.]{%
    \includegraphics[width=0.28\linewidth]{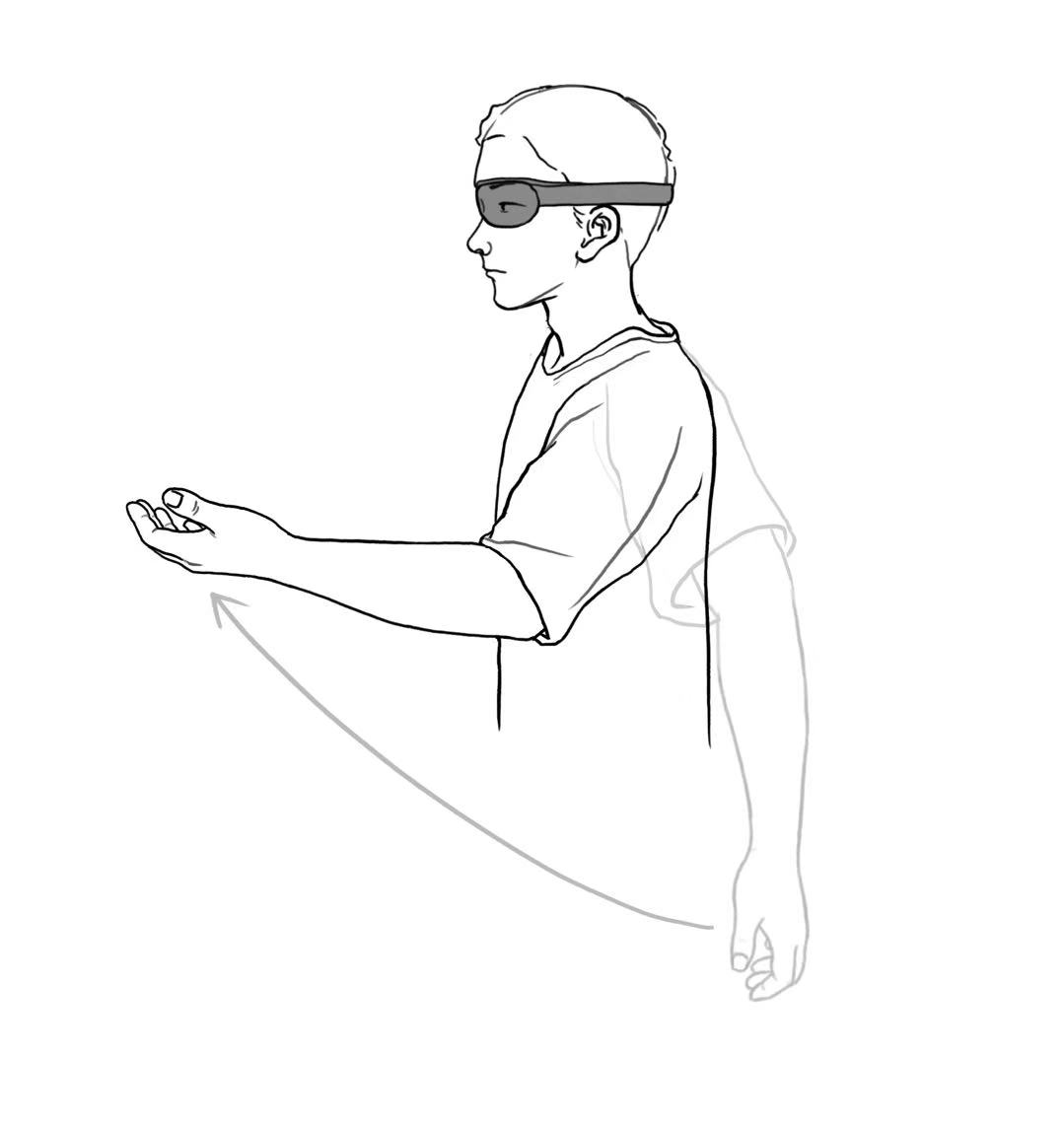}}
  \\
  \subfloat[Dart-like forearm flick.]{%
    \includegraphics[width=0.28\linewidth]{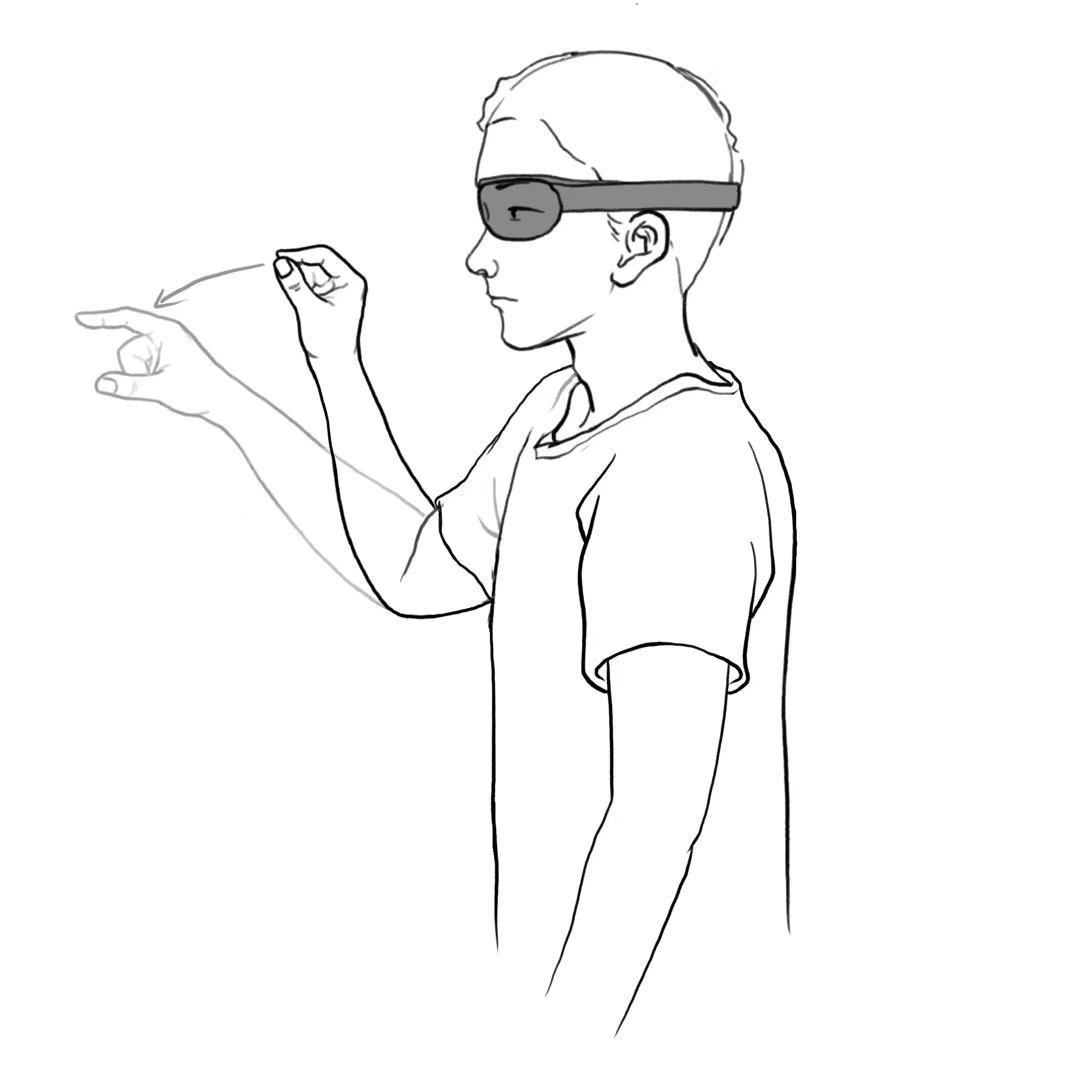}}
  \hfill
  \subfloat[Card-dealing forward motion.]{%
    \includegraphics[width=0.28\linewidth]{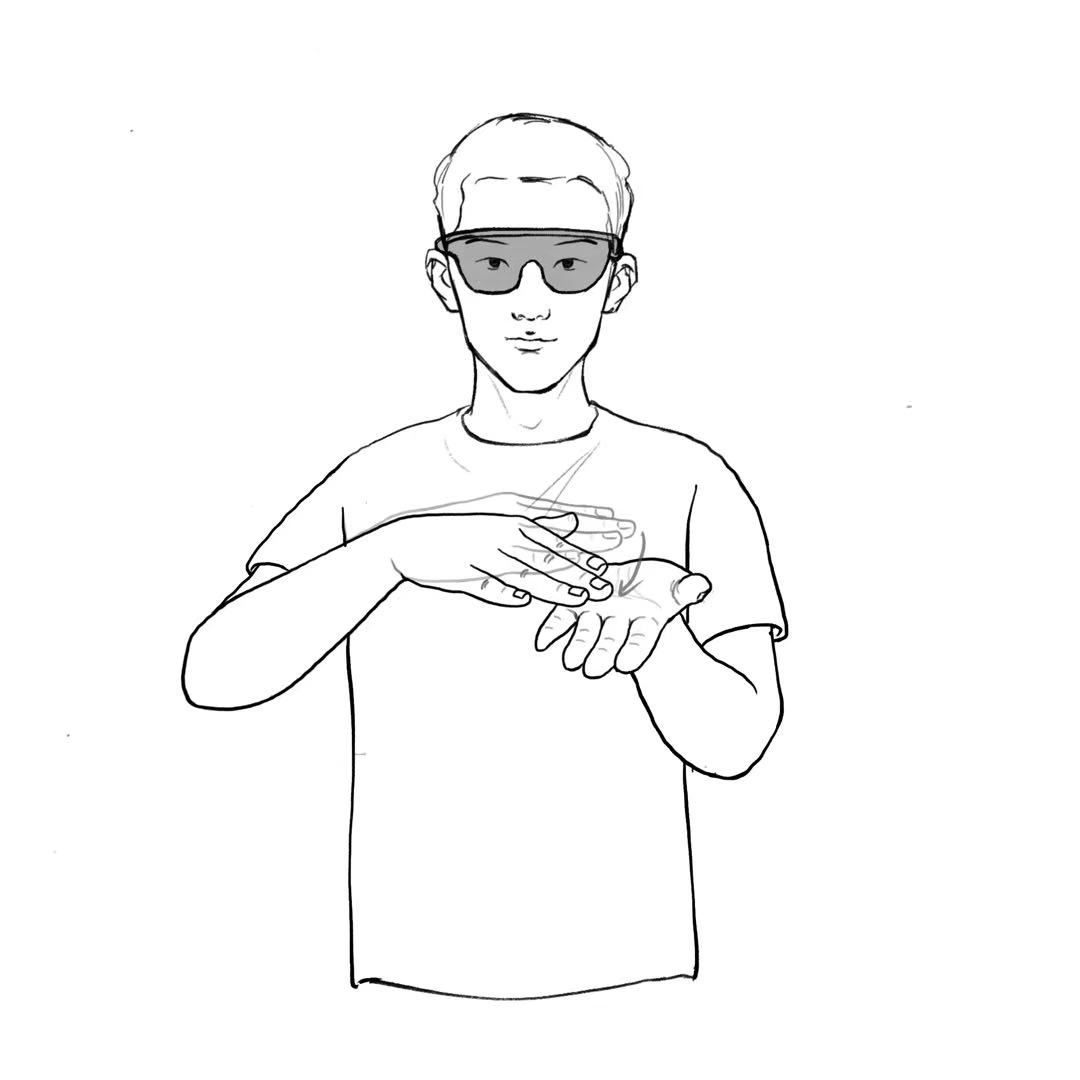}}
  \hfill
  \subfloat[Low-level push alternative.]{%
    \includegraphics[width=0.28\linewidth]{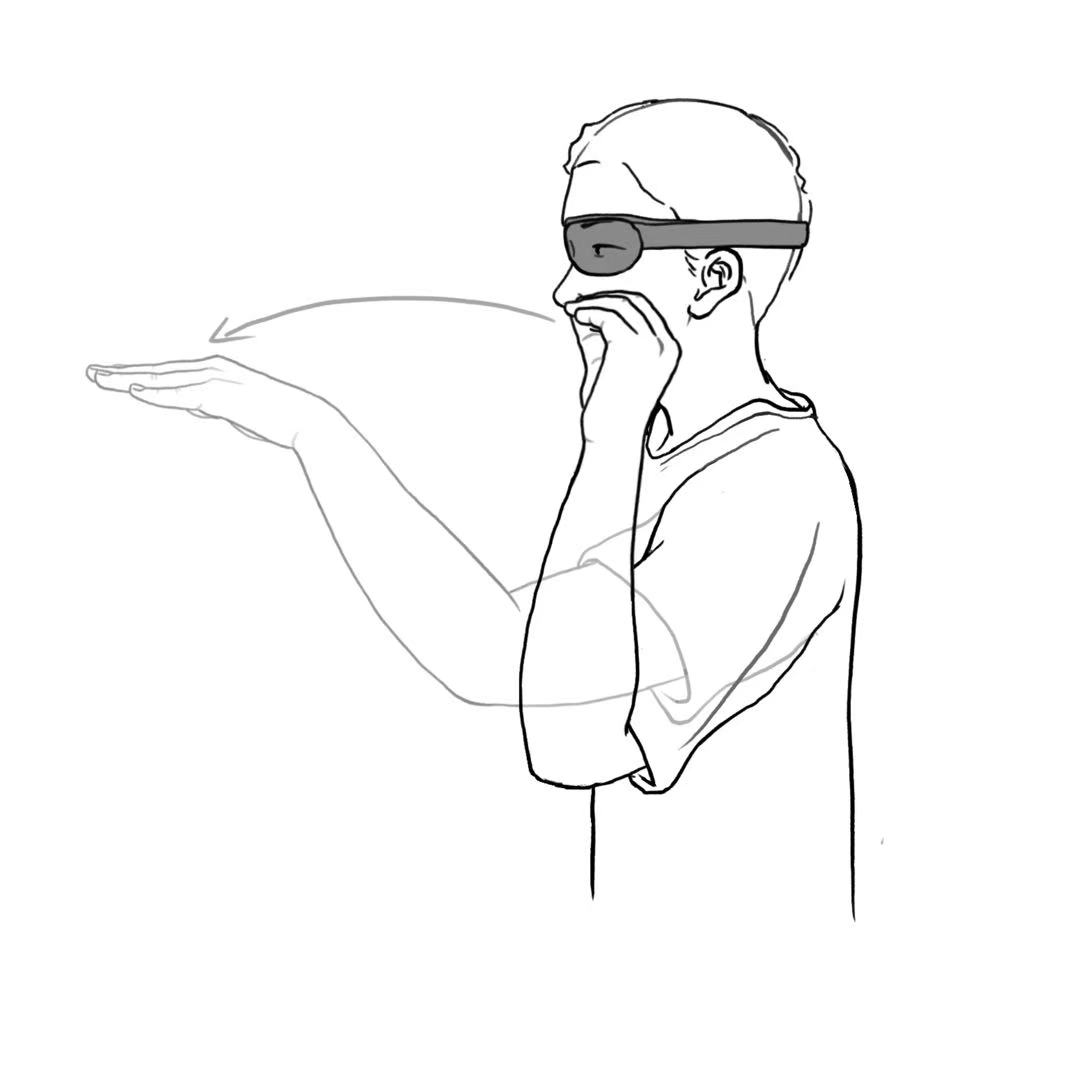}}
  \caption{\textbf{PT reasoning around "Throw".} 
  Round~1 revealed intuitive overhand and push-shot mimicry; Round~2 replaced 
  these with elbow-anchored, low-amplitude variants that redirect rotation to 
  the forearm and reduce shoulder load.}
  \label{fig:throw}
\end{figure}

\paragraph{Case study: Throw.}
Nearly all participants performed a baseball-style overhand arc in Round~1
(Figure~\ref{fig:throw}a), combining shoulder external rotation, torso drive,
and a final wrist snap. PTs characterized this as one of the highest-risk
patterns across all intents: without a physical object to absorb force, users
tend to over-rotate the shoulder and hyper-extend the elbow, producing sharp,
unsustainable peaks in joint load.

In Round~2, PTs reconstructed \emph{Throw} into compact, elbow-anchored
variants that preserved the semantic intent of “sending away’’ while
eliminating problematic rotations. Examples included underhand tosses
(Figure~\ref{fig:throw}c), forward “card-dealing’’ motions
(Figure~\ref{fig:throw}f), and small-arc “dart’’ flicks
(Figure~\ref{fig:throw}d). In each variant, rotation was absorbed through
forearm supination–pronation, with the spine held neutral and the shoulder
acting purely as a stabilizer.

\textbf{Principle summary.}  
If a gesture requires rotation, \emph{let only the forearm rotate}. All other
joints—particularly the shoulder and wrist—should stabilize rather than drive
the movement. Sustainable rotation in AR emerges from \textbf{elbow-led,
forearm-based substitutes} rather than intuitive but risky full-arm mimicry.

\subsubsection{\colorbox{canvasSL!65}{\strut Principle 2: Shoulder-Line Workspace Constraint}}

A dominant pattern across Round~1 was the emergence of  
\emph{"gorilla-arm"} postures---arms drifting away from the torso, 
lifting above shoulder height, or extending forward in large sweeping arcs. 
Although these movements often appear expressive or "natural" in XR systems, 
PTs emphasized that they are \emph{biomechanically unsustainable}: 
deltoid and supraspinatus activation increases sharply once the upper arm 
abducts or elevates, and users fatigue within seconds during repeated interaction.

Across participants, PTs converged on a shared sustainable workspace: 
a \textbf{chest-level, elbow-anchored envelope of approximately $\pm 30^\circ$}. 
Within this range, the \textbf{shoulder serves as a stabilizer}, 
while the \textbf{elbow} generates motion through compact horizontal translations 
and micro-pivots. 
PTs consistently framed this configuration as the default workspace for 
everyday AR---safe, repeatable, and socially unobtrusive.
\begin{figure}[!h]
  \centering
  \subfloat[Unsustainable XR default: elevated "gorilla-arm" posture outside the shoulder-line envelope.]
  {\includegraphics[width=0.45\linewidth]{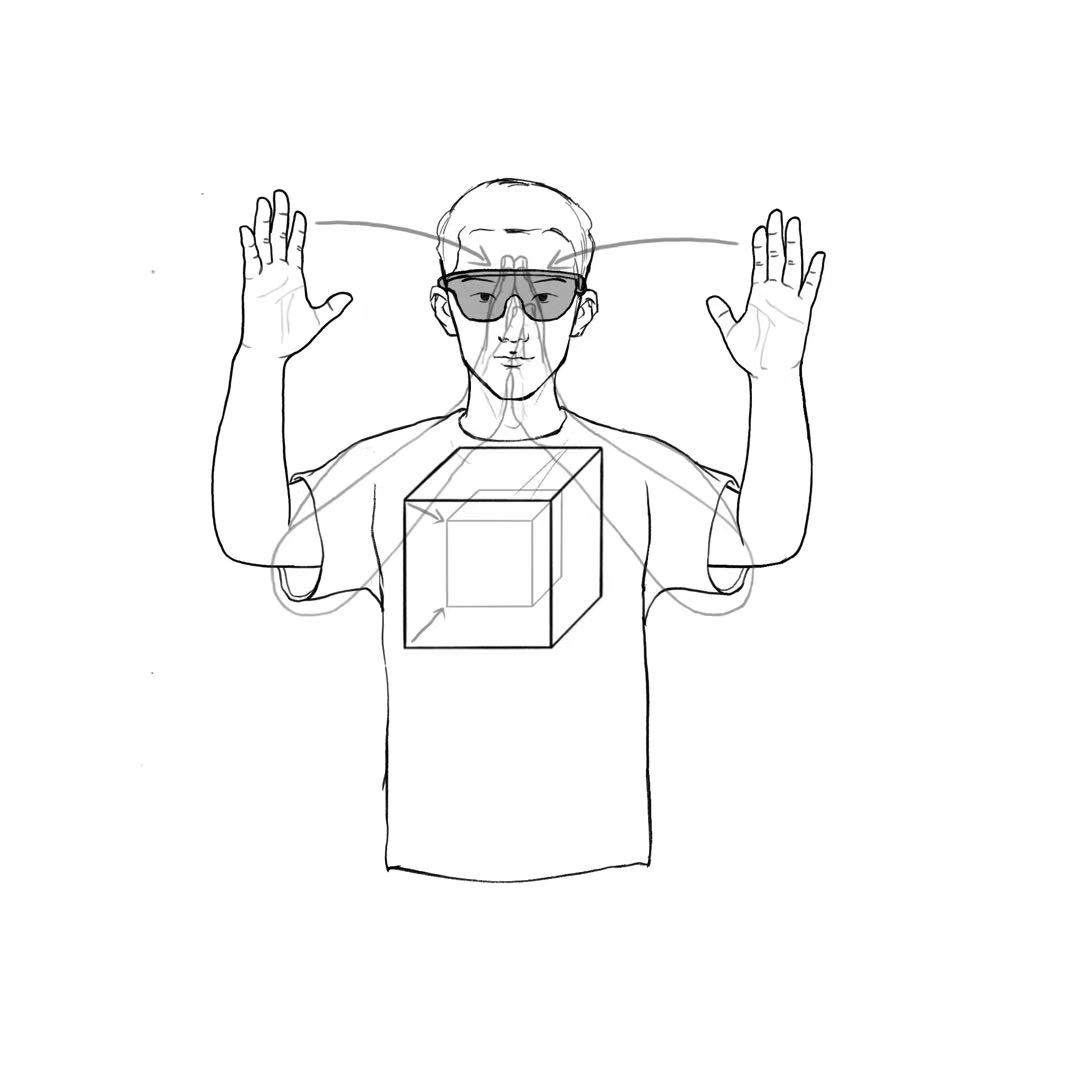}}
  \hfill
  \subfloat[PT-recommended posture: chest-level, elbow-anchored motion within the $\pm 30^\circ$ workspace.]
  {\includegraphics[width=0.45\linewidth]{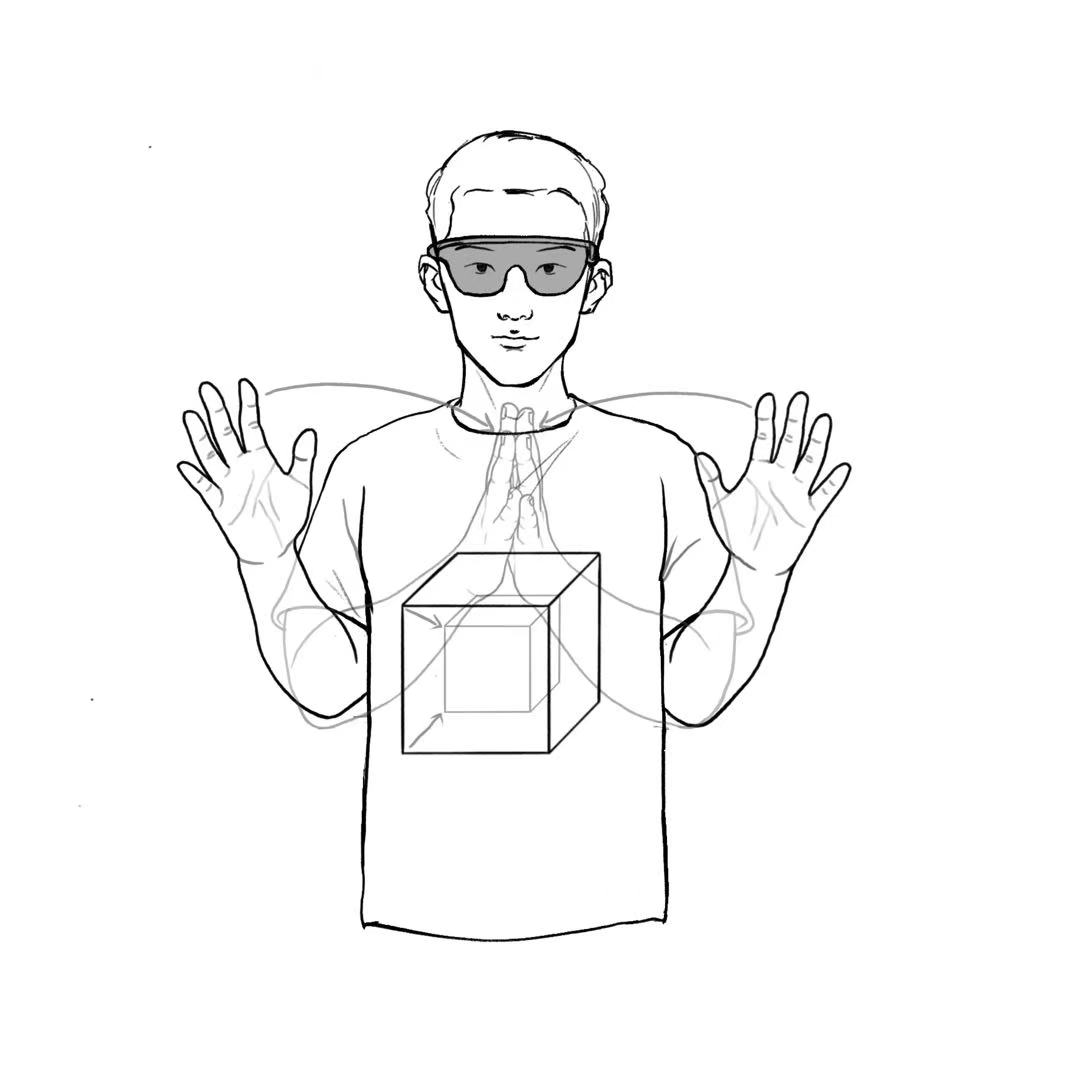}}
  \caption{Comparison of unsustainable versus PT-recommended workspace postures. 
  PTs uniformly redirected users from high-elevation, abducted arm positions to 
  a compact chest-level workspace anchored by the elbow.}
  \label{fig:shoulder-logic}
\end{figure}

\paragraph{Case study: \emph{Enlarge / Shrink}.}

In Round~1, participants enacted \emph{Enlarge} and \emph{Shrink} through 
high-amplitude bilateral motions: wide arm spreading, elevated forward reaches, 
and large cross-body contractions (Figure~\ref{fig:shoulder-logic}a). 
These patterns routinely pushed the upper arm outside the shoulder-line envelope, 
inducing shoulder abduction and elevation—two of the fastest paths to deltoid 
fatigue.

In Round~2, PTs re-authored these gestures as \textbf{elbow-driven scalings}:  
short horizontal \emph{lateral slides} for \emph{Enlarge}, and 
compact \emph{elbow-anchored compressions} for \emph{Shrink}, all performed at 
chest level and within the approximately $\pm 30^\circ$ workspace 
(Figure~\ref{fig:shoulder-logic}b). 
The result preserved the semantic structure of scaling while eliminating 
high-risk shoulder activation.

PTs highlighted four explicit prohibitions:
\begin{enumerate}[itemsep=1pt,leftmargin=1.3em]
    \item \textbf{Shoulder abduction}: arms drifting outward during scaling,
    \item \textbf{Over-shoulder elevation}: raising the arms to create “big” gestures,
    \item \textbf{Posterior reach}: pulling the elbows behind the frontal plane,
    \item \textbf{Cross-midline sweeps}: compressing or expanding across the torso midline.
\end{enumerate}

These constraints formed the basis of the \textbf{shoulder-line rule} 
embedded in the Everyday-AR Golden Ergonomic Canvas: 
\emph{the shoulder stabilizes; the elbow drives}. 
This reframes XR interaction away from theatrical, large-amplitude motions and toward sustainable embodied ergonomics.

\subsubsection{\colorbox{canvasPD!65}{\strut Principle 3: Proximal–Distal Stability}}

A central thread across PT reasoning was that distal joints (fingers, wrist)
fatigue rapidly when they are required to both \emph{move} and \emph{stabilize}
the hand in mid-air. Round~1 gestures such as single-finger pinches, isolated
index taps, and wrist-led dragging place the entire stabilization burden on the
forearm flexors and shoulder elevators—muscles ill-suited for prolonged static
control. PTs repeatedly described these as ``two-minute gestures’’: motions that
feel initially precise yet become unsustainable almost immediately without a
support surface.

To remedy this, PTs shifted nearly all precision tasks toward
proximal anchoring: letting the \emph{elbow} and \emph{upper arm} create
a stable base from which fine movements can emerge. This anchoring reduces
shoulder co-contraction, distributes load across the larger muscle groups of the
upper arm, and transforms fragile distal actions into repeatable, everyday
input forms.
\begin{figure}[t]
  \centering

  \subfloat[Round~1: distal-dominant single-finger tracing.]{%
    \includegraphics[width=0.30\linewidth]{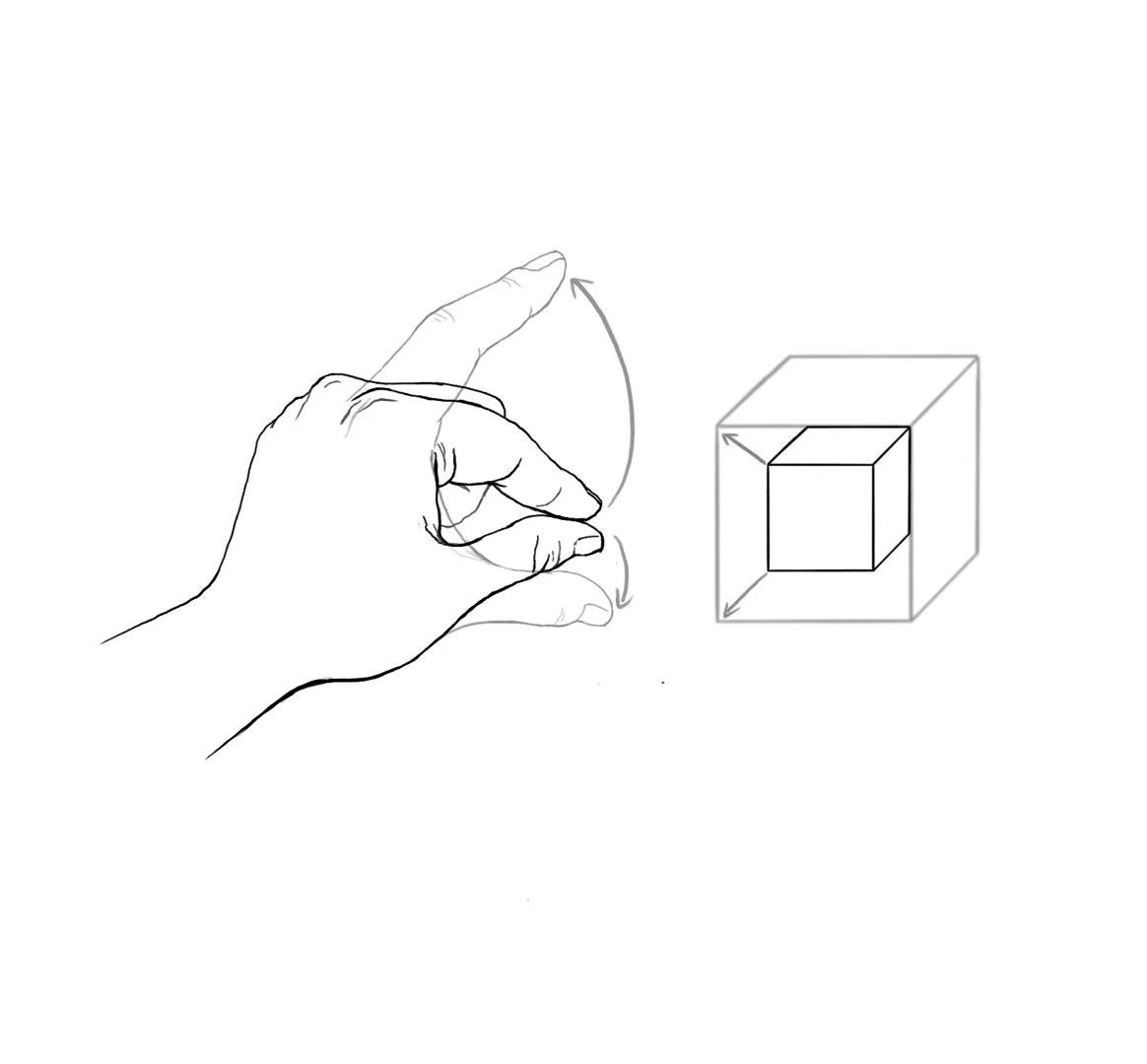}
  }
  \hfill
  \subfloat[Round~2: multi-finger grip replacing isolated index taps.]{%
    \includegraphics[width=0.30\linewidth]{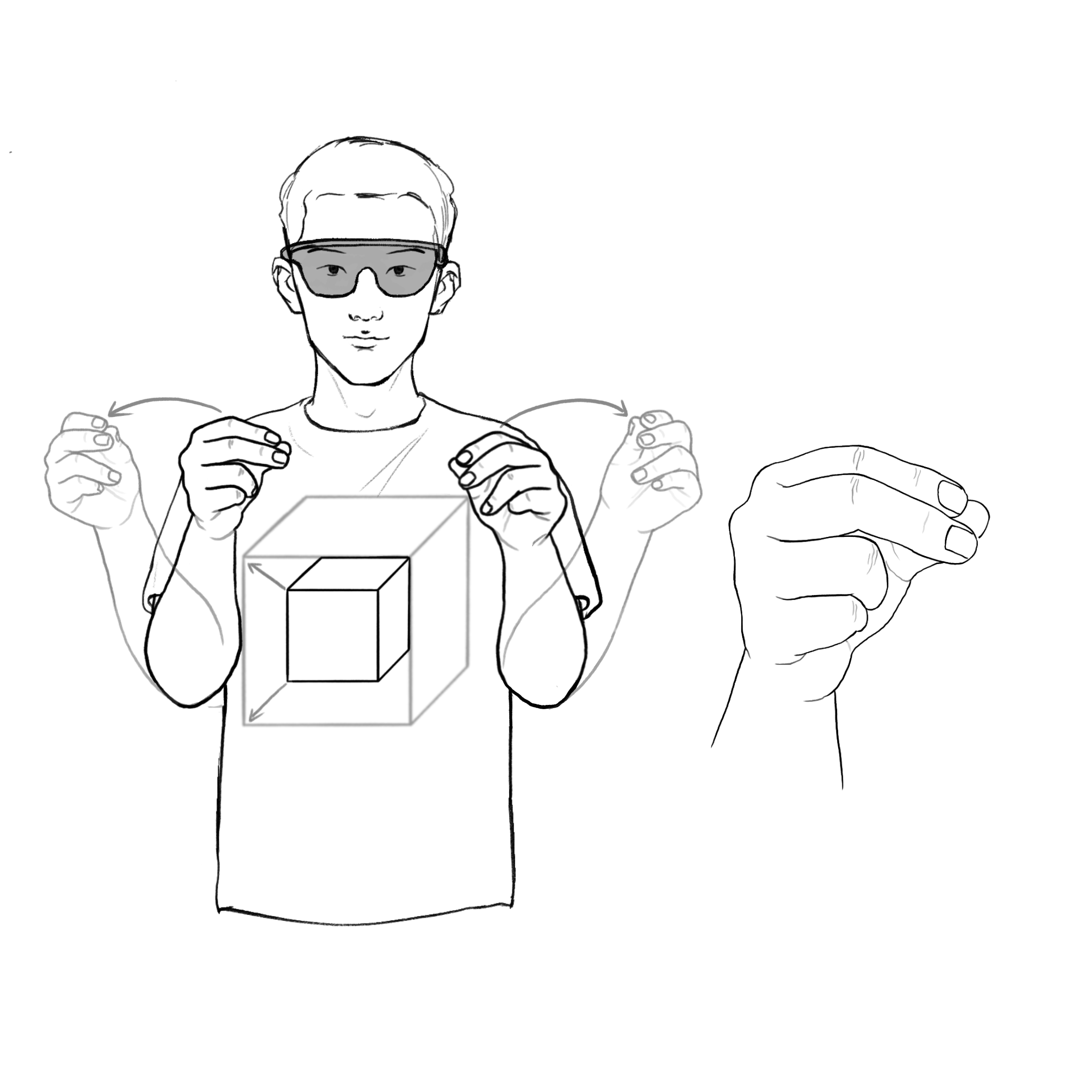}
  }
  \hfill
  \subfloat[Round~2: elbow-anchored micro-drag with tripod/pseudo-pen posture.]{%
    \includegraphics[width=0.30\linewidth]{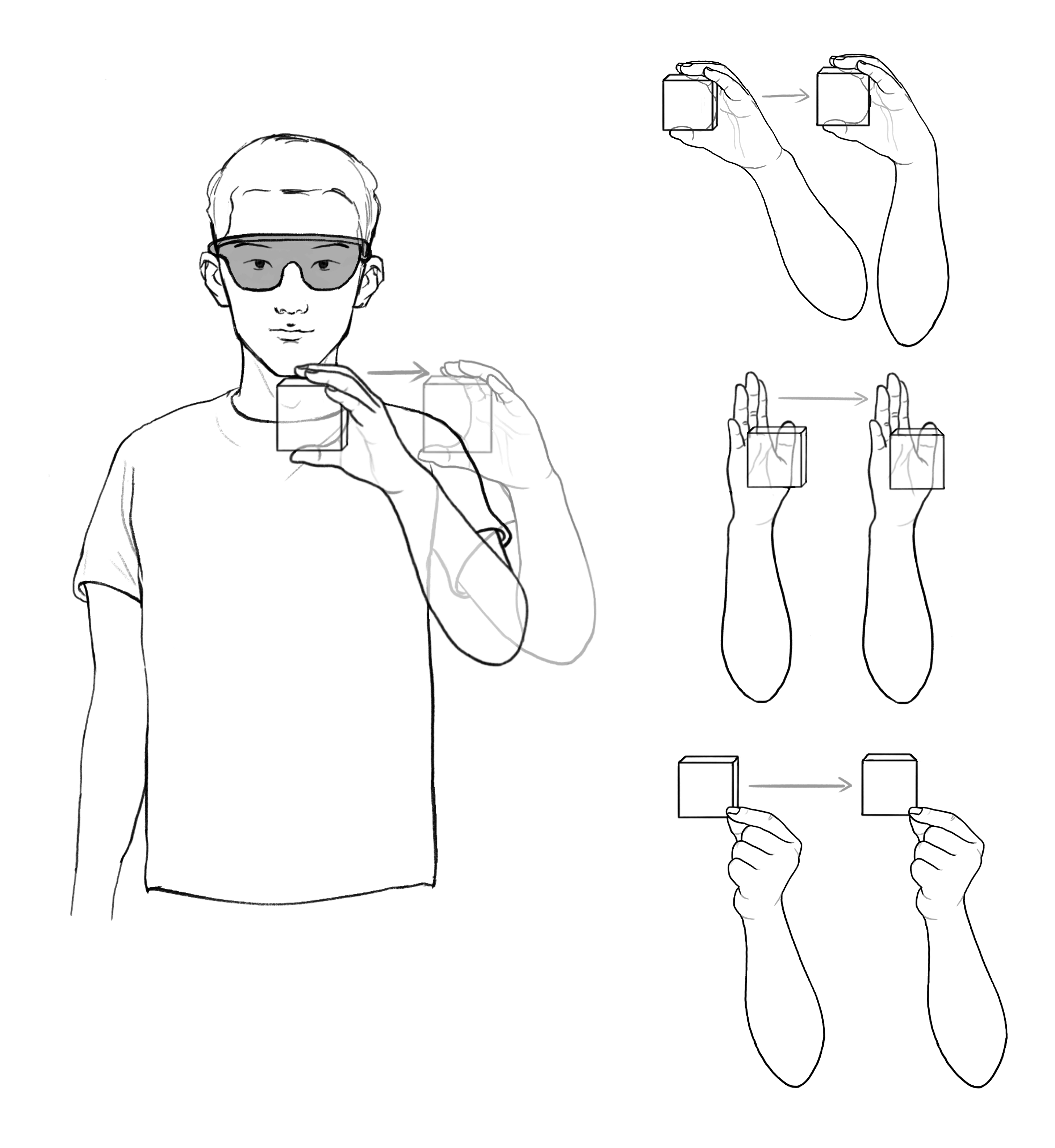}
  }
\caption{PT substitution logic for \emph{Enlarge / Move-a-little}: shifting effort from distal-only gestures (single-finger, wrist-led) to proximal anchoring via elbow-driven stabilization, multi-finger load distribution, and tripod-like grips.}

  \label{fig:proximal-distal}
\end{figure}
\paragraph{Case study: \emph{Enlarge / Move a little}.}
In Round~1, participants overwhelmingly traced virtual objects using isolated
index-tip movements or wrist-driven arcs—highly distal patterns that collapsed
under static load (Figure~\ref{fig:proximal-distal}a). PTs classified these as
high-risk for fatigue, noting that unsupported wrist flexion requires continuous
co-contraction of small stabilizers.

In Round~2, therapists decomposed these tasks into a
\textbf{proximal–distal action chain}:

\begin{enumerate}[itemsep=2pt,leftmargin=1.3em]
    \item \textbf{Proximal anchoring}: the elbow stays close to the torso to
    offload stabilization from the wrist.
    \item \textbf{Multi-finger load distribution}: small, repetitive actions are
    performed using two–three fingers or a palm-based push rather than a single
    fingertip (Figure~\ref{fig:proximal-distal}b).
    \item \textbf{Fine movement atop a stable base}: precision occurs through
    micro-pivots at the elbow or tripod-style pseudo-pen grips that maintain
    stability (Figure~\ref{fig:proximal-distal}c).
\end{enumerate}

Across PTs, the logic was consistent: \emph{fine motor accuracy is not
the problem—unsupported fine motor accuracy is}. Sustainable AR interaction
requires a gross-motor scaffold that handles stabilization so that distal joints
can perform short, low-amplitude motions. This transforms drawing, dragging,
and selection from ``two-minute gestures’’ into input forms viable for high-
frequency everyday use.

\subsubsection{\colorbox{canvasFG!65}{\strut Principle 4: Fine--Gross Motor Integration}}

Across Round~1--2, therapists emphasized that \emph{fine-motor precision is 
only sustainable when supported by gross-motor stability}. In mid-air 
interaction, the wrist and shoulder cannot simultaneously stabilize and 
perform fine control; precision becomes viable only as the \emph{final step} 
of a short, well-supported action sequence. PTs operationalized this through 
two recurring strategies: establishing a stable gross-motor base before 
precision, and distributing motor--cognitive load bilaterally when the action 
is high-risk. We illustrate this principle through one representative case.

\begin{figure}[t]
  \centering
  \subfloat[Round~1 natural pattern: users reach and pull in one unstable 
  motion, elevating the shoulder and forcing wrist stabilization.]{%
    \includegraphics[width=0.43\linewidth]{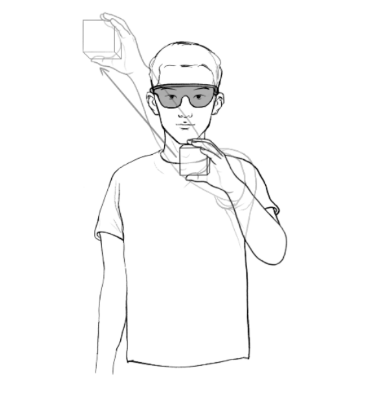}
    \label{fig:movealot-natural}}
  \hfill
  \subfloat[Round~2 PT substitute: a gross-motor step brings the cube into the 
  chest-level workspace, enabling a brief, stabilized fine confirmation.]{%
    \includegraphics[width=0.53\linewidth]{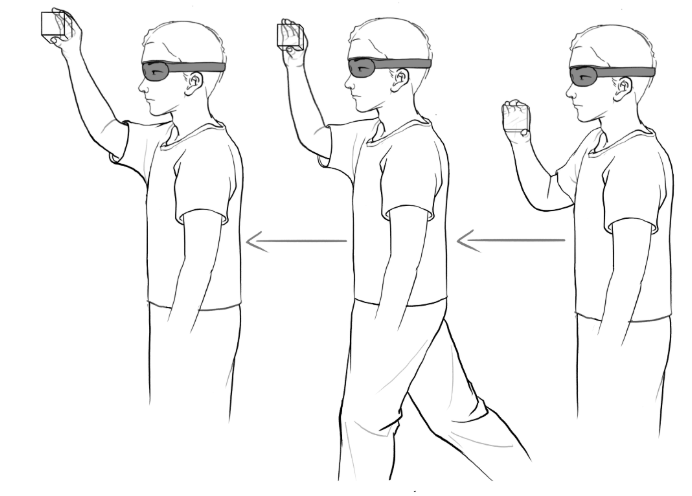}
    \label{fig:movealot-pt}}
  \caption{PT re-authoring of \emph{Move a Lot}: unstable whole-arm pulling 
  (left) is replaced by a two-stage gross-to-fine sequence (right).}
  \label{fig:movealot}
\end{figure}
\paragraph{Case study: \emph{Move a Lot}---gross-motor repositioning before fine action.}

In Round~1, participants attempted to retrieve distant objects through a single 
combined motion: lifting the arm, reaching outward, and pulling with an 
unsupported wrist (Figure~\ref{fig:movealot-natural}). PTs characterized this as 
an unstable "all-at-once" pattern that elevates the shoulder, overloads distal 
stabilizers, and degrades precision.

In Round~2, therapists decomposed the gesture into \textbf{a two-stage 
gross-to-fine sequence} (Figure~\ref{fig:movealot-pt}):  
(1) a small \emph{gross-motor step} brings the object into the chest-level 
workspace, followed by  
(2) a brief \emph{elbow-supported confirmation} once the body is fully stabilized.  
This re-authoring captures the core logic of Fine--Gross Motor Integration:  
\emph{precision only emerges after gross-motor stability is re-established}.  
Related patterns appeared in reactive gestures such as \emph{Dodge}, where PTs 
similarly replaced torso bending with a short lower-body step to restore 
balance before any fine action.
Related patterns appeared in reactive gestures such as \emph{Dodge}, where PTs 
similarly replaced torso bending with a short lower-body step to restore 
balance before any fine action.

Critically, this gross-to-fine restructuring mirrors the broader contrast 
shown in Figure~\ref{fig:natural-ergonomic}: Round~1 gestures relied on 
floating, distal-dominant patterns, whereas PT-informed variants grounded fine 
actions within stable, stepwise gross-motor scaffolds—a protective, 
load-distributing strategy that preserved endurance, clarity, and safety in 
everyday AR use.

\subsection{PT-Informed Natural--Ergonomic Substitution Patterns}
\label{sec:substitution-patterns}

\begin{figure*}[!h]
  \centering
  \includegraphics[height=0.9\textheight]{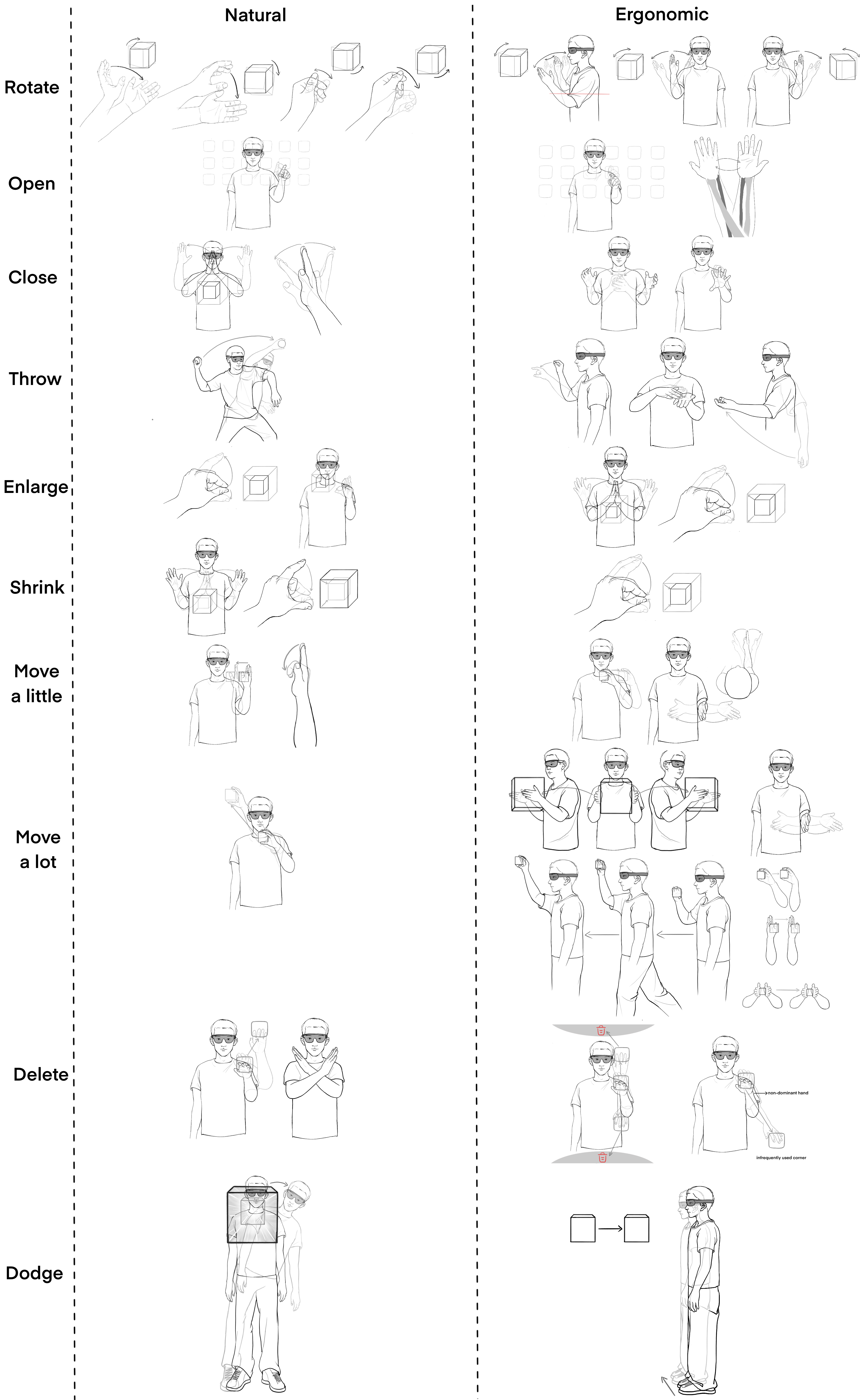}%
  \caption{\textbf{Natural versus PT-informed ergonomic substitutions.}
  The left column shows instinctive "natural" gestures elicited in Round~1; 
  the right column shows PT-guided ergonomic variants synthesized from Rounds~2--3.}
  \label{fig:natural-ergonomic}
\end{figure*}

Figure~\ref{fig:natural-ergonomic} consolidates the transformation of gesture 
intents across the three study rounds. 
The left column depicts the instinctive \emph{natural} gestures that participants 
performed when first encountering each intent in Round~1; 
the right column shows the corresponding \emph{PT-informed ergonomic substitutions} 
that therapists converged on through think-aloud reasoning in Rounds~2--3. 
We present ten representative gesture intents selected to highlight the most distinct cross-intent patterns identified in our analysis. These examples serve to foreground the analytic contrasts discussed in the Results. For completeness, Appendix~\ref{appendix:gesture-libraries} provides the full set of all fifteen PT-informed ergonomic gesture variants as part of the final gesture library.

Across intents, PTs identified recurring failure modes in the natural 
gestures: shoulder-driven arcs that rapidly increased load, 
wrist- or finger-isolated micro-movements that required continuous mid-air 
stabilization, torsional trunk rotation used for spatial reference, 
and socially ambiguous expressive motions unsuitable for everyday contexts. 
Rather than merely shrinking amplitude, PTs systematically \emph{relocated} control 
to safer joints, following the principles introduced in 
Section~\ref{sec:principles}: rotation was redirected from the spine and wrist to the 
forearm, large shoulder-led sweeps became elbow-anchored translations within a 
chest-level envelope, and single-finger "two-minute gestures" were replaced by 
multi-finger or palm-supported precision built on proximal anchoring.

Taken together, the substitution library in Figure~\ref{fig:natural-ergonomic} 
reveals three cross-intent patterns. 
First, expressive full-arm motions (e.g., \emph{Rotate}, \emph{Throw}, 
\emph{Move-a-lot}) were consistently recast as compact elbow-led gestures in which 
semantic meaning is conveyed by direction rather than magnitude. 
Second, precision tasks (e.g., \emph{Open}, \emph{Close}, \emph{Move-a-little}) 
shifted from unsupported distal control to stabilized configurations that distribute 
load across the hand and forearm. 
Third, cognitively heavy or high-stakes intents such as \emph{Delete} were 
reframed as short \emph{action chains} that separate selection, anchoring, and 
confirmation, often using bimanual coordination to reduce error and fatigue.

These patterns form the concrete bridge between the abstract biomechanical 
principles in Section~\ref{sec:principles} and the spatial specification of the 
Everyday-AR Golden Ergonomic Canvas in Section~\ref{sec:golden-canvas}. 
They show that sustainable AR-glasses gestures do not emerge from simply making 
natural motions smaller, but from \emph{reassigning} control to joints and 
sequences that respect endurance, stability, and social readability while 
preserving the core intent of the gesture.

\subsection{Everyday-AR Golden Ergonomic Canvas}
\label{sec:golden-canvas}

The Everyday-AR Golden Ergonomic Canvas synthesizes the spatial and postural constraints
that therapists repeatedly converged upon when redesigning intuitive gestures
for everyday AR glasses. Whereas the four biomechanical principles
(Sec.~4.2) capture \emph{why} specific motion patterns succeed or fail, the
Canvas specifies \emph{where} sustainable interaction occurs, and \emph{how}
joint sequencing should unfold within that region. It serves as a visual and
conceptual anchor that unifies therapist reasoning across fine-gross motor 
integration, social legibility, and long-term ergonomic safety.

\vspace{0.5em}
\noindent\textbf{Shoulder paradox as a design inflection point.}
Across all 15 gesture intents, Round~1 enactments relied heavily on the
shoulder—wide sweeps, elevated reaches, and cross-body arcs described as
"natural" starting points. However, during Round~2, therapists consistently flagged
these patterns as the fastest to fatigue and the most socially conspicuous.
Although anatomically proximal, the shoulder proved unreliable for repeated
interaction. This tension—between the shoulder’s expressive familiarity and its
poor endurance—became the pivot that redirected motion toward the elbow and
forearm, ultimately shaping the core of the Canvas.

\begin{figure*}[t]
  \centering
  \includegraphics[width=\linewidth]{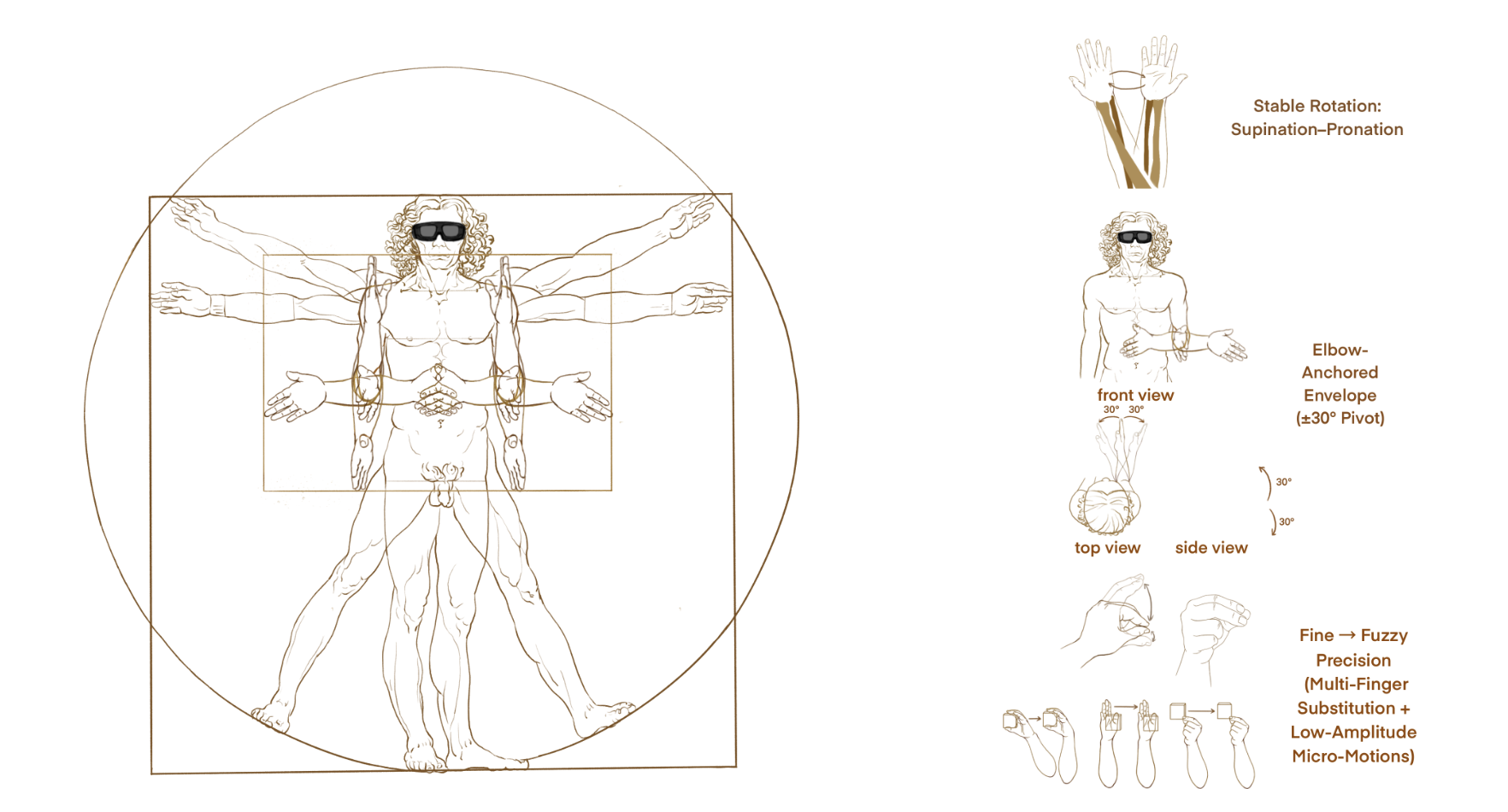}
  \caption{\textbf{Golden Ergonomic Canvas and its operational logic.}
  (Left) The Everyday-AR Golden Ergonomic Canvas: a chest-level, elbow-anchored workspace
  that bounds sustainable interaction to a narrow, repeatable region. The 
  $\pm 30^{\circ}$ envelope reflects therapist-led corrections across all 15 
  gesture intents, and the subtle lower-limb stance variation captures
  micro-adjustments of base-of-support used to maintain gross-motor stability
  (Principle~4).  
  (Right) Operational panels illustrating how therapists enacted the Canvas:
  forearm supination--pronation replacing wrist/spine rotation; front/top/side 
  views of the elbow-pivot envelope; and multi-finger fuzzy precision supporting 
  low-amplitude fine-motor tasks.}
  \label{fig:canvas}
\end{figure*}

\vspace{0.4em}
\noindent\textbf{Elbow-centered core.}
Therapists treated an elbow-anchored pivot at the mid-rib as the most reliable configuration for repeated interaction. Wrist-only gestures were fragile and difficult to stabilize, while shoulder-initiated gestures fatigued quickly.
The elbow served as a stable middle point, proximal enough to anchor and distal enough to express. This formed the geometric foundation of the Canvas.


\vspace{0.4em}
\noindent\textbf{The sustainable $\pm 30^{\circ}$ envelope.}
The Canvas constrains everyday AR interaction to a narrow 3D volume: roughly
shoulder-width in the horizontal plane and a 0–30° elevation window in the
vertical. Therapist substitutions repeatedly corrected gestures back into this
region, citing reduced fatigue, social coherence, and minimized compensatory
trunk rotation. The envelope is not a prescriptive posture, but a bounded 
workspace where gestures remain repeatable, legible, and biomechanically safe.

\vspace{0.4em}
\noindent\textbf{Operationalizing the Canvas.}
The right-hand panels (Fig.~\ref{fig:canvas}) visualize the joint-level
strategies therapists used to enact the Canvas. Forearm
supination–pronation substituted for wrist or spine rotation; small precision
adjustments were performed only after establishing a stable proximal "base step"; and multi-finger grips replaced single-digit taps to avoid distal 
floating-wrist strain. These operational rules do not introduce additional
dimensions; rather, they instantiate the Canvas during real-time gesture
substitution.

\subsection{Everyday Contexts for AR Glasses}
\label{sec:everyday-contexts}
While the Everyday-AR Golden Ergonomic Canvas defines the biomechanical envelope for 
sustainable AR interaction, therapists emphasized that gestures must also fit 
the \emph{social atmosphere} in which they are performed. Physical feasibility 
alone is insufficient: visibility, audience presence, and the perceived purpose 
of a setting strongly shape which gesture variants feel appropriate.

PTs further emphasized that social constraints can 
\emph{systematically override biomechanical ideals}. In many everyday situations, especially high-visibility or shared environments, users may avoid large-amplitude, elbow-supported, or bilateral movements even when these are the most ergonomic options, simply because such motions appear too conspicuous or socially inappropriate. This trade-off motivates the need to 
treat everyday contexts not as descriptive backdrops but as fundamental 
design parameters for sustainable AR interaction.
To capture this dimension, Round~3 asked therapists to generate—in their own 
words—the everyday contexts in which AR gestures would realistically occur (e.g.,
“on the subway,” “with friends,” “in my office,” “on the couch”). These 
participant-defined contexts were then sorted and annotated during the 
think-aloud session, revealing consistent patterns in social exposure and 
situational expectations.

\paragraph{From emergent stages to dramaturgical framing.}
Open coding first surfaced three visibility-based categories:  
(1) high-exposure settings,  
(2) moderate semi-public settings, and  
(3) low-exposure private or restorative settings.  
Further analysis showed that therapists consistently distinguished 
\emph{individual} versus \emph{public} cases within each visibility level.

To formalize these patterns, we aligned the emergent visibility gradient with 
Goffman’s dramaturgical framing—\textbf{front-stage}, \textbf{back-stage}, and 
\textbf{off-stage}. This produced the analysis-derived matrix in 
Table~\ref{tab:stage-matrix}, which synthesizes therapists’ reasoning without 
being shown directly to participants.

\begin{table}[t]
\centering
\caption{Dramaturgical matrix integrating PT-generated contexts with 
Goffman's front/back/off-stage framing (analysis-derived).}
\label{tab:stage-matrix}

\renewcommand{\arraystretch}{1.55}
\setlength{\tabcolsep}{4.5pt}

\begin{tabular}{p{0.2\linewidth} p{0.44\linewidth} p{0.24\linewidth}}
\toprule
\textbf{Stage} & \textbf{Individual} & \textbf{Public} \\
\midrule

\rowcolor{black!5}
\textbf{Front-stage} &
F1: coffee chat, party &
F2: conference presentation \\

\rowcolor{black!2}
\textbf{Back-stage} &
B1: personal workspace, cooking &
B2: office \\

\rowcolor{black!5}
\textbf{Off-stage} &
O1: at home, casual self-use &
O2: subway commute \\

\bottomrule
\end{tabular}
\end{table}

\paragraph{Six everyday contexts.}
Across these dramaturgical stages, therapists converged on six recurring 
\textbf{everyday contexts} that jointly modulate ergonomic cost, social 
visibility, and acceptable gesture size:
\begin{itemize}[leftmargin=1.3em]
  \item \textbf{O1 Amusement (Off-stage Individual)} — relaxed personal use, 
        permitting both micro-motions and energetic play.
  \item \textbf{O2 Transit (Off-stage Public)} — crowded settings requiring 
        compressed, non-disruptive gestures.
  \item \textbf{B1 Utility (Back-stage Individual)} — quick, task-focused actions 
        such as selecting or glancing.
  \item \textbf{B2 Workplace (Back-stage Public)} — semi-public office 
        environments prioritizing clarity and precision.
  \item \textbf{F1 Social (Front-stage Individual)} — casual interactions 
        where moderately expressive gestures are acceptable.
  \item \textbf{F2 Presentation (Front-stage Public)} — formal presentations 
        where large, highly legible gestures are expected.
\end{itemize}

\paragraph{Motor-skill interpretation.}
These contexts reveal a visibility-driven gradient in preferred motor skills:
\begin{itemize}[leftmargin=1.3em]
  \item \textbf{Fine motions} dominate off-stage or crowded settings (O1, O2).
  \item \textbf{Fine–gross hybrids} appear in back-stage contexts (B1, B2).
  \item \textbf{Gross, expressive gestures} are reserved for front-stage settings (F1, F2).
\end{itemize}
Amusement (O1) uniquely permits both extremes: quiet micro-gestures during relaxation and full-range movements during playful interactions.
A fine-grained view of how these motor-skill preferences manifest across the 15 gesture intents is provided in Appendix~C (Fig.~\ref{appendix:context-heatmap}), which visualizes per-intent distributions across all six dramaturgical contexts. 
This heatmap complements the macro gradient reported here by revealing the micro-variability that PTs considered when judging gesture appropriateness.

\paragraph{Relation to the Golden Canvas.}
These dramaturgical contexts define the \emph{social envelope} that complements 
the Everyday-AR Golden Ergonomic Canvas’s \emph{biomechanical envelope}. While the Canvas 
specifies sustainable joint strategies (e.g., elbow anchoring and shoulder 
minimization), the dramaturgical matrix clarifies how gesture variants should
expand, compress, or shift joint emphasis depending on visibility, audience 
presence, and situational purpose. Designing everyday AR gestures therefore 
requires optimizing both the \emph{physical} and the \emph{social} dimensions 
of action.

\subsection{Summary of Findings}

Across three rounds of PT engagement, our analysis revealed a coherent
body of clinical reasoning: Rounds~1--2 established the biomechanical
basis for gesture sustainability, while Round~3 extended this foundation into
the social and dramaturgical domain. Together, these layers reshape how
everyday AR-glasses gestures should be designed, evaluated, and socially
situated.

First, Round~1--2 open coding surfaced four cross-cutting biomechanical principles—joint-rotation substitution, shoulder-line workspace constraint, proximal–distal stability, and fine–gross motor integration. Despite differences in clinical background, PTs consistently converged on these logics when
diagnosing why intuitive gestures fail and how to re-author them into sustainable forms. The coding book ((Figure~\ref{fig:table3}) provides the explicit analytic categories used in our axial coding and clarifies how these principles were derived.

Second, these principles coalesced into the Everyday-AR Golden Ergonomic Canvas: an
elbow-anchored, shoulder-protected workspace positioned within a chest-level
$\pm 30^{\circ}$ envelope. The Canvas resolves the \emph{Shoulder Paradox}—a
proximal joint that is simultaneously expressive and biomechanically
fragile—by recentering gesture control on the elbow while preserving social
legibility.

Third, Round~3 expanded sustainability beyond biomechanics. Therapists
emphasized that gestures must also be \emph{socially calibrated}. Through a
dramaturgical matrix derived from Goffman’s frontstage/backstage/off-stage
framing and therapists’ distinctions between individual versus public use, we
identified six everyday contexts that shape gesture size, subtlety, and
acceptable effort. These contexts modulate how variants of the Everyday-AR Golden Ergonomic Canvas should be selected: compressed fine-motor forms in transit, hybrid forms in
backstage utility, and expressive gross gestures only in frontstage
performance.

Taken together, these findings show that sustainable AR-glasses interaction is
neither purely ergonomic nor purely social—it is the alignment of a
biomechanical envelope with a dramaturgical envelope. The
Everyday-AR Golden Ergonomic Canvas defines what the body can sustain; the Stage-Aware
Context Matrix defines what the situation demands. Sustainable everyday AR interaction therefore emerges when both envelopes are considered together.

This synthesis sets the stage for our Discussion, where we use these findings to
rethink gesture design, reframe everyday AR ergonomics, and articulate
implications for future wearable interaction systems.

\section{Discussion}

\subsection{Gesture as Performance: Revisiting Goffman for Everyday AR}

Our findings show that gestures for AR glasses are never merely biomechanical inputs; they function as \emph{performances} embedded within social settings.
Drawing on Goffman’s dramaturgical framing~\cite{Goffman1959} and gesture research by Goldin-Meadow~\cite{goldinmeadow1999gesture}, we argue that everyday AR places gesture production within a continuous negotiation between backstage discretion and frontstage visibility. 

Where prior HCI work on social acceptability has focused on whether gestures are publicly tolerable or embarrassing~\cite{Rico2010Social},
our findings highlight a different layer: \emph{staging}. Users implicitly modulate amplitude, posture, and expressivity depending on whether a gesture is
meant to maintain system interaction quietly or to support communicative presence. This reframes gesture design as a form of \emph{social choreography},
in which interaction techniques must simultaneously support ergonomic safety and the management of impressions in everyday life.

While social staging shapes the visibility and acceptability of gestures, a second challenge emerges from a different source: the widespread assumption that intuitive or natural gestures are inherently optimal.

\subsection{Rethinking Intuitiveness: From Naturalness to Responsibility}

Gesture elicitation studies often prize “naturalness,” assuming that culturally familiar movements are optimal~\cite{Wobbrock2009UserDefined,ruiz2011user}. Our PTs
challenged this assumption: gestures that felt natural—single-finger pinches, wrist twists, or shoulder-led pointing—proved biomechanically fragile when repeated. This aligns with VR ergonomics research showing that prolonged “natural” motions frequently induce strain~\cite{hincapie2014consumed}. We therefore argue for a shift from \emph{naturalness} (short-term familiarity) to \emph{sustainability} (joint safety, fatigue resistance, adaptability), resonating with Ability-Based Design~\cite{wobbrock2011ability}.

Naturalness is not only biomechanically misleading—it is also socially and culturally contingent. Our therapists drew on different embodied metaphors:
younger PTs invoked smartphone swipe gestures to express actions such as \emph{Delete}, whereas older PTs referenced sports-derived schemas, including a
“cut-throat’’ motion, to convey the same intent. Cross-cultural research further shows that gestures such as beckoning, pointing, waving, and thumbs-up differ in
meaning, politeness, and appropriateness across regions. Treating intuitiveness as a universal baseline therefore risks overfitting gesture vocabularies to a narrow set of bodily histories and cultural assumptions.

PT-informed translation suggests that sustainable gesture design requires attention to multiple layers of context: the biomechanics of safe repetition, the situational demands of everyday use, and the demographic or regional
practices that shape gesture meaning. Considering these factors together helps designers anticipate fatigue while supporting gesture vocabularies that remain usable and interpretable across diverse settings and populations.

\subsection{Technical Futures and Divergent Values}
Gesture recognition for AR is fragmented. Vision-based pipelines (e.g., Snap Spectacles) focus on fine-motor hand tracking~\cite{nguyen2023hands}, while sensor-based systems such as Meta’s EMG wristbands~\cite{frl2021emg,kaifosh2025neuromotor} use bio-signals. These approaches encode divergent values.

Computer vision often privileges \emph{external performance}---tracking visible outcomes or finger joints---which can obscure internal strain if the task succeeds. By contrast, EMG/IMU abstracts muscle activity and intent, aligning with backstage discretion, yet entangling the body’s intimate physiological signals and intrinsic communicative intent with data infrastructures~\cite{koelle2020methodologies}.

History shows that communication norms evolve with technology: early cellphone users speaking into headsets were once mistaken for talking to themselves. Today, small gestures (a nod, a finger raise) help onlookers interpret whether someone is conversing via earbuds. Ubiquitous AR will demand a similar renegotiation: gestures must both control systems and manage how interactions are read by others.

Our Everyday-AR Golden Ergonomic Canvas is recognizer-agnostic. Its dimensions---fatigue resistance, amplitude scaling, joint anchoring, and social legibility---apply regardless of sensing pipeline. By decoupling gesture vocabularies from any one recognizer, we foreground sustainability and communication as enduring criteria. Future AR should be judged not only on accuracy or device size, but on how well it balances bodily endurance, visibility, privacy, and the social intelligibility of gestures. Advancing AR gestures as sustainable, socially intelligible performances will require joint stewardship between researchers, designers, and platform developers.

\subsection{Design Implications for Everyday AR Gesture Design}
\label{sec:design-implications}

The PTs in our study did not merely correct unsafe movements; they consistently reframed gestures through this layered lens. They balanced proximal stability with communicative expressiveness, redesigned “natural” actions that were biomechanically misleading, repositioned the shoulder as a stabilizer rather than an actuator, and anticipated how gestures would be interpreted across different social situations and sensing modalities.

\begin{figure*}[!h]
  \centering
  \includegraphics[width=\textwidth]{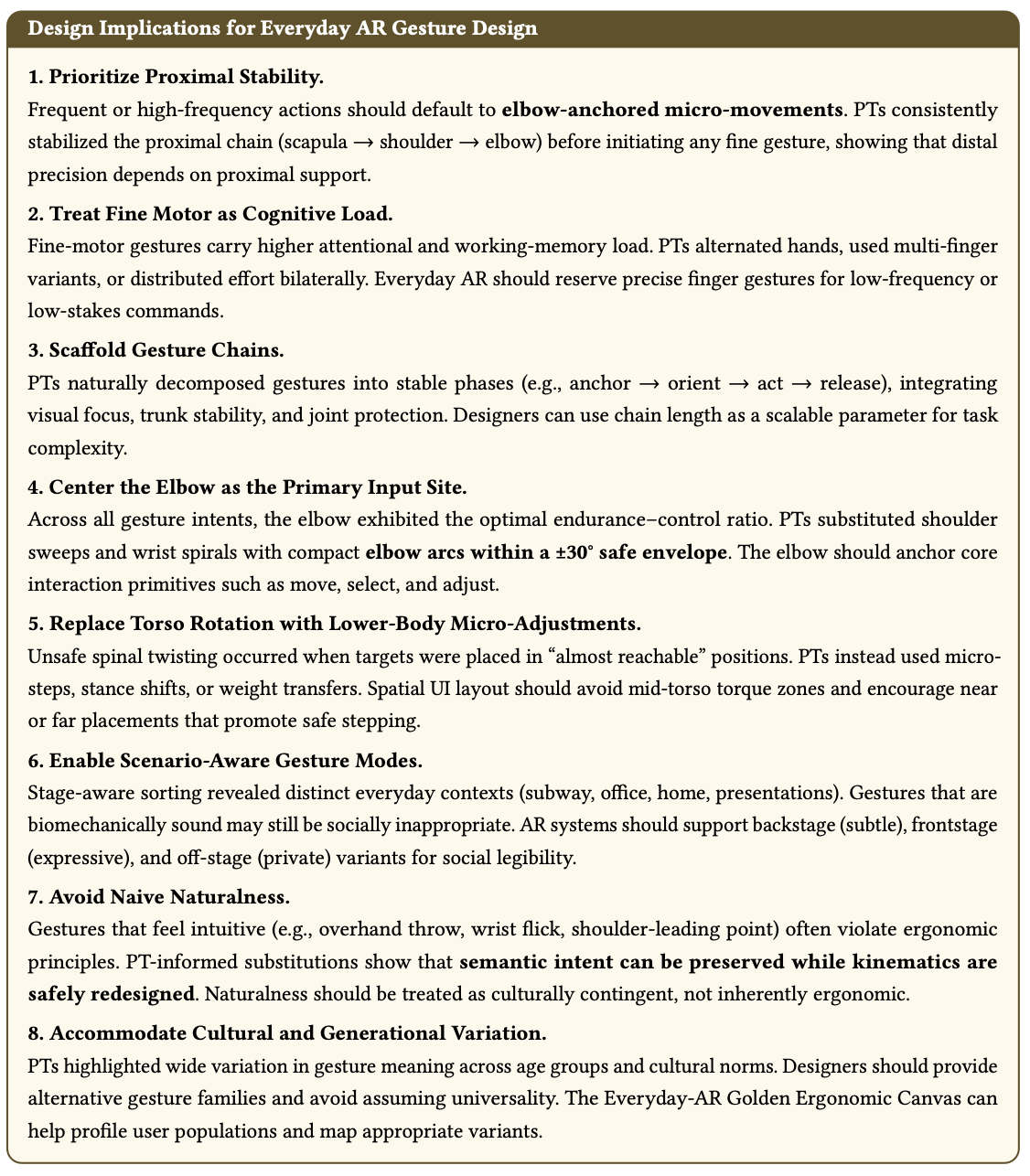}
  \caption{Design implications for everyday AR gesture design derived from PT reasoning. The figure synthesizes six core principles—ranging from proximal stability to socially legible expressiveness—into a unified scaffold for constructing sustainable, low-fatigue mid-air gesture vocabularies in wearable AR.}
  \Description{A conceptual diagram summarizing PT-informed design implications for everyday AR gesture design.The figure presents the Everyday-AR Golden Ergonomic Canvas as a unified scaffold, organizing multiple principles such as proximal stability, controlled shoulder involvement, fine-motor load management, gesture chaining, fatigue awareness, and social legibility.Together, these elements illustrate how biomechanical, cognitive, and social considerations jointly inform sustainable mid-air gesture vocabularies for wearable AR.
}

  \label{fig:wide}
\end{figure*}

The Everyday-AR Golden Ergonomic Canvas consolidates these perspectives by offering a repeatable postural scaffold for mid-air interaction—one that protects the body, respects cultural variation, and supports socially coherent expression. 
Building on this integrated view, we distill the following \textit{design implications} for constructing gesture vocabularies in wearable AR. Figure~\ref{fig:wide} visualizes these implications as an integrated design scaffold, translating PT reasoning into a set of actionable principles for everyday AR gesture vocabularies.

\section{Limitations and Future Work}

Our work is a design-theoretic and qualitative contribution; we do not present a controlled user study comparing fatigue, performance, or recognizer accuracy between gesture sets. Such studies require a stable gesture specification and theoretical scaffold; our contribution provides that precursor. Once gesture specifications are formalized, future work will incorporate end-user evaluations to assess real-world performance, fatigue, and social fit.

Our study therefore focuses on the design logic of sustainable gesture forms rather than on the recognition performance of any particular AR hardware pipeline. To keep elicitation and therapist reasoning hardware-agnostic, we did not constrain participants by the field-of-view (FOV), occlusion patterns, or joint-angle thresholds of specific commercial devices. This methodological choice reflects the current landscape of AR sensing: vision-based glasses such as Snap Spectacles rely on camera-based hand tracking with limited FOV, whereas emerging neuromotor interfaces such as Meta’s EMG-based Orion prototype interpret muscle activation and intent through wrist-worn electrodes. These pipelines differ substantially in their affordances and failure modes.

Such heterogeneity introduces practical constraints. Vision-based systems may lose track of large-amplitude or off-axis gestures, while EMG-based pipelines require calibration, exhibit muscle-specific drift, and introduce wearing burden. Our findings do not resolve these engineering challenges. Instead, they provide a shared \emph{biomechanical reference frame}---the Golden Ergonomic Canvas---from which recognizer limitations and sensing trade-offs can be systematically examined.

Future research should therefore integrate recognizer performance with ergonomic criteria, studying how FOV constraints, EMG fidelity, and vision--IMU fusion interact with the chest-level $\pm 30^\circ$ envelope identified in therapist reasoning. As sensing pipelines continue to diversify, we anticipate recognizer-aware adaptations of our framework that jointly optimize gesture recognizability, joint protection, and social legibility.

Beyond these scope limitations, our findings offer a \emph{recognizer-agnostic reference frame for sustainable gesture design in lightweight optical AR glasses}, rather than a universal solution across all platforms and cultures. We highlight three directions for extending and challenging this framework.

\paragraph{Longitudinal and situated studies.}
As AR glasses become lightweight and socially acceptable, gestures will migrate from labs and homes into the flow of everyday life.
Future work should deploy systems in the wild to examine how users sustain gestures across hours or days, negotiate visibility in public, and balance endurance with impression management.
Such studies would illuminate not only ergonomic viability but also the lived politics of discretion, dignity, and social legibility.

\paragraph{Cultural and generational diversity.}
Our interviews showed that what feels ``natural'' is historically and culturally contingent: younger PTs referenced smartphone metaphors, while older PTs drew on sports or paper-based routines.
Expanding gesture research across generations and cultures can surface multiple grammars of AR interaction, avoiding the imposition of a single normative standard.

\paragraph{Designing for sustainability and communication.}
Finally, future work should translate our Everyday-AR Golden Ergonomic Canvas into practical design tools.
Recognizer-agnostic prototyping environments, ergonomic risk scoring, and stage-aware simulations could help designers evaluate fatigue, clarity, and social fit early in ideation, fostering gesture vocabularies that are technically robust, ergonomically sound, and socially coherent.

Together, these directions frame future research not only as technical extension but as value-driven inquiry.
The challenge ahead is to craft gesture vocabularies that safeguard endurance, dignity, and communicative clarity as AR becomes embedded in everyday life.

\section{Conclusion}

As lightweight AR glasses move into everyday environments, mid-air gestures must 
support more than technical recognition—they must sustain bodies, respect social 
settings, and adapt to diverse cultural and technological contexts. Our study 
shows that intuitive gestures alone cannot meet these demands. What feels 
natural often draws on expressive motor metaphors whose biomechanical support 
was once provided by objects, surfaces, or resisted forces. In mid-air, these 
movements become fragile: expressive, visible, and semantically clear—but 
unsustainable.

By engaging licensed physical therapists as expert diagnosticians, we reframed 
gesture design as a process of \emph{clinical reasoning–driven redesign}. Through this lens, 
fragility becomes a diagnosable condition, and gesture redesign becomes an 
exercise in reorganizing joint recruitment, stabilizing the proximal chain, and 
balancing communicative intent with bodily endurance. The reasoning underlying 
these translations—distilled into our four principles, and the Everyday-AR Golden Ergonomic Canvas—offers a vocabulary for understanding how 
gestures operate simultaneously as biomechanical actions and social 
performances.

The Everyday-AR Golden Ergonomic Canvas, in particular, provides a bounded workspace for 
everyday interaction: a chest-level, elbow-anchored posture that preserves 
semantic clarity while minimizing fatigue. It does not prescribe what gestures 
should be, but supports designers in constructing gesture vocabularies that 
remain sustainable, legible, and adaptable across scenarios, populations, and 
recognition pipelines. In doing so, it reframes naturalness not as familiarity, 
but as \emph{responsible naturalness}—the alignment of meaning, endurance, and 
context.

Future AR systems will face diverging sensing capabilities and evolving 
communication norms, from fine-motor CV pipelines to EMG- and IMU-based 
intent recognition. Rather than anchoring gesture vocabularies to any one 
technology, our work argues for a durable foundation rooted in clinical 
reasoning and everyday social life. The challenge ahead is shared: to build AR 
interactions that respect the whole person—body, culture, and context—while 
advancing the expressive potential of mid-air control. By treating gestures as both embodied labor and social performance, we offer a path toward sustainable, socially coherent interaction as AR becomes woven into 
daily experience.

\begin{acks}
We thank Snap Research for providing Spectacles devices that supported this study, and we are grateful to Jesse McCulloch and Steven Xu for their technical coordination.
We also thank the Northeastern University CHI Crunch community, led by Casper Harteveld, for constructive feedback and academic support throughout the development of this work.
We are grateful to the licensed physical therapists across the United States who generously shared their time and professional expertise.
\end{acks}

\balance

\bibliographystyle{ACM-Reference-Format}
\bibliography{reference}

@book{marr2021extended,
  author    = {Marr, Bernard},
  title     = {Extended reality in practice: 100+ amazing ways virtual, augmented and mixed reality are changing business and society},
  publisher = {John Wiley \& Sons},
  year      = {2021}
}

@inproceedings{wu2025ghostgait,
  author    = {Wu, Wei and Xu, Binyan and Harteveld, Casper},
  title     = {Ghost Gait: Cultic Feedback and Meme-Driven Accountability in Wearable AR Fitness Play},
  year      = {2025},
  isbn      = {9798400720239},
  publisher = {Association for Computing Machinery},
  address   = {New York, NY, USA},
  booktitle = {Companion Proceedings of the Annual Symposium on Computer-Human Interaction in Play},
  series    = {CHI PLAY Companion '25},
  pages     = {102--106},
  numpages  = {5},
  doi       = {10.1145/3744736.3749345},
  url       = {https://doi.org/10.1145/3744736.3749345}
}

@inproceedings{xu2025ptmovementlogics,
  author    = {Xu, Binyan and Wu, Wei and Kweon, Soonhyeon and Harteveld, Casper and Chukoskie, Leanne},
  title     = {Designing Embodied AR Games Through PT Movement Logics: A Spectacles-Based Study},
  year      = {2025},
  isbn      = {9798400720239},
  publisher = {Association for Computing Machinery},
  address   = {New York, NY, USA},
  booktitle = {Companion Proceedings of the Annual Symposium on Computer-Human Interaction in Play},
  series    = {CHI PLAY Companion '25},
  pages     = {52--57},
  numpages  = {6},
  doi       = {10.1145/3744736.3749341},
  url       = {https://doi.org/10.1145/3744736.3749341}
}

@inproceedings{Fuchs1998OVST,
  author = {Rolland, Jannick P. and Fuchs, Henry},
title = {Optical Versus Video See-Through Head-Mounted Displays in Medical Visualization},
year = {2000},
issue_date = {June 2000},
publisher = {MIT Press},
address = {Cambridge, MA, USA},
volume = {9},
number = {3},
issn = {1054-7460},
url = {https://doi.org/10.1162/105474600566808},
doi = {10.1162/105474600566808},
journal = {Presence: Teleoper. Virtual Environ.},
month = jun,
pages = {287–309},
numpages = {23}
}

@misc{RayBanMetaAI,
  author    = {{Ray-Ban and Meta}},
  title     = {Ray-Ban Meta AI Glasses Product Page},
  howpublished = {\url{https://www.ray-ban.com/meta-ai-glasses}},
  year      = {2024},
  note      = {Accessed: 2025-09-12}
}

@misc{snapAWE2025,
  author    = {{Snap Inc.}},
  title     = {Spectacles Announcements at AWE 2025},
  howpublished = {\url{https://ar.snap.com/awe2025}},
  year      = {2025}
}

@inproceedings{Wobbrock2009UserDefined,
  author    = {Wobbrock, Jacob O. and Morris, Meredith Ringel and Wilson, Andrew D.},
  title     = {User-defined Gestures for Surface Computing},
  booktitle = {Proceedings of the SIGCHI Conference on Human Factors in Computing Systems},
  series    = {CHI '09},
  year      = {2009},
  pages     = {1083--1092},
  publisher = {Association for Computing Machinery},
  address   = {New York, NY, USA},
  doi       = {10.1145/1518701.1518866},
  isbn      = {978-1-60558-246-7}
}

@book{Goffman1959,
  author    = {Goffman, Erving},
  title     = {The Presentation of Self in Everyday Life},
  publisher = {Anchor},
  year      = {1959}
}

@article{pfeuffer2024gazepinch,
  author    = {Pfeuffer, K. and Gellersen, H. and Gonzalez-Franco, M.},
  title     = {Design principles and challenges for gaze+ pinch interaction in XR},
  journal   = {IEEE Computer Graphics and Applications},
  volume    = {44},
  number    = {3},
  pages     = {74--81},
  year      = {2024},
  publisher = {IEEE}
}

@article{xia2022iterative,
author = {Xia, Haijun and Glueck, Michael and Annett, Michelle and Wang, Michael and Wigdor, Daniel},
title = {Iteratively Designing Gesture Vocabularies: A Survey and Analysis of Best Practices in the HCI Literature},
year = {2022},
issue_date = {August 2022},
publisher = {Association for Computing Machinery},
address = {New York, NY, USA},
volume = {29},
number = {4},
issn = {1073-0516},
url = {https://doi.org/10.1145/3503537},
doi = {10.1145/3503537}
}

@inproceedings{HincapieRamos2014CE,
  author    = {Juan David Hincapi{\'e}-Ramos and Xiang Guo and Paymahn Moghadasian and Pourang Irani},
  title     = {Consumed Endurance: A Metric to Quantify Arm Fatigue of Mid-Air Interactions},
  booktitle = {Proceedings of the SIGCHI Conference on Human Factors in Computing Systems (CHI '14)},
  year      = {2014},
  pages     = {1063--1072},
  publisher = {ACM},
  doi       = {10.1145/2556288.2557130}
}

@inproceedings{Rico2010Social,
  author    = {Julie Rico and Stephen Brewster},
  title     = {Usable Gestures for Mobile Interfaces: Evaluating Social Acceptability},
  booktitle = {Proceedings of the SIGCHI Conference on Human Factors in Computing Systems (CHI '10)},
  year      = {2010},
  pages     = {887--896},
  publisher = {ACM},
  doi       = {10.1145/1753326.1753458}
}

@inproceedings{Serrano2014HandToFace,
  author    = {Marcos Serrano and Barrett Ens and Pourang Irani},
  title     = {Exploring Hand-to-Face Input for Head-Worn Displays},
  booktitle = {Proceedings of the SIGCHI Conference on Human Factors in Computing Systems (CHI '14)},
  year      = {2014},
  pages     = {3181--3190},
  publisher = {ACM},
  doi       = {10.1145/2556288.2556984}
}

@misc{reutersSnap2025,
  author    = {{Reuters}},
  title     = {Snap to launch smart glasses for users in 2026},
  howpublished = {\url{https://www.reuters.com/technology/snap-launch-consumer-ar-glasses-2026-2025-06-01}},
  note      = {Accessed: 2025-09-06},
  year      = {2025}
}

@misc{metaOrion,
  author    = {{Meta}},
  title     = {Introducing Orion: Our First True AR Glasses},
  howpublished = {\url{https://about.meta.com/news/orion-ar-glasses}},
  note      = {Accessed: 2025-09-06},
  year      = {2024}
}

@misc{uploadvrOrion2025,
  author    = {{UploadVR}},
  title     = {Meta Orion AR Glasses Expected for Developers in 2026},
  howpublished = {\url{https://uploadvr.com/meta-orion-ar-glasses-2026}},
  note      = {Accessed: 2025-09-06},
  year      = {2025}
}

@misc{MetaAIGlassesSite,
  author    = {{Meta}},
  title     = {Meta AI Glasses Features},
  howpublished = {\url{https://about.meta.com/ai-glasses}},
  note      = {Accessed: 2025-09-06},
  year      = {2024}
}

@article{duval2022spellcasters,
author = {Duval, Jared and Thakkar, Rutul and Du, Delong and Chin, Kassandra and Luo, Sherry and Elor, Aviv and El-Nasr, Magy Seif and John, Michael},
title = {Designing Spellcasters from Clinician Perspectives: A Customizable Gesture-Based Immersive Virtual Reality Game for Stroke Rehabilitation},
year = {2022},
issue_date = {September 2022},
publisher = {Association for Computing Machinery},
address = {New York, NY, USA},
volume = {15},
number = {3},
issn = {1936-7228},
url = {https://doi.org/10.1145/3530820},
doi = {10.1145/3530820},
journal = {ACM Trans. Access. Comput.},
month = aug,
articleno = {26},
numpages = {25}
}

@article{wobbrock2011ability,
author = {Wobbrock, Jacob O. and Kane, Shaun K. and Gajos, Krzysztof Z. and Harada, Susumu and Froehlich, Jon},
title = {Ability-Based Design: Concept, Principles and Examples},
year = {2011},
issue_date = {April 2011},
publisher = {Association for Computing Machinery},
address = {New York, NY, USA},
volume = {3},
number = {3},
issn = {1936-7228},
url = {https://doi.org/10.1145/1952383.1952384},
doi = {10.1145/1952383.1952384},
}

@article{li2025vibring,
  author    = {Li, B. and Huang, X. and Xiao, R.},
  title     = {VibRing: A Wearable Vibroacoustic Sensor for Single-Handed Gesture Recognition},
  journal   = {Proceedings of the ACM on Human-Computer Interaction},
  volume    = {9},
  number    = {4},
  pages     = {1--25},
  year      = {2025},
  publisher = {Association for Computing Machinery}
}

@book{casesmith2014occupational,
  title={Occupational Therapy for Children},
  author={Case-Smith, Jane and O’Brien, Jeanette},
  publisher={Elsevier},
  year={2014}
}

@book{bruininks2005bot2,
  title={Bruininks-Oseretsky Test of Motor Proficiency (BOT-2)},
  author={Bruininks, Robert},
  year={2005}
}

@misc{APTAeducation2020,
  title={Physical Therapist Education Framework},
  author={{American Physical Therapy Association}},
  year={2020},
  howpublished={\url{https://www.apta.org/}},
  note={Accessed: 2025-08-26}
}

@misc{WHOrehab2017,
  title={Rehabilitation in Health Systems},
  author={{World Health Organization}},
  year={2017},
  howpublished={\url{https://www.who.int/publications/i/item/9789241549974}},
  note={Accessed: 2025-08-26}
}

@misc{glasswiki,
  title   = {Google Glass},
  year    = {2025},
  howpublished = {\url{https://en.wikipedia.org/wiki/Google_Glass}},
  note    = {Accessed 2025-09-05}
}

@article{time2015glass,
  author  = {Satariano, Adam},
  title   = {Google Will Stop Selling Glass Next Week},
  journal = {TIME},
  year    = {2015},
  url     = {https://time.com/3669927/google-glass-explorer-program-ends/}
}

@misc{mslearn_hands,
  title   = {Hand and Finger Tracking on HoloLens 2},
  howpublished = {\url{https://learn.microsoft.com/windows/mixed-reality/design/hand-and-finger-tracking}},
  year    = {2024},
  note    = {Accessed 2025-09-05}
}

@misc{mslearn_eyes,
  title   = {Eye Tracking on HoloLens 2},
  howpublished = {\url{https://learn.microsoft.com/windows/mixed-reality/design/eye-tracking}},
  year    = {2024},
  note    = {Accessed 2025-09-05}
}

@misc{verge2022ml2,
  author  = {Robertson, Adi},
  title   = {Magic Leap 2 Is Rolling Out to Healthcare Partners},
  howpublished = {\url{https://www.theverge.com/2022/1/12/22880090/magic-leap-2-ar-headset-rollout-healthcare-companies}},
  year    = {2022}
}

@misc{ml22022pr,
  title   = {Magic Leap 2 Now Available to Customers as the Most Immersive AR Headset for Enterprise},
  howpublished = {\url{https://www.prnewswire.com/news-releases/magic-leap-2-now-available-to-customers-as-the-most-immersive-augmented-reality-headset-for-enterprise-301637447.html}},
  year    = {2022},
  note    = {Accessed 2025-09-05}
}

@article{ong2021xrtelehealth,
  author  = {Ong, T. and colleagues},
  title   = {Extended Reality for Telehealth During COVID-19: Review},
  journal = {Frontiers in Virtual Reality},
  year    = {2021},
  url     = {https://pmc.ncbi.nlm.nih.gov/articles/PMC8315161/}
}

@article{dinh2023artelemed,
  author  = {Dinh, A. and colleagues},
  title   = {Augmented Reality in Real-Time Telemedicine and Telementoring: Systematic Review},
  journal = {JMIR mHealth and uHealth},
  year    = {2023},
  url     = {https://mhealth.jmir.org/2023/1/e45464}
}

@article{jms2024mror,
  title   = {Mixed Reality in the Operating Room: A Systematic Review},
  journal = {Journal of Medical Systems},
  volume  = {48},
  year    = {2024},
  url     = {https://link.springer.com/article/10.1007/s10916-024-02095-7}
}

@misc{snap2021spectacles,
  title   = {Introducing the Next Generation of Spectacles},
  howpublished = {\url{https://newsroom.snap.com/introducing-the-next-generation-of-spectacles}},
  year    = {2021}
}

@misc{meta2024rbai,
  title   = {New Ray-Ban~|~Meta Styles and Meta AI with Vision},
  howpublished = {\url{https://about.fb.com/news/2024/04/new-ray-ban-meta-smart-glasses-styles-and-meta-ai-updates/}},
  year    = {2024}
}

@misc{frl2021emg,
  author  = {{Facebook Reality Labs}},
  title   = {Inside Facebook Reality Labs: Wrist-based Interaction for the Next Computing Platform},
  year    = {2021},
  howpublished = {https://tech.facebook.com/reality-labs/2021/3/inside-facebook-reality-labs-wrist-based-interaction-for-the-next-computing-platform/}
}

@article{feg2020ergonomics,
  author={Federation of European Ergonomics},
  title={Ergonomic guidelines for AR/VR},
  journal={Ergonomics},
  year={2020}
}

@book{dutta2024immersive,
  author    = {Dutta, S.},
  title     = {Immersive Realm of Extended Reality},
  year      = {2024},
  publisher = {BPB Publications},
  address   = {Delhi}
}

@inproceedings{bowman2021everyday,
  author    = {Douglas A. Bowman},
  title     = {User Experience Considerations for Everyday Augmented Reality (Keynote)},
  booktitle = {ISMAR},
  year      = {2021},
  pages     = {16--16},
  publisher = {IEEE}
}

@inproceedings{tran2025wearable,
author = {Tran, Tram Thi Minh and Brown, Shane and Weidlich, Oliver and Yoo, Soojeong and Parker, Callum},
title = {Wearable AR in Everyday Contexts: Insights from a Digital Ethnography of YouTube Videos},
year = {2025},
isbn = {9798400713941},
publisher = {Association for Computing Machinery},
address = {New York, NY, USA},
url = {https://doi.org/10.1145/3706598.3713572},
doi = {10.1145/3706598.3713572}
}

@inproceedings{Burnett2013Swipe,
  author    = {Gary Burnett and Elizabeth Crundall and David Large and Glyn Lawson and Lee Skrypchuk},
  title     = {A Study of Unidirectional Swipe Gestures on In-Vehicle Touch Screens},
  booktitle = {Proceedings of the International Conference on Automotive User Interfaces and Interactive Vehicular Applications (AutomotiveUI)},
  year      = {2013},
  pages     = {22--29},
  address   = {New York, NY, USA},
  publisher = {ACM}
}

@inproceedings{Garzotto2013Touchless,
  author    = {Franca Garzotto and Matteo Valoriani},
  title     = {Touchless Gestural Interaction with Small Displays: A Case Study},
  booktitle = {Proceedings of the Biannual Conference of the Italian Chapter of SIGCHI (CHItaly)},
  year      = {2013},
  address   = {New York, NY, USA},
  publisher = {ACM}
}

@inproceedings{Hinckley2011SensorSynaesthesia,
  author    = {Ken Hinckley and Hyunyoung Song},
  title     = {Sensor Synaesthesia: Touch in Motion and Motion in Touch},
  booktitle = {Proceedings of the SIGCHI Conference on Human Factors in Computing Systems (CHI)},
  year      = {2011},
  pages     = {801--810},
  address   = {New York, NY, USA},
  publisher = {ACM}
}

@inproceedings{Annenberg1989Gesture,
  author    = {Dannenberg, Roger B. and Amon, Dale},
  title     = {A Gesture Based User Interface Prototyping System},
  booktitle = {Proceedings of the 2nd Annual ACM SIGGRAPH Symposium on User Interface Software and Technology (UIST '89)},
  year      = {1989},
  pages     = {127--132},
  doi       = {10.1145/73660.73676}
}

@inproceedings{Ashbrook2010MAGIC,
author = {Ashbrook, Daniel and Starner, Thad},
title = {MAGIC: a motion gesture design tool},
year = {2010},
isbn = {9781605589299},
publisher = {Association for Computing Machinery},
address = {New York, NY, USA},
url = {https://doi.org/10.1145/1753326.1753653},
doi = {10.1145/1753326.1753653},
booktitle = {Proceedings of the SIGCHI Conference on Human Factors in Computing Systems},
pages = {2159–2168},
numpages = {10},
location = {Atlanta, Georgia, USA},
series = {CHI '10}
}

@inproceedings{Dey2004aCAPpella,
  author    = {Anind K. Dey and Raffay Hamid and Chris Beckmann and Ian Li and Daniel Hsu},
  title     = {a CAPpella: Programming by Demonstration of Context-Aware Applications},
  booktitle = {Proceedings of the SIGCHI Conference on Human Factors in Computing Systems (CHI)},
  year      = {2004},
  pages     = {33--40},
  address   = {New York, NY, USA},
  publisher = {ACM}
}

@inproceedings{Lu2013GestureStudio,
  author    = {Hao L{\"u} and Yang Li},
  title     = {Gesture Studio: Authoring Multi-Touch Interactions through Demonstration and Declaration},
  booktitle = {Proceedings of the SIGCHI Conference on Human Factors in Computing Systems (CHI)},
  year      = {2013},
  pages     = {257--266},
  address   = {New York, NY, USA},
  publisher = {ACM}
}

@inproceedings{Nielsen2004Procedure,
  author    = {Michael Nielsen and Moritz St{\"o}rring and Thomas B. Moeslund and Erik Granum},
  title     = {A Procedure for Developing Intuitive and Ergonomic Gesture Interfaces for HCI},
  booktitle = {International Gesture Workshop (IGW)},
  year      = {2004},
  pages     = {409--420},
  address   = {Berlin, Heidelberg},
  publisher = {Springer}
}

@inproceedings{Pyryeskin2012Comparing,
  author    = {Dmitry Pyryeskin and Mark Hancock and Jesse Hoey},
  title     = {Comparing Elicited Gestures to Designer-Created Gestures for Selection Above a Multitouch Surface},
  booktitle = {Proceedings of the ACM Conference on Interactive Tabletops and Surfaces (ITS)},
  year      = {2012},
  pages     = {1--10},
  address   = {New York, NY, USA},
  publisher = {ACM}
}

@mastersthesis{schan2019smartglasses,
  author       = {Schan, C.},
  title        = {Use of smart glasses in order picking operations},
  school       = {Norwegian University of Science and Technology (NTNU)},
  year         = {2019},
  type         = {Master's thesis}
}

@incollection{gajsek2019smart,
  author       = {Gajšek, B. and Vujica Herzog, N.},
  title        = {Smart glasses in sustainable manual order picking systems},
  booktitle    = {Sustainable Logistics and Production in Industry 4.0: New Opportunities and Challenges},
  pages        = {219--241},
  publisher    = {Springer International Publishing},
  year         = {2019},
  address      = {Cham}
}

@techreport{murauer2019fullshift,
  author       = {Murauer, C. S.},
  title        = {Full shift usage of smart glasses in order picking processes considering a methodical approach of continuous user involvement},
  institution  = {Unknown Institution},
  year         = {2019},
  note         = {Report or thesis; publication venue unspecified}
}

@misc{matveiuk2019xr,
  author       = {Matveiuk, K.},
  title        = {Role of XR wearables in intralogistics field: Insight into AR applications},
  year         = {2019},
  note         = {Unpublished or grey literature; publication venue not specified}
}

@techreport{booth2019af,
  author       = {Booth, J. A.},
  title        = {The Use of Virtual and Augmented Realities in Air Force Training},
  year         = {2019},
  institution  = {United States Air Force},
  note         = {Technical report}
}

@article{zari2023magic,
  author       = {Zari, G. and Condino, S. and Cutolo, F. and Ferrari, V.},
  title        = {Magic Leap 1 versus Microsoft HoloLens 2 for the visualization of 3D content obtained from radiological images},
  journal      = {Sensors},
  volume       = {23},
  number       = {6},
  pages        = {3040},
  year         = {2023},
  publisher    = {MDPI},
  doi          = {10.3390/s23063040}
}

@article{caruso2021eye,
  author       = {Caruso, T. J. and Hess, O. and Roy, K. and Wang, E. and Rodriguez, S. and Palivathukal, C. and Haber, N.},
  title        = {Integrated eye tracking on Magic Leap One during augmented reality medical simulation: a technical report},
  journal      = {BMJ Simulation \& Technology Enhanced Learning},
  doi =         {10.1136/bmjstel-2020-000782},
  volume       = {7},
  number       = {5},
  pages        = {431},
  year         = {2021},
  publisher    = {BMJ},
}

@inproceedings{andersson2020developing,
  author={Andersson, Henrik Bjelke and Børresen, Thomas and Prasolova-Førland, Ekaterina and McCallum, Simon and Estrada, Jose Garcia},
  booktitle={2020 IEEE Conference on Virtual Reality and 3D User Interfaces Abstracts and Workshops (VRW)}, 
  title={Developing An AR Application For Neurosurgical Training: Lessons Learned For Medical Specialist Education}, 
  year={2020},
  pages={407-412},
  address      = {Atlanta, GA, USA},
  publisher    = {IEEE},
  organization = {IEEE},
  doi={10.1109/VRW50115.2020.00087}}

@article{nickel2022telestration,
  author    = {Nickel, F. and Cizmic, A. and Chand, M.},
  title     = {Telestration and augmented reality in minimally invasive surgery: An invaluable tool in the age of COVID-19 for remote proctoring and telementoring},
  journal   = {JAMA Surgery},
  volume    = {157},
  number    = {2},
  pages     = {169--170},
  year      = {2022},
  publisher = {American Medical Association},

}

@online{snap2026specs,
  title        = {Snap to Launch New Lightweight, Immersive Specs in 2026},
  year         = {2025},
  month        = sep,
  url          = {https://www.snap.com/en-US/news/post/snap-to-launch-new-lightweight-immersive-specs-2026},
  note         = {Accessed: 2025-09-10},
  organization = {Snap Inc.}
}

@online{metaOrionTech,
  title        = {Orion: Emerging Technology},
  year         = {2025},
  url          = {https://www.meta.com/emerging-tech/orion/?srsltid=AfmBOooO16OcJ4_sG9HUPnv6119TMEDyhnlbf6SyPII1h3qmHzHRb6bQ},
  note         = {Accessed: 2025-09-10},
  organization = {Meta}
}

@online{metaOrionNews,
  title        = {Introducing Orion: Our First True Augmented Reality Glasses},
  year         = {2024},
  month        = sep,
  url          = {https://about.fb.com/news/2024/09/introducing-orion-our-first-true-augmented-reality-glasses/},
  note         = {Accessed: 2025-09-10},
  organization = {Meta}
}

@article{kaifosh2025neurointerface,
  author    = {Kaifosh, P. and Reardon, T. R. and CTRL-labs at Reality Labs},
  title     = {A generic non-invasive neuromotor interface for human-computer interaction},
  journal   = {Nature},
  year      = {2025},
  publisher = {Springer Nature},
  doi       = {10.1038/s41586-025-09255-w}
}

@inproceedings{bonner2023filters,
  author    = {Bonner, J. and Mathis, F. and O'Hagan, J. and McGill, M.},
  title     = {When filters escape the smartphone: Exploring acceptance and concerns regarding augmented expression of social identity for everyday AR},
  booktitle = {Proceedings of the 29th ACM Symposium on Virtual Reality Software and Technology},
  year      = {2023},
  month     = oct,
  pages     = {1--14},
  publisher = {ACM},
  address = {New York, NY, USA},
}

@article{haider2025ar,
  author    = {Haider, U. and Rasheed, S. and Ahmed, T. and Ahmed, F. and Nawaz, A. and Rajper, A. and ur Rasool, A.},
  title     = {Applications of Augmented Reality in Industrial Manufacturing in the Era of Industry 5.0},
  journal   = {International Journal of Engineering and Applied Physics},
  volume    = {5},
  number    = {1},
  pages     = {1127--1135},
  year      = {2025},
  publisher = {International Journal of Engineering and Applied Physics},

}

@inproceedings{alrawi2025remote,
  author    = {AlRawi, L. N. and Wang, L. and Socha, T.},
  title     = {Remote Operation Support Based on Wearable Augmented Reality: Challenges and Field Deployment},
  booktitle = {SPE/IADC Drilling Conference and Exhibition},
  address = {Stavanger, Norway},
  pages     = {D021S009R003},
  year      = {2025},
  month     = feb,
  publisher = {Society of Petroleum Engineers (SPE)},
  doi       = {10.2118/223659-MS}
}

@book{simeone2023everyday,
  title={Everyday virtual and augmented reality},
  author={Simeone, Adalberto and Weyers, Benjamin and Bialkova, Svetlana and Lindeman, Robert William},
  year={2023},
  publisher={Springer}
}

@article{goldinmeadow1999gesture,
  author  = {Susan Goldin-Meadow},
  title   = {The role of gesture in communication and thinking},
  journal = {Trends in Cognitive Sciences},
  year    = {1999},
  volume  = {3},
  number  = {11},
  pages   = {419--429},
  doi     = {10.1016/S1364-6613(99)01397-2}
}

@article{kaifosh2025neuromotor,
  author    = {Kaifosh, Patrick and Reardon, Thomas R. and CTRL-labs at Reality Labs},
  title     = {A generic non-invasive neuromotor interface for human--computer interaction},
  journal   = {Nature},
  year      = {2025},
  doi       = {10.1038/s41586-025-09255-w},
  note      = {Advance online publication}
}

@inproceedings{nguyen2023hands,
  author    = {Thanh Nguyen and Yvonne Rogers},
  title     = {Hands in the Air: Design Challenges for Mid-air Gestural Interfaces},
  booktitle = {Proceedings of the ACM Conference on Human Factors in Computing Systems (CHI '23)},
  pages     = {1--15},
  year      = {2023},
  publisher = {ACM},
  doi       = {10.1145/3544548.3581374}
}

@article{koelle2020methodologies,
  author  = {Marion Koelle and Daniel Buschek and Susanne Boll},
  title   = {Exploring Social Acceptability of Physiological Sensing},
  journal = {Proceedings of the ACM on Interactive, Mobile, Wearable and Ubiquitous Technologies},
  year    = {2020},
  volume  = {4},
  number  = {3},
  pages   = {1--26},
  doi     = {10.1145/3411823}
}

@book{aurelia2024immersive,
    author={Aurelia, Sagaya},
    title={Immersive Technologies: Navigating the Impacts, Challenges, and Opportunities},
    publisher={CRC Press},
    year={2024},
}

@article{schuermans2022msk,
  author    = {Schuermans, J. and Van Hootegem, A. and Van den Bossche, M. and Van Gendt, M. and Witvrouw, E. and Wezenbeek, E.},
  title     = {Extended reality in musculoskeletal rehabilitation and injury prevention -- A systematic review},
  journal   = {Physical Therapy in Sport},
  volume    = {55},
  pages     = {229--240},
  year      = {2022},
  publisher = {Elsevier}
}

@article{morimoto2022spinexr,
  author    = {Morimoto, T. and Kobayashi, T. and Hirata, H. and Otani, K. and Sugimoto, M. and Tsukamoto, M. and Mawatari, M.},
  title     = {XR (extended reality: virtual reality, augmented reality, mixed reality) technology in spine medicine: Status quo and quo vadis},
  journal   = {Journal of Clinical Medicine},
  volume    = {11},
  number    = {2},
  pages     = {470},
  year      = {2022},
  publisher = {MDPI}
}

@article{burke2025higheredxr,
  author    = {Burke, D. and Crompton, H. and Nickel, C.},
  title     = {The Use of Extended Reality (XR) in Higher Education: A Systematic Review},
  journal   = {TechTrends},
  pages     = {1--14},
  year      = {2025},
  publisher = {Springer}
}

@article{alhakamy2024xrindustry,
author = {Alhakamy, A’aeshah},
title = {Extended Reality (XR) Toward Building Immersive Solutions: The Key to Unlocking Industry 4.0},
year = {2024},
issue_date = {September 2024},
publisher = {Association for Computing Machinery},
address = {New York, NY, USA},
volume = {56},
number = {9},
issn = {0360-0300},
url = {https://doi.org/10.1145/3652595},
doi = {10.1145/3652595},
journal = {ACM Comput. Surv.},
month = apr,
articleno = {237},
numpages = {38}
}

@inproceedings{ali2023xrhealthcare,
  author={Ali, Shah Mahsoom and Aich, Satyabrata and Athar, Ali and Kim, Hee-Cheol},
  booktitle={2023 25th International Conference on Advanced Communication Technology (ICACT)}, 
  title={Medical Education, Training and Treatment Using XR in Healthcare}, 
  address      = {Pyeongchang, South Korea},
  pages        = {388--393},
  publisher    = {IEEE}, 
  year={2023},
  doi={10.23919/ICACT56868.2023.10079321}}

@article{luo2024xrealityenglish,
  author    = {Luo, S. and Zou, D. and Kohnke, L.},
  title     = {A Systematic Review of Research on xReality (XR) in the English Classroom: Trends, Research Areas, Benefits, and Challenges},
  journal   = {Computers \& Education: X Reality},
  volume    = {4},
  pages     = {100049},
  year      = {2024},
  publisher = {Elsevier}
}

@article{mizrahi2020neuromechanical,
  author    = {Mizrahi, J.},
  title     = {Neuro-mechanical aspects of playing-related mobility disorders in orchestra violinists and upper strings players: A review},
  journal   = {European Journal of Translational Myology},
  volume    = {30},
  number    = {3},
  pages     = {9095},
  year      = {2020}
}

@article{stergiou2006optimal,
  author    = {Stergiou, N. and Harbourne, R. T. and Cavanaugh, J. T.},
  title     = {Optimal movement variability: A new theoretical perspective for neurologic physical therapy},
  journal   = {Journal of Neurologic Physical Therapy},
  volume    = {30},
  number    = {3},
  pages     = {120--129},
  year      = {2006}
}

@article{schwartz2022attunement,
  author    = {Schwartz, B. and Rubel, J. A. and Deisenhofer, A. K. and Lutz, W.},
  title     = {Movement-based patient--therapist attunement in psychological therapy and its association with early change},
  journal   = {Digital Health},
  volume    = {8},
  pages     = {20552076221129098},
  year      = {2022}
}

@inproceedings{hincapie2014consumed,
  title={Consumed endurance: A metric to quantify arm fatigue of mid-air interactions},
  author={Hincapié-Ramos, Juan David and Guo, Xian and Moghadasian, Parastoo and Irani, Pourang},
  booktitle={CHI},
  pages={1063--1072},
  year={2014}
}

@inproceedings{ruiz2011user,
  title={User-defined motion gestures for mobile interaction},
  author={Ruiz, Jaime and Li, Yang and Lank, Edward},
  booktitle={CHI},
  pages={197--206},
  year={2011}
}

@inproceedings{hook2009softn,
author = {Schiphorst, Thecla},
title = {soft(n): toward a somaesthetics of touch},
year = {2009},
isbn = {9781605582474},
publisher = {Association for Computing Machinery},
address = {New York, NY, USA},
url = {https://doi.org/10.1145/1520340.1520345},
doi = {10.1145/1520340.1520345},

}

@article{schiphorst2013moving,
  title={Moving and making strange: An embodied approach to movement-based interaction design},
  author={Schiphorst, Thecla},
  journal={ACM Transactions on Computer-Human Interaction},
  volume={20},
  number={1},
  pages={7:1--7:25},
  year={2013},
  publisher={ACM},
  doi={10.1145/2442106.2442113}
}

@inproceedings{hook2017embodied,
author = {van Dijk, Jelle and Hummels, Caroline},
title = {Designing for Embodied Being-in-the-World: Two Cases, Seven Principles and One Framework},
year = {2017},
isbn = {9781450346764},
publisher = {Association for Computing Machinery},
address = {New York, NY, USA},
url = {https://doi.org/10.1145/3024969.3025007},
doi = {10.1145/3024969.3025007}
}

@article{metaNature2025,
  title={Vision-based continuous hand- and wrist-tracking for wearable AR},
  author={Meta Reality Labs},
  journal={Nature},
  year={2025},
  doi={10.1038/s41586-025-09255-w}
}

\clearpage
\appendix
\section{Appendix A: Materials and Supplementary Analyses}

\subsection{Corpus Statistics and Counting}
\label{appendix:corpus-prevalence}
\begin{figure}[h]
  \centering

  \begin{subfigure}[t]{\columnwidth}
    \centering
    \includegraphics[width=0.95\linewidth]{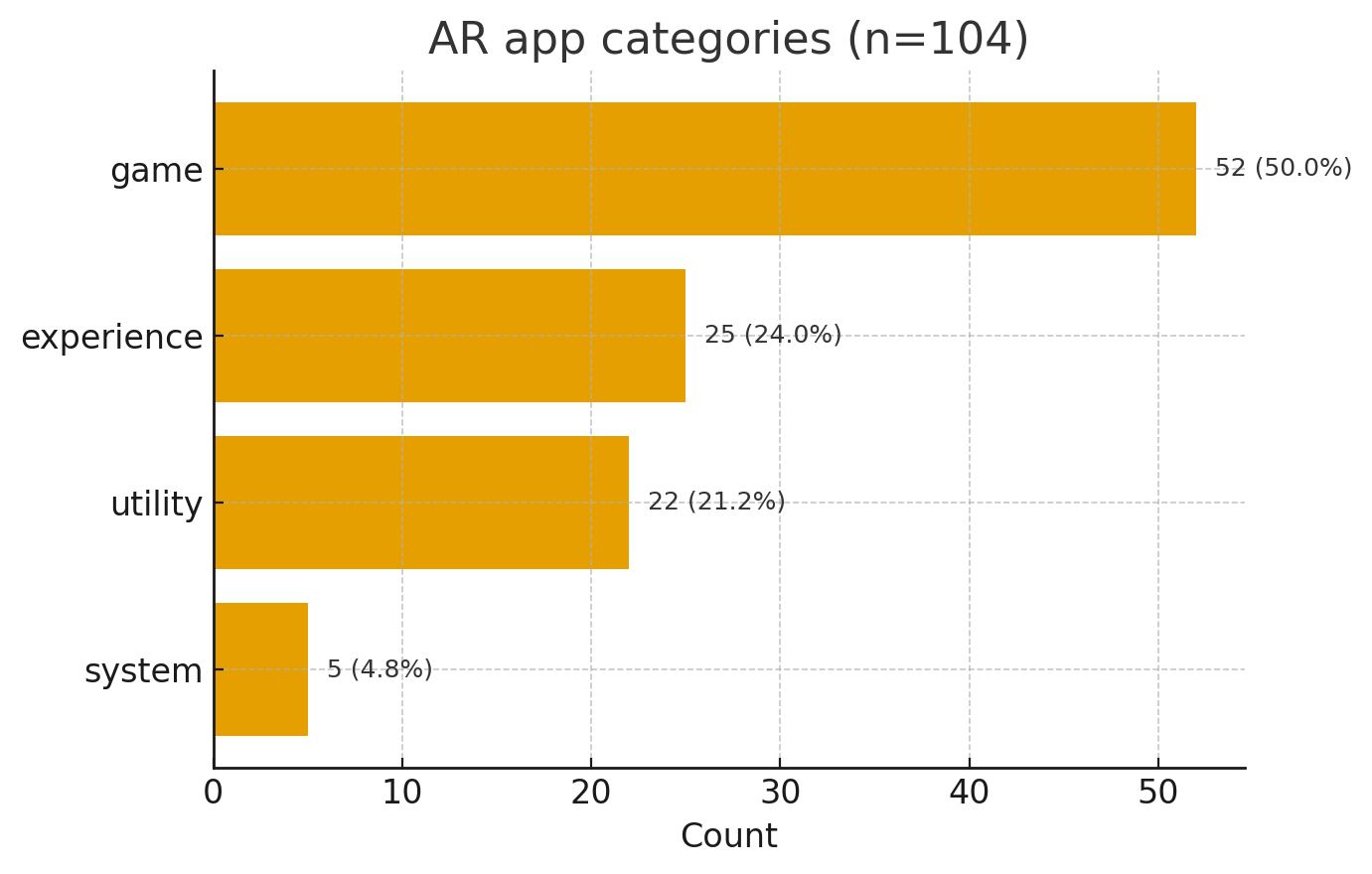}
    \caption{Distribution of app categories (n=104).}
    \label{fig:appcats_appendix}
  \end{subfigure}

  \vspace{0.8em}

  \begin{subfigure}[t]{\columnwidth}
    \centering
    \includegraphics[width=0.95\linewidth]{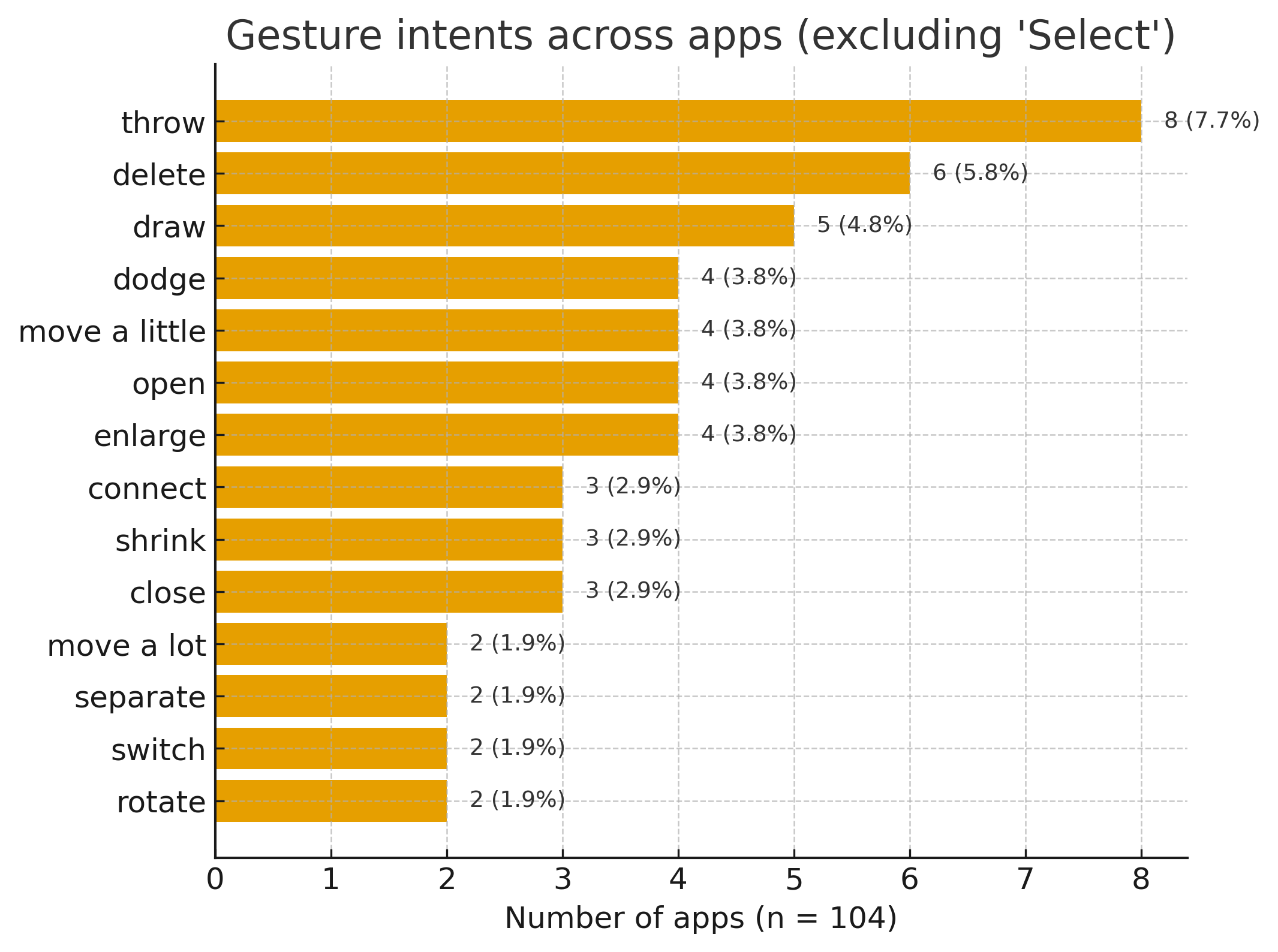}
    \caption{Prevalence of gesture intents (excluding \emph{Select}).}
    \label{fig:intents_noselect_bar_appendix}
  \end{subfigure}

  \caption{Corpus overview for Snap Spectacles AR applications (August~2025).
  Top: distribution of app categories (n=104).
  Bottom: prevalence of gesture intents across 104 apps, excluding \emph{Select} as a ubiquitous baseline.}
  \label{fig:corpus_overview_appendix}
\end{figure}

Across the 104 applications we reviewed, the most common non-overlapping gesture intents were 
\emph{Throw} (8/104; 7.7\%), \emph{Delete} (6/104; 5.8\%), and \emph{Draw} (5/104; 4.8\%), with other intents (e.g., \emph{Rotate}, \emph{Shrink}, \emph{Connect}) each occurring in roughly 2--4\% of apps. 
Because \emph{Select} appeared in 96\% of apps and often overlaps other actions, we treat it as a baseline and exclude it from the comparative prevalence analysis. 
Near-identical listings (e.g., variants with the same title and publisher) were de-duplicated prior to analysis.

\paragraph{Counting strategy.}
Each application could support multiple gesture intents. To reflect design coverage per referent without inflating totals, we applied two rules:
\begin{enumerate}[leftmargin=1.5em,itemsep=2pt]
  \item \textbf{Baseline treatment of \emph{Select}.} Given its ubiquity and overlap, \emph{Select} was recorded independently as an overlapping baseline.
  \item \textbf{Unique counts for other intents.} All remaining intents were counted once per app (non-overlapping), regardless of within-app repetitions.
\end{enumerate}
A reproducibility note and a de-identified appraisal sheet (app list with per-intent flags) are available upon request.

\vspace{0.5\baselineskip}

\subsection{Example Gesture Cues}
\label{appendix:cue-gallery}

\begin{figure*}[!htb]
  \centering

  \begin{minipage}{0.45\textwidth}
    \centering
    \includegraphics[width=\linewidth]{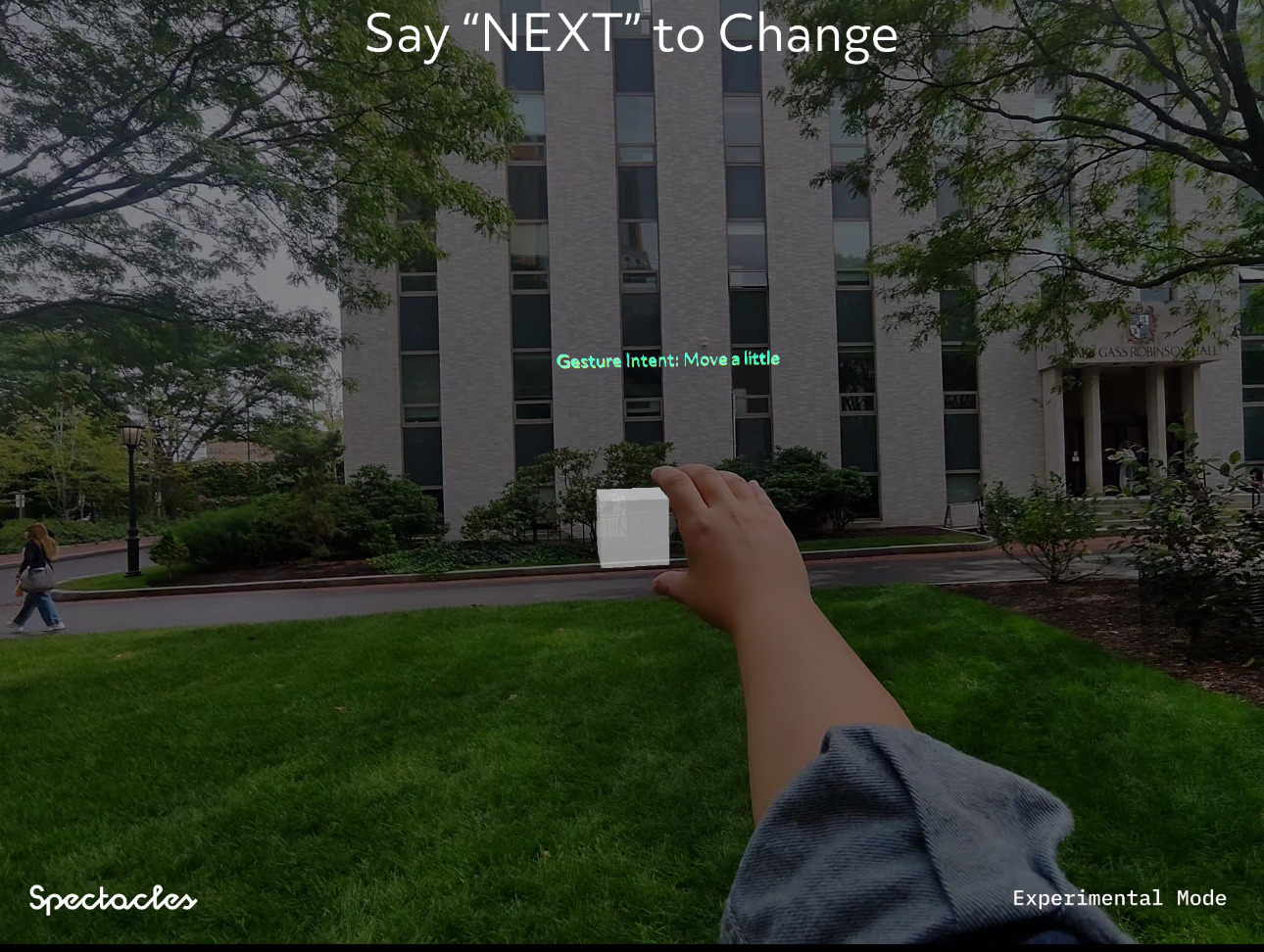}
  \end{minipage}
  \hfill
  \begin{minipage}{0.45\textwidth}
    \centering
    \includegraphics[width=\linewidth]{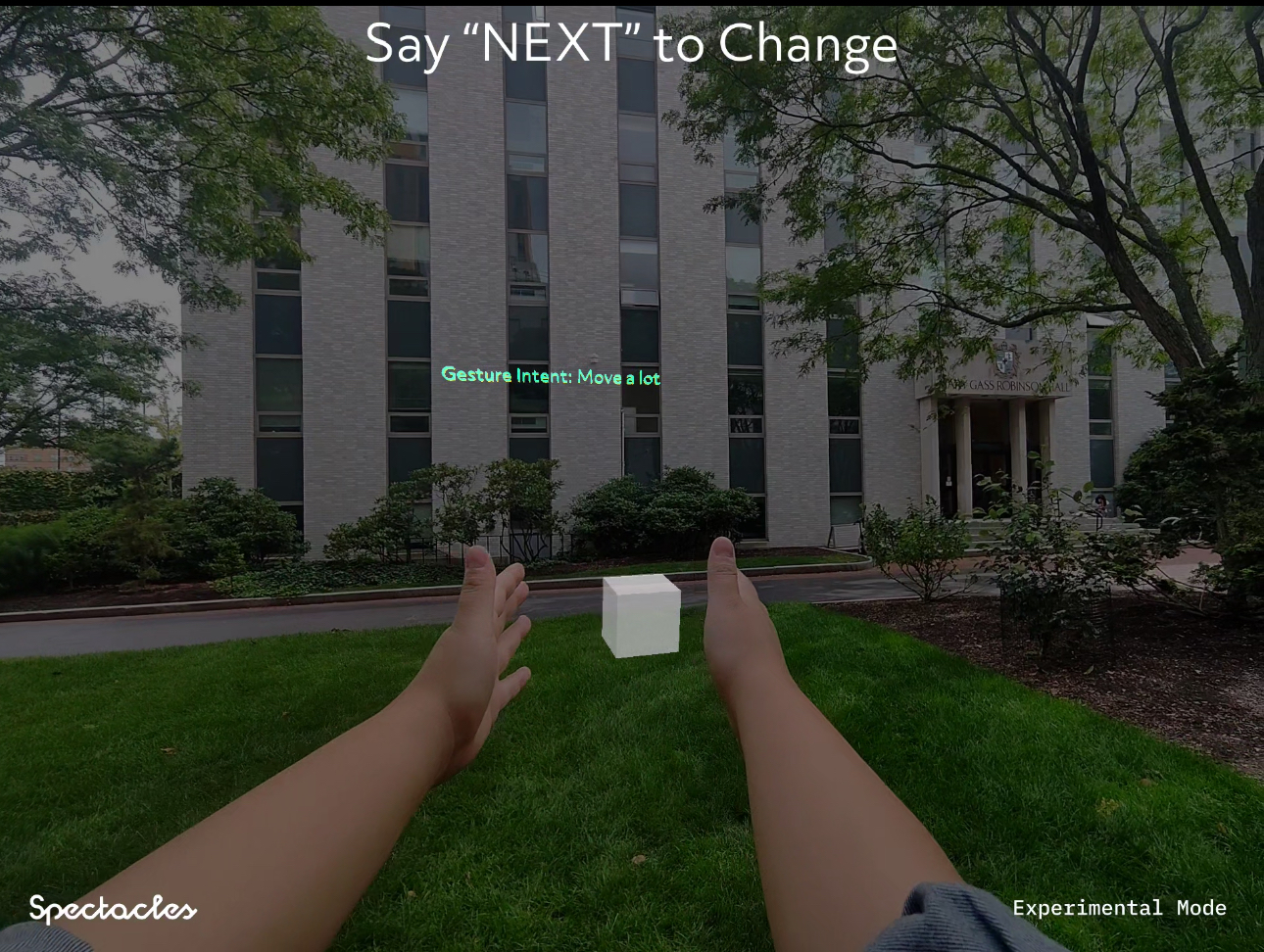}
  \end{minipage}

  \vspace{0.8em}

  \begin{minipage}{0.45\textwidth}
    \centering
    \includegraphics[width=\linewidth]{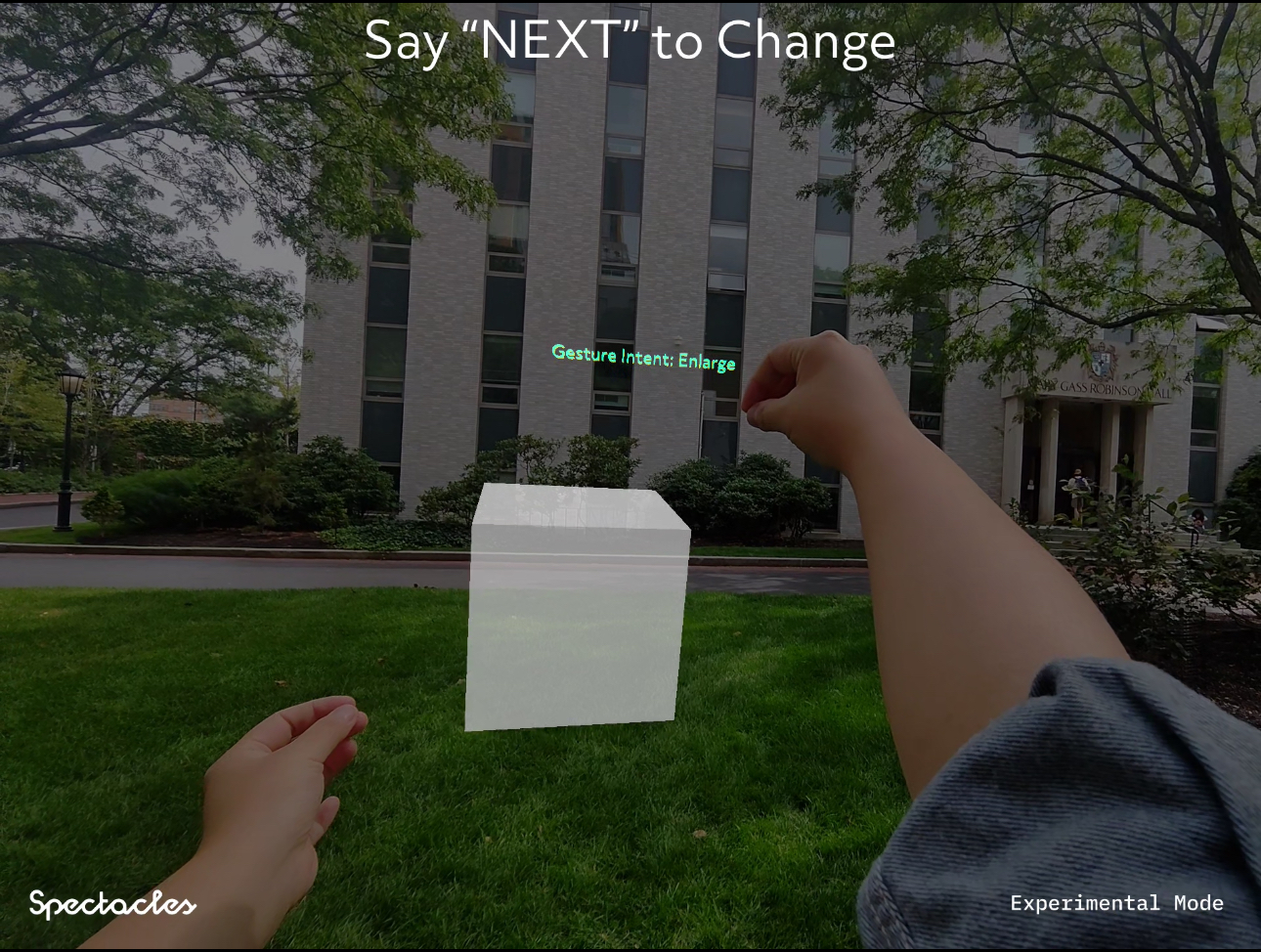}
  \end{minipage}
  \hfill
  \begin{minipage}{0.45\textwidth}
    \centering
    \includegraphics[width=\linewidth]{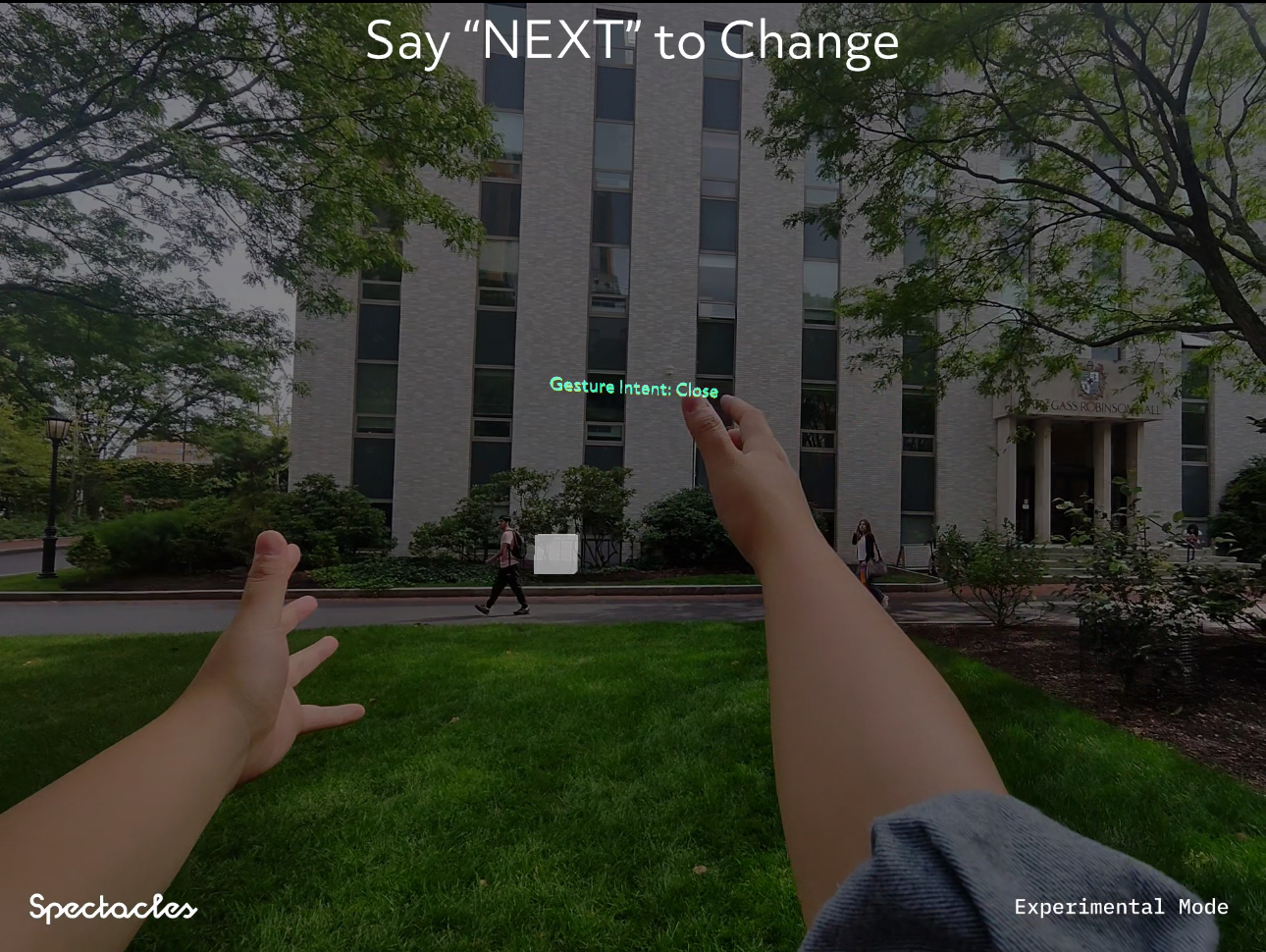}
  \end{minipage}

  \vspace{0.8em}

  \begin{minipage}{0.45\textwidth}
    \centering
    \includegraphics[width=\linewidth]{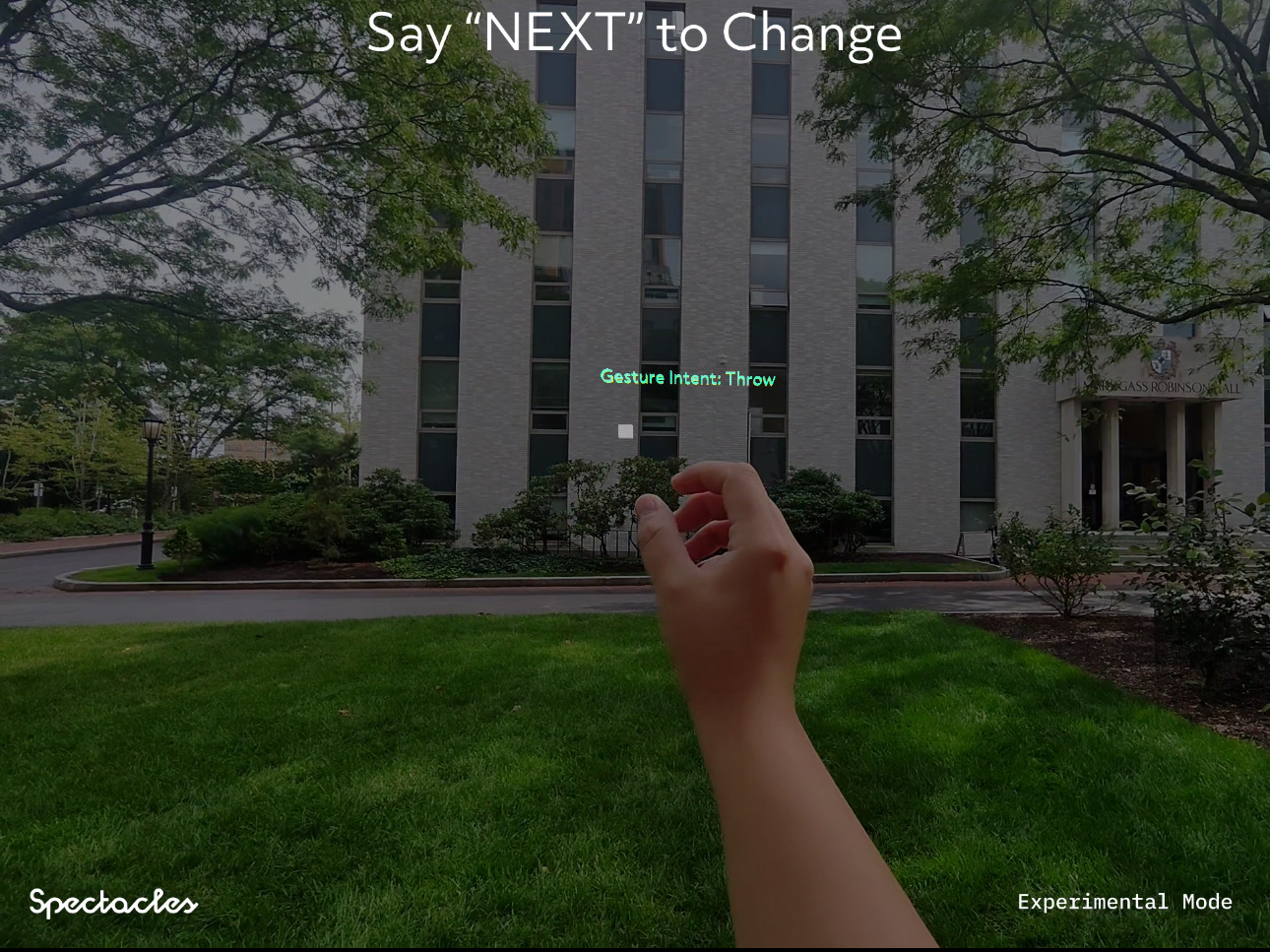}
  \end{minipage}

  \caption{Example subset of gesture cues used during elicitation.}
  \label{fig:all_cues_subset}
\end{figure*}

To support consistent elicitation across participants, we constructed a set of 15 gesture cues, each rendered directly in Spectacles as a minimal 3D animation paired with a concise intent label. These cues conveyed only the high-level action goal---not a prescribed trajectory---allowing participants to perform gestures that felt natural within their habitual motor patterns. Five examples from the full set are shown in Fig.~\ref{fig:all_cues_subset}: \emph{Move a Little}, \emph{Move a Lot}, \emph{Enlarge}, \emph{Throw}, and \emph{Close}. During Round~1, these prompts elicited users’ baseline gestures; in Round~2, the same 
cues enabled therapists to identify fatigue risks, diagnose proximal--distal load issues, and propose PT-informed substitutions. Participants advanced between cues using the voice command ``Say `NEXT' to Change,'' ensuring that cue navigation did not contaminate the gestures being elicited.

\section{Appendix B: Full Gesture Libraries}
\label{appendix:gesture-libraries}
To complement the summary themes reported in Section~4, this appendix presents the complete set of PT-authored \emph{ergonomic gesture variants} across the 15 intents examined in our elicitation study.  
Each library figure illustrates all substitutions proposed during Rounds~1--2, without frequency annotations, providing a visual catalogue of gestures that therapists considered ergonomically sustainable for everyday AR use.

\begin{figure*}[t]
  \centering
  \includegraphics[width=\linewidth]{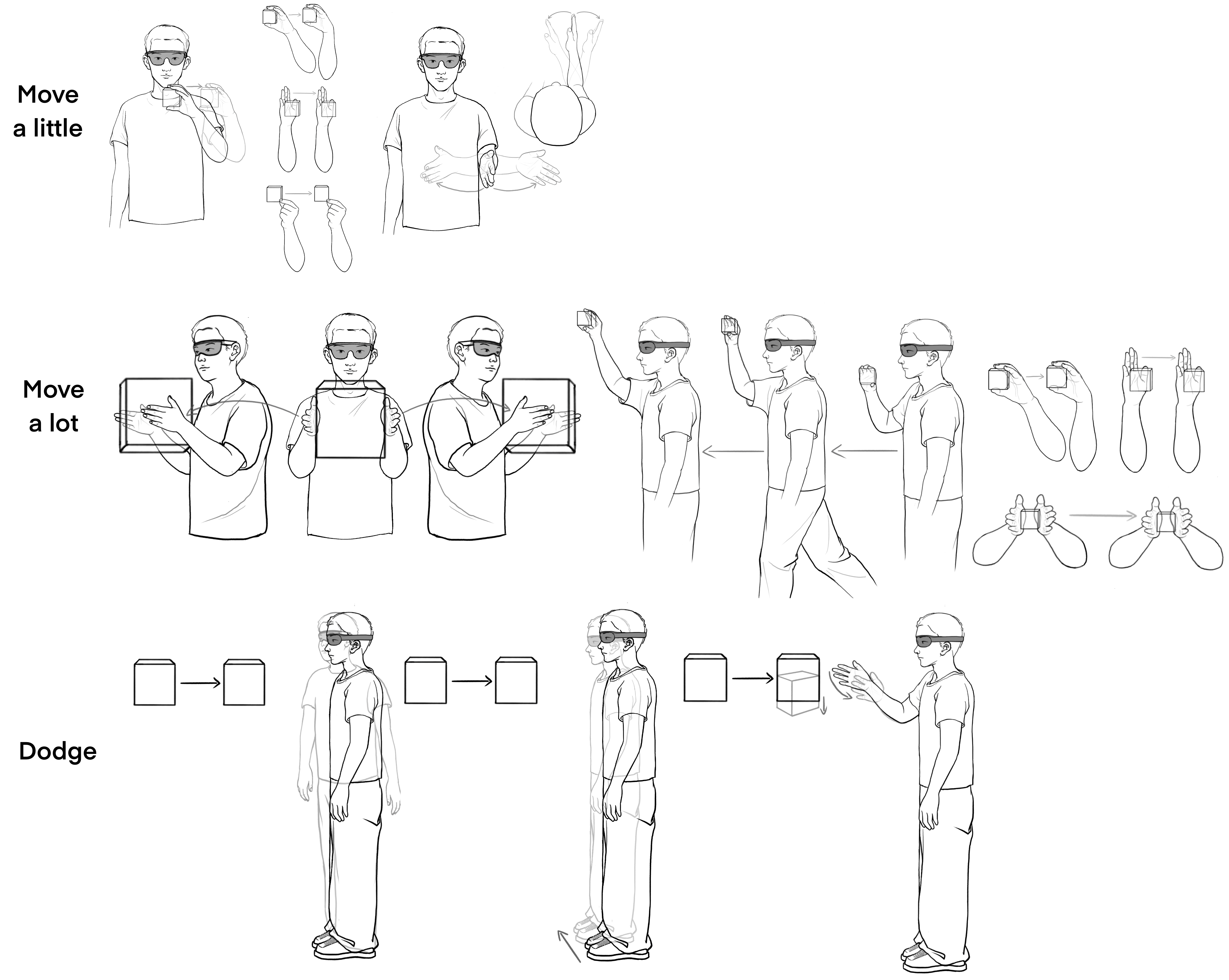}
  \caption{Ergonomic gesture variants for \textbf{Move a Little}, \textbf{Move a Lot}, \textbf{Dodge} intents.  
  Each illustration depicts a distinct ergonomic substitution surfaced during the elicitation rounds.}
  \label{fig:library03}
\end{figure*}
\begin{figure*}[t]
  \centering
  \includegraphics[width=\linewidth]{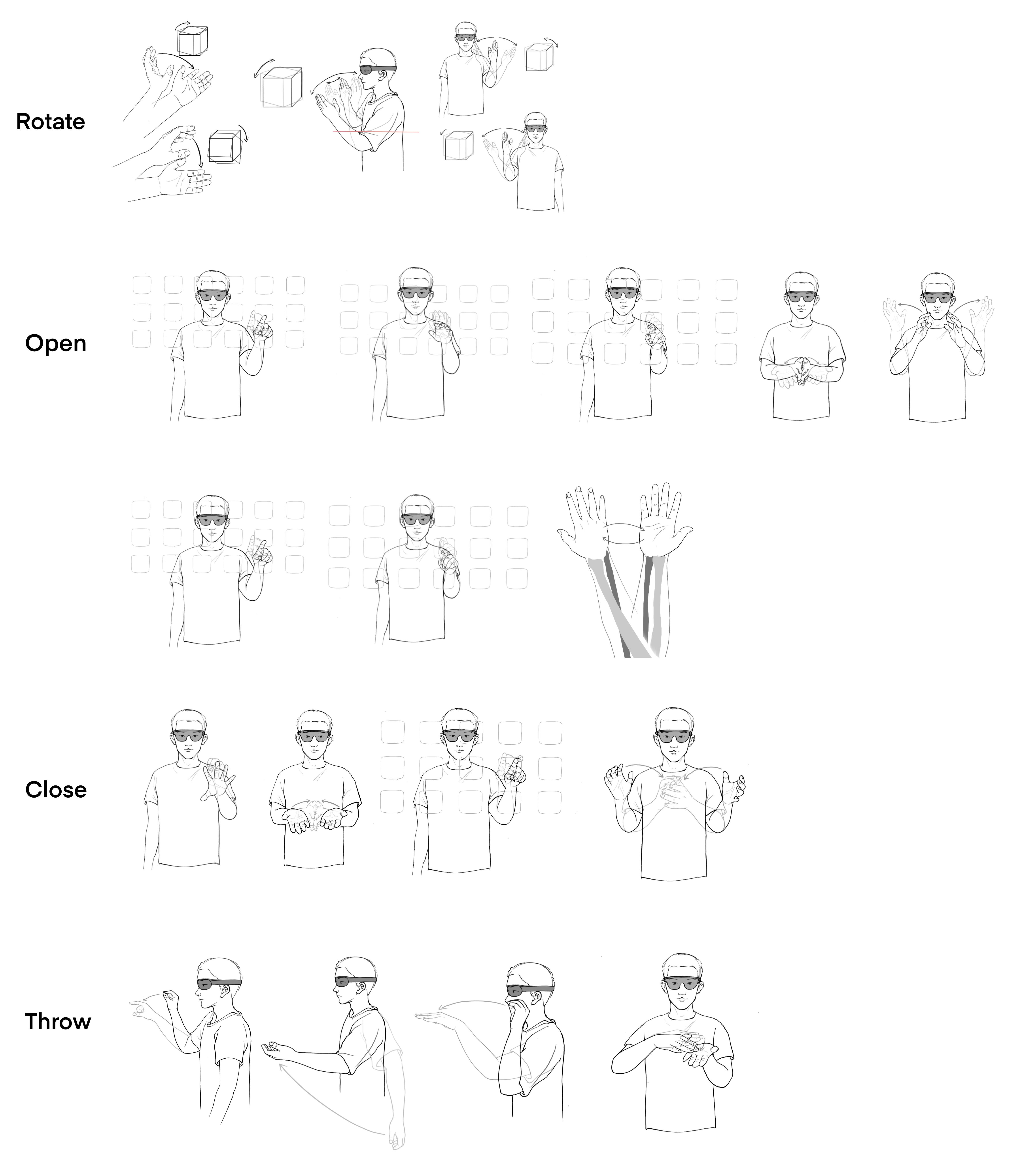}
  \caption{Ergonomic gesture variants for \textbf{Rotate}, \textbf{Open}, \textbf{Close}, and \textbf{Throw} intents.  
  Each sketch shows a PT-proposed alternative; numbers below each sketch indicate how many participants selected or refined that option.}
  \label{fig:library01}
\end{figure*}

\begin{figure*}[t]
  \centering
  \includegraphics[width=\linewidth]{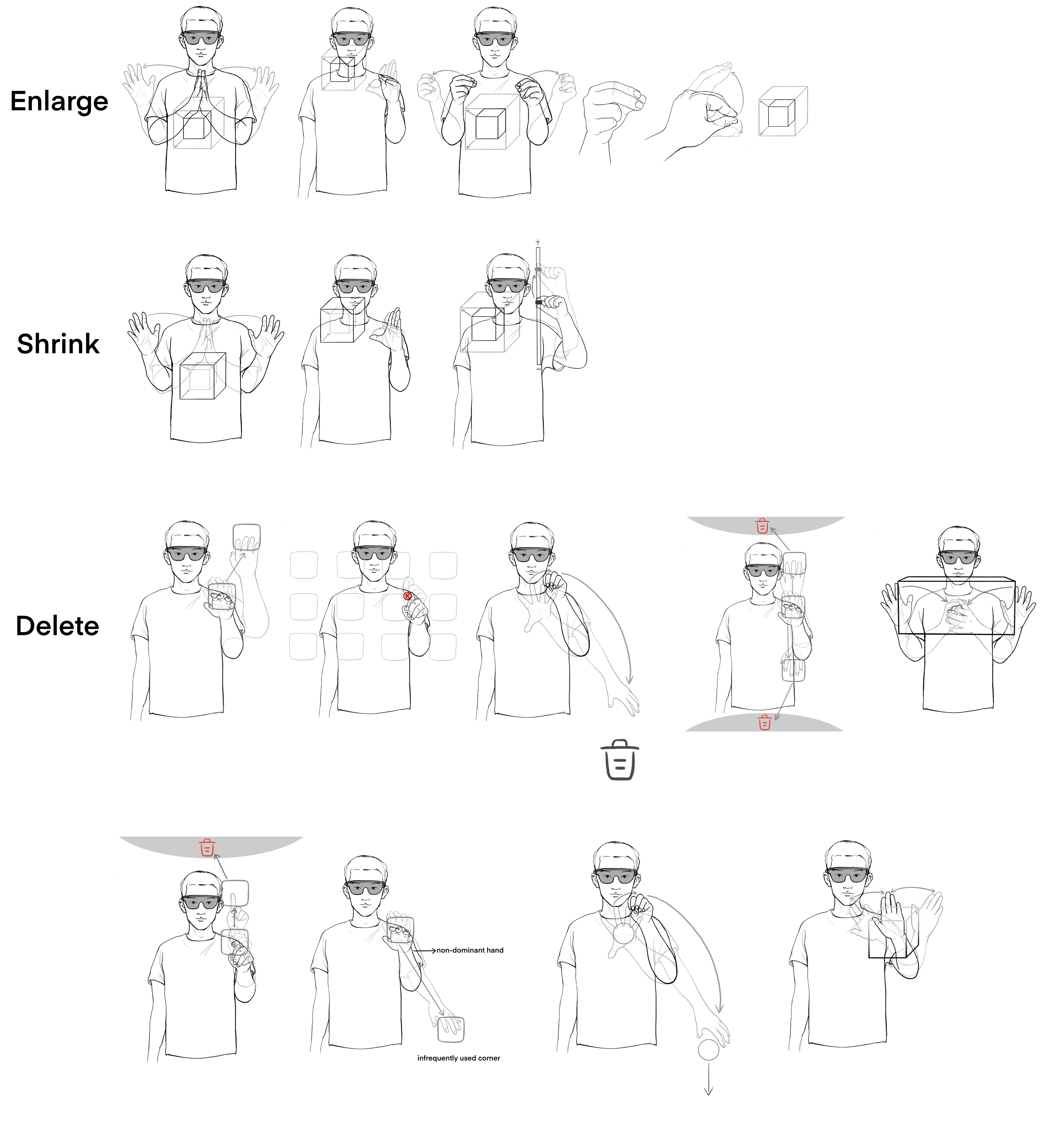}
  \caption{Ergonomic gesture variants for \textbf{Enlarge}, \textbf{Shrink}, and \textbf{Delete} intents.  
  Frequencies report how often each alternative was proposed across the ten PTs.}
  \label{fig:library02}
\end{figure*}

\begin{figure*}[t]
  \centering
  \includegraphics[width=\linewidth]{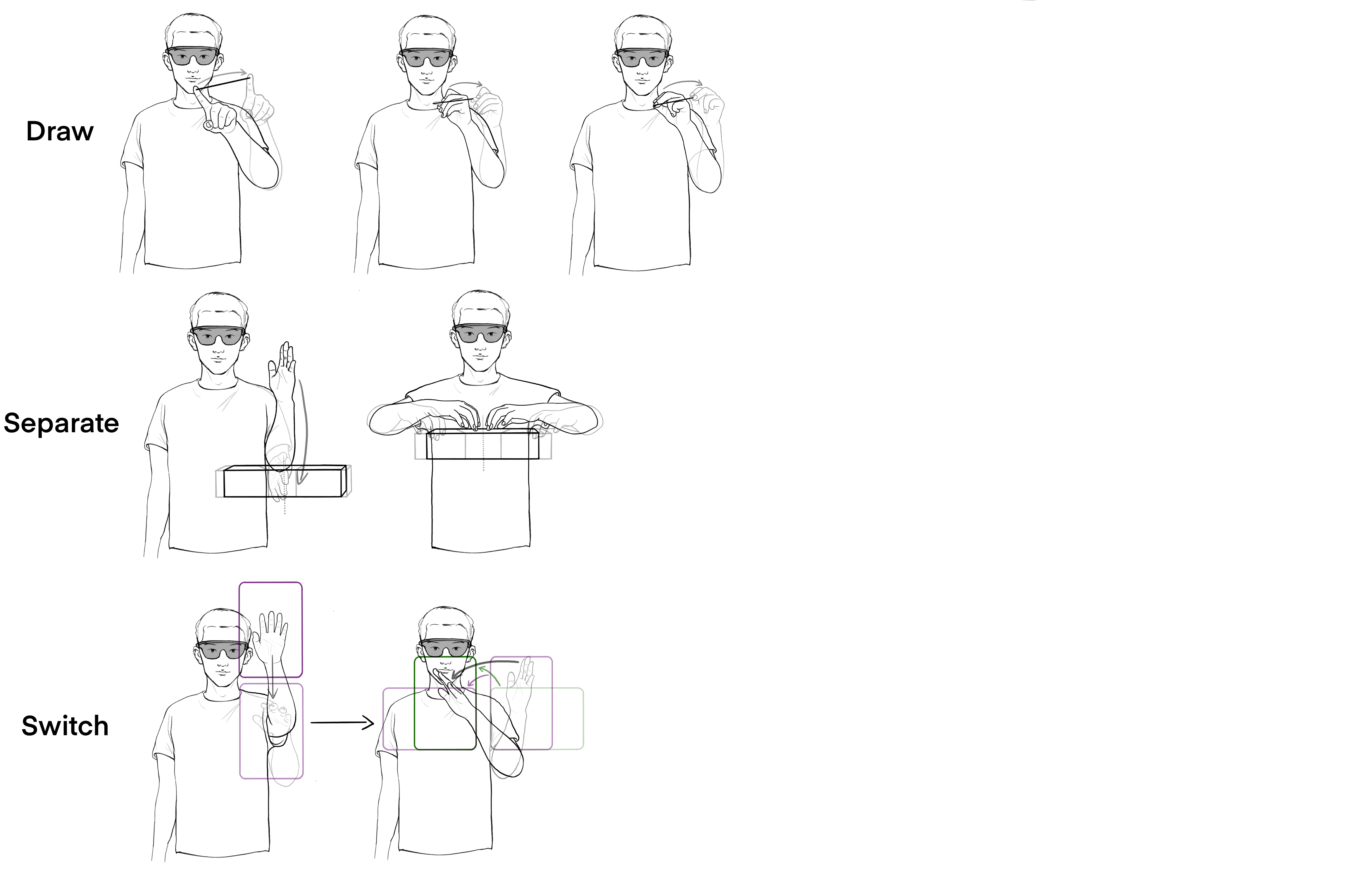}
  \caption{Ergonomic gesture variants for \textbf{Draw}, \textbf{Separate}, and \textbf{Switch} intents.  
  Each illustration depicts a distinct ergonomic substitution surfaced during the elicitation rounds.}
  \label{fig:library04}
\end{figure*}

\clearpage
\onecolumn
\section{Appendix C: Context–Motor Distributions}
\label{appendix:context-heatmap}

\begin{figure}[h]
  \centering
  \includegraphics[width=0.95\columnwidth]{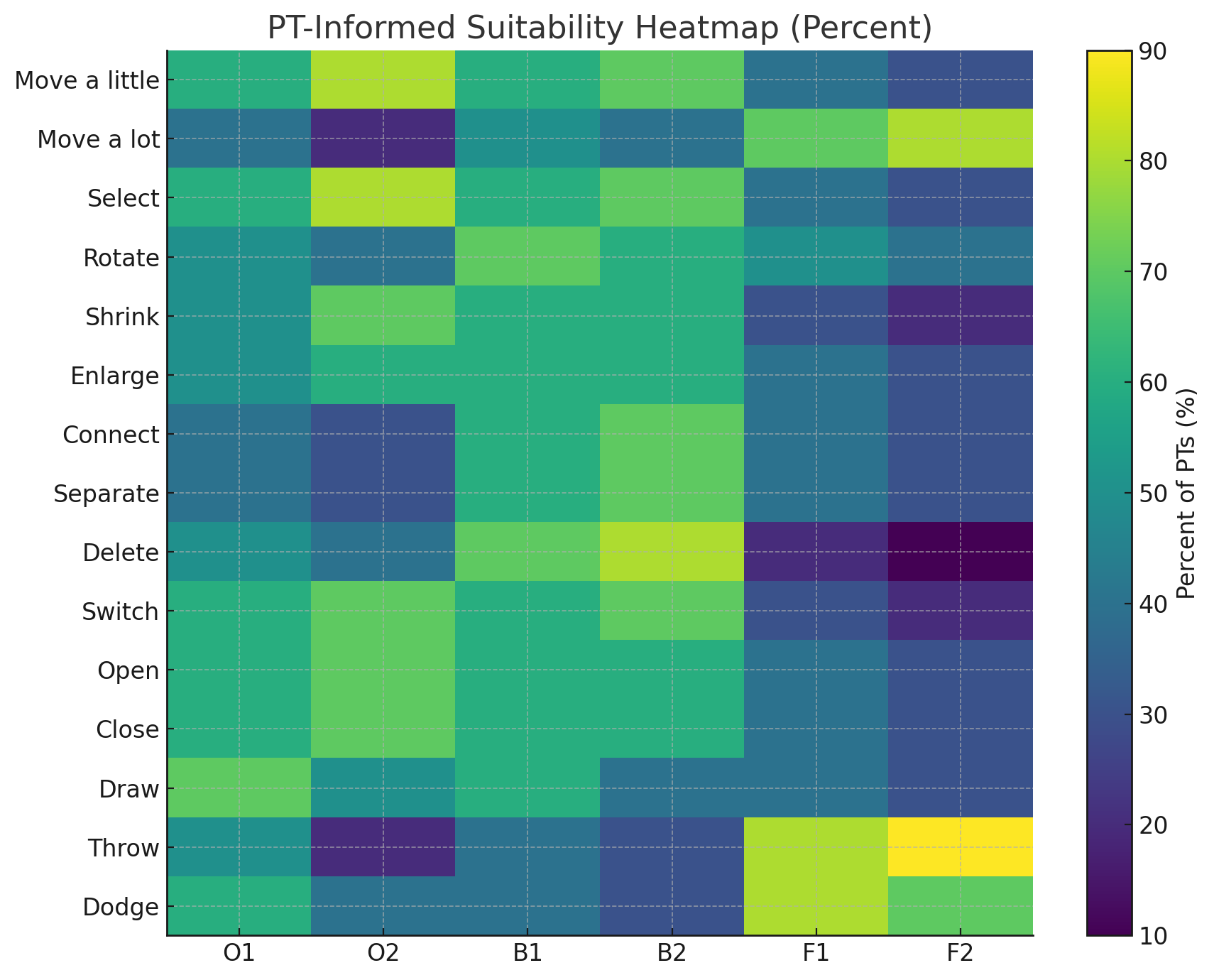}
  \caption{Full distributions of 15 gesture intents across six dramaturgical contexts.
  Lighter shading indicates stronger consensus among PTs about where a gesture is most
  appropriate. The heatmap reveals per-intent variability that underlies the broader 
  front/back/off-stage gradient.}
  \label{fig:context-motor-heatmap}
  \Description{A heatmap showing 15 gesture intents (rows) and six dramaturgical contexts
  (columns). Lighter cells indicate higher PT agreement.}
\end{figure}

To complement the macro gradient described in the main text, 
Figure~\ref{fig:context-motor-heatmap} presents the full distribution
of \emph{all 15 gesture intents} across the six dramaturgical contexts 
(O1--O2, B1--B2, F1--F2). 
Lighter cells indicate higher agreement among therapists during the 
Stage-Aware Card Sorting task. 
This figure shows how some gestures (e.g., \emph{Throw}) were strongly 
frontstage, while others (e.g., \emph{Move a little}) remained off-stage 
even in public settings. 
The heatmap also highlights heterogeneity within \emph{Amusement} (O1), 
which bifurcates into quiet fine-motor play and expansive gross-motor 
fitness-like gestures.

The heatmap provides a reusable reference for designers who wish to 
understand not only the general stage gradient but also the 
\emph{micro-variability} of each gesture intent. 
This supports a layered approach to gesture design: beginning with the 
broader ergonomic guidance outlined in the Everyday-AR Golden Ergonomic Canvas and 
refining gesture variants according to the contextual likelihoods 
captured here.

\clearpage

\section {Appendix D: Deprecated Intermediate Representation: 
Eight-Dimension Ergonomic Table}
\label{app:ergonomic-table}
During early synthesis, we explored a dimensional approach to 
organizing therapist reasoning, resulting in the eight-dimension 
table reproduced in Table~\ref{tab:golden_canvas_appendix}. 
This representation was useful for structuring preliminary 
observations, but ultimately proved too granular and 
checklist-oriented for the everyday-AR context. 
\begin{table*}[h]
\setlength{\tabcolsep}{2pt}
\small
\centering
\caption{Deprecated intermediate construct: early eight-dimension 
representation of therapist reasoning. Retained for transparency; 
dimension labels now aligned with the unified Everyday-AR Golden Ergonomic Canvas 
in Section~\ref{sec:golden-canvas}.}
\label{tab:golden_canvas_appendix}
\begin{tabular}{|p{0.18\linewidth}|p{0.32\linewidth}|p{0.22\linewidth}|p{0.22\linewidth}|}
\hline
\textbf{Dimension} & \textbf{PT-Informed Rule} & \textbf{Metric} & \textbf{Example Gesture} \\
\hline
\textbf{Posture Envelope} &
Keep the default functional workspace in the chest--navel zone and within comfortable shoulder angles; bring high or distant targets down and closer rather than requiring overhead reaching. &
Shoulder elevation $<90^\circ$, elbow flexion $\approx 90^\circ$ for default tasks; warn if the hand is held above shoulder level for $>$3--5\,s. &
To enlarge an object positioned high in the scene, first drag it down into the chest-height envelope, then scale. \\
\hline
\textbf{Range-of-Motion Band} &
Encode task magnitude as small/medium/large movement bands; high-frequency tasks stay in the small band, while large excursions are reserved for occasional actions or explicit training. &
Small movements $<15$\,cm; large movements $>30$\,cm in arc or translation. &
“Move a little’’ = short push or tilt; “move a lot’’ = wide sweep across the field of view. \\
\hline
\textbf{Joint Load \& Fatigue} &
For frequent or long-duration actions, minimise excursion and allow anti-fatigue substitutions; track cumulative repetitions and static holds to prevent overuse. &
$>$15 repetitions/min or $>$10 consecutive large-amplitude repetitions triggers an alternate form; sustained elevation above shoulder level for $>$3--5\,s raises a fatigue flag. &
Enlarge or shrink with a three-finger pinch or brief bimanual scaling instead of repeated large-arm reaches. \\
\hline
\textbf{Stability \& Support} &
Use the forearm, elbow, or torso as an anchor when precision is required; avoid prolonged free-air spirals and sustained wrist torque without proximal support. &
Ratio of wrist to shoulder angular velocity; proportion of task time with some segment supported or anchored. &
Rotate a virtual knob with a small anchored forearm arc instead of a large, unsupported finger-traced circle in mid-air. \\
\hline
\textbf{Visual Occlusion} &
Keep critical targets and the real floor within a comfortable gaze window; avoid placing virtual panels or effects that block the view of feet, obstacles, or the therapist during movement. &
Percentage of task time with targets inside a central $20^\circ$--$30^\circ$ visual cone; duration for which the ground plane is occluded while stepping or walking. &
When the user looks down to step, automatically lower or fade floating panels so they do not cover the floor or real-world obstacles. \\
\hline
\textbf{Social Visibility} &
Provide stage-aware variants: subtle micro-gestures for public or crowded spaces, broader whole-body gestures for private or clinic use; flag gestures that may appear risky or inappropriate. &
Stage toggle; stance (BOS) flags for unstable or large gestures; risk tags for throw-like or aggressive motions. &
Throw: soft wrist flick on a subway platform; wide overhand arc at home or in the clinic. \\
\hline
\textbf{Cognitive Layer} &
Structure tasks into clear phases (select $\rightarrow$ confirm $\rightarrow$ execute); use dwell or press-and-hold for high-consequence actions and ensure explicit cancel paths. &
Confirmation dwell 300--500\,ms; explicit dismiss gesture available for all destructive actions. &
Delete: hold the object, then scoop and drop into a bin with a brief confirmation pause before commit. \\
\hline
\textbf{Safety Limits} &
Use bimanual gestures or side-switching for large or infrequent tasks; avoid overloading a single limb or fragile joint and respect device- or clinic-specific safety constraints. &
Mirror-switch cost $<$1\,s; balance of repetitions across sides; flags when large-amplitude or high-load gestures exceed configured per-session limits. &
Rotate a large object or confirm a system-level action using a bimanual gesture that encourages a symmetric stance and shared load. \\
\hline
\end{tabular}
\end{table*}

As our axial coding progressed, PT feedback and gesture-by-gesture 
analysis converged toward a more integrated spatial envelope 
(elbow-anchored workspace; shoulder-protected arcs) combined with 
a small set of substitution logics (joint anchoring, amplitude 
reduction, grip grammar, and stage-aware variants). 

This led to the \textit{Everyday-AR Golden Ergonomic Canvas} presented in 
Section~\ref{sec:golden-canvas}---a unified framework better aligned with PT reasoning, 
gesture sustainability, and everyday social legibility. 
We include the earlier dimensional table here for methodological 
transparency.
\clearpage
\twocolumn
\section {Appendix E: Study Protocol}
\label{app:pt-protocol}

\subsection*{Title}
Exploring Physical Therapists' Perspectives on AR Gesture Design Through Embodied Play

\subsection*{Round 1: Intuitive Execution (Embodied Discovery)}

\textbf{Goal.}
Observe physical therapists' instinctive movement choices when interacting with AR content on lightweight AR glasses, and capture their initial reasoning about body mechanics and comfort. 

\textbf{Procedure.}
\begin{enumerate}
  \item Participants wear Snap Spectacles and are briefly introduced to the on-device gesture-intent generator (15 gesture intents derived from the Spectacles corpus).
  \item For each trial, Spectacles display a 10\,s looping cube animation (e.g., rotating cube for \emph{Rotate}) with an overlaid text label (\emph{``Gesture Intent: Rotate''}). Built-in gesture shortcuts are disabled.
  \item The 15 gesture intents are presented in randomized order. Participants see a fixed instruction (\emph{``say NEXT to change''}), and the researcher advances cues from a paired phone when the participant says ``next''.
  \item For each intent, participants perform whatever gesture first comes to mind. They are instructed to:
  \begin{itemize}
    \item Verbally think aloud while performing the gesture (what they are doing, why they chose this form, how it feels).
    \item Optionally name the gesture in their own words (e.g., ``I am swiping left with my whole arm'').
  \end{itemize}
  \item No constraints are imposed on movement: participants may use arms, hands, torso, or whole-body adjustments as they wish.
  \item Data capture includes: fixed-position video of full-body movement, first-person audio/video from Spectacles, and structured observer notes.
\end{enumerate}

\subsection*{Round 2: PT-Grounded Substitution and Ergonomic Reflection}

\textbf{Goal.}
Use clinical reasoning to re-author intuitive gestures into more sustainable, joint-safe forms, emphasizing fatigue, joint load, and accessibility.

\textbf{Procedure.}
\begin{enumerate}
  \item Participants revisit the same 15 gesture intents using the identical Spectacles cue set (same randomized list, cube animations, and \emph{``say NEXT''} instruction).
  \item For each intent, the participant first recalls or briefly re-performs their Round~1 gesture, then:
  \begin{itemize}
    \item Modifies the gesture to be more ergonomic and sustainable, considering joint health, fatigue, range-of-motion limits, and accessibility for different patient groups.
    \item Thinks aloud using clinical language where possible (e.g., proximal vs.\ distal loading, shoulder vs.\ elbow recruitment, balance and base-of-support).
    \item Explains why certain intuitive forms may be risky or unsustainable and how their proposed substitution mitigates those risks.
  \end{itemize}
  \item Observers record gesture forms, articulated principles, fatigue or risk flags (e.g., distal overuse, awkward wrist torque, overhead reach), and any context-dependent substitutions mentioned spontaneously.
  \item All interactions and commentary are recorded via video, first-person capture, and notes.
\end{enumerate}

\subsection*{Round 3: Context Elicitation and Stage-Aware Card Sorting}

\textbf{Goal.}
Situate gesture designs within everyday contexts, relating ergonomic choices to social visibility and staging (front/back/off-stage; individual/public).

\textbf{Procedure.}
\begin{enumerate}
  \item Participants are asked to freely enumerate everyday contexts in which they could realistically imagine using AR glasses for each of the 15 gesture intents (e.g., cooking at home, commuting on the subway, working in the clinic or office, socializing with friends, giving a presentation).
  \item Each context is written onto a card or placed as a labeled region on a card-sorting board.
  \item Participants then:
  \begin{itemize}
    \item Place gesture intents (or representative gestures from Rounds~1--2) into these context regions, indicating where and how each gesture would likely be performed.
    \item Discuss which gestures must change across contexts (e.g., compressing amplitude in crowded transit, using more expressive forms in presentations) and why.
    \item Reflect on social visibility, audience, and setting (front-stage, back-stage, off-stage; individual vs.\ public use) when justifying these adjustments.
  \end{itemize}
  \item The researcher prompts clarification as needed (e.g., when participants mention progression potential, such as scaling range or complexity as users gain strength or confidence).
  \item The entire card-sorting exercise is video- and audio-recorded, with observers noting key rationales about ergonomics, social legibility, and context-sensitive gesture variants.
\end{enumerate}

\end{document}